\documentclass[preprint,aps]{revtex4}
\usepackage{amssymb,amsmath}
\usepackage{graphicx}
\usepackage{dcolumn}
\usepackage{bm}

\begin{document}
\preprint{ }

\title{The order, shape and critical point for the quark-gluon plasma phase transition}
\author{Ismail Zakout$^{1,2,3}$, 
Carsten Greiner$^{1}$ and J\"urgen Schaffner-Bielich$^{1}$}
\affiliation{$^{1}$ Institut f\"ur Theoretische Physik,\\
$^2$ Frankfurt Institute for Advanced Studies, J. W. Goethe 
Universit\"at, 
D-60054 Frankfurt am Main, Germany\\
$^{3}$ Jefferson Lab, Harvard University, Cambridge MA
02318, USA}


\date{\today} 
\begin{abstract}
The order, shape and critical point for the phase transition between
the hadronic matter and quark-gluon plasma are considered 
in a thermodynamical consistent approach.
The hadronic phase is taken as Van der Waals gas 
of all the known hadronic mass spectrum particles $m_H\le 2.0$ GeV 
as well as 
Hagedorn bubbles which correspond hadronic states with mass 
spectrum $m_H> 2.0$ GeV.
The density of states for Hagedorn bubbles is derived by calculating
the microcanonical ensemble for a bag of quarks and gluons 
with specific internal color-flavor symmetry. 
The mixed-grand and microcanonical ensembles 
are derived for massless and massive flavors.
We find Hagedorn bubbles are strongly suppressed 
in the dilute hadronic matter and they appear 
just below the line of the phase transition. 
The order of the phase transition depends 
on Hagedorn bubble's internal color-flavor structure 
and the volume fluctuation as well.
On the other hand, the highly compressed hadronic matter undergoes 
a smooth phase transition from the gas of known mass spectrum hadrons 
to another one dominated by Hagedorn bubbles with specific internal color-flavor 
structure before the phase transition to quark-gluon plasma takes place at last.
The phase transition is found a first order for the intermediate 
and large chemical potentials. 
The existence of the tri-critical point depends on the modification 
of the bubble's internal structure specified by a phenomenological 
parameter $\gamma\propto\mu_B$ in the medium. 
\end{abstract}
\maketitle


\section{Introduction}

The thermodynamical description of the strongly interacting hadronic matter gas
can be approximated by a free gas with a modified level density.  
This new level density is given by the statistical bootstrap
equation and its solution is the asymptotic mass spectral function for 
hadronic fireballs with mass exceeding 2 GeV.  
A solution to this equation exists only for some range 
of parameters~\cite{Hagedorn65a}.  However, the
bootstrap model with internal symmetry~\cite{Redlich79a} of the fireball
provides subsidiary variables and allows for new types of the phase
transition~\cite{Antoniou05a,Kapoyannis98a,Kapoyannis98b}.  
The hadronic fireballs are bags of confined quark and gluon components 
and stand for highly excited exotic hadrons; we denote them in the following 
as Hagedorn bubbles. 
Furthermore, these Hagedorn states have received much attention 
to understand the phase transition 
in terms of the AdS black hole duality~\cite{Aharony03,Aharony05} 
and the gauge field~\cite{Dumitru04,Dumitru05}.

Quarks and gluons are confined due to color confinement while the low-lying
hadronic mass spectrum is generated by the broken chiral
symmetry.  The hadron's constituent quarks are massive even for light flavors.
However, the constituent quark mass decreases slowly with respect to
temperature and then drops quickly to its current mass at the critical
temperature.  It also decreases with respect to the baryonic
chemical potential.
The decreasing of u,d- quark's constituent mass near the phase
transition is not fully understood.  
Furthermore, the internal color symmetry of the bound state
remains to be that of a color singlet even for finite temperature and
chemical potential due to color confinement.  
When the temperature reaches the critical one, color is expected to be liberated 
and chiral symmetry is restored.  
QCD predicts a phase transition from the hadronic
gas phase to a deconfined quark-gluon plasma phase.  
The order of the phase
transition for hot and dense hadronic matter, however, remains unclear whether it is first,
second, higher order, or just a rapid but smooth crossover.  The internal
structure of the hadronic quark and gluon bag remains 
in a total color singlet state.  
In more realistic calculations, the internal structure is
imposed in the partition function for Hagedorn bubbles~\cite{Muller1982a}.  
In the standard MIT bag model, the quarks and gluons are confined 
within a sharp surface, which represent a boundary 
between different media: a perturbative vacuum inside the bag 
where the quarks and gluon can be approximated as an ideal gas, 
due to the asymptotic freedom of QCD, and
a nonperturbative vacuum outside the bag where free quarks 
do not exist due to confinement.

The hadronic phase consists of the whole hadronic mass spectrum
including resonances of all the known particles.  
Hard core repulsive forces can be represented by an excluded volume.  
Effects from strong interactions are included by adding a free gas 
of Hagedorn bubbles which are bags of constituent quarks and gluons with 
specific internal color-flavor structure.  
Despite of complexity of the internal color-flavor structure, 
Hagedorn bubbles remain to be in an overall color singlet state.  
%
With increasing baryonic density at low temperature, the bubble size grows 
but retains its own internal color symmetry.  
This means that volume fluctuation is expected to be suppressed 
whenever the bags start to overlap with each other 
for a large chemical potential.
When the temperature increases, the surface 
is smeared out until finally the bubbles dissociate 
at the critical temperature.  
Therefore, it is expected the bubble volume fluctuation increases
when the temperature increases.

The (grand) canonical ensemble and its Laplace transform to the micro-canonical
ensemble for gluonic bags or glueballs was derived by
Kapusta~\cite{Kapusta81a} without imposing any color constraint.  
The internal symmetry constraint was originally introduced 
for the statistical bootstrap
model~\cite{Redlich79a}.  
The bootstrap density of states can be derived from the
MIT bag model.  In the hadronic phase, the highly excited fireballs derived
from the bootstrap equation are Hagedorn bubbles (e.g. gas of bags) of 
confined quarks and gluons in a color singlet state or colorless charge.

%
%
The colorless bubble is usually less restricted than the color 
singlet state. 
The color charges are set to zero for the colorless state 
$n_{\mbox{C}i}=
\frac{\partial \Omega_V}{\partial i\phi_{\mbox{C}i}}=0$,
where $\Omega_V$ is the bubble grand potential density 
and $\phi_{\mbox{C}i}$ are the color parameters in 
the $SU(N_c)$ representation.
In the less strict constraint 
the color charges $n_3$ and $n_8$ 
are set to zero for the gas of bags
instead for any individual bag.
The asymptotic approximation for the colorless charge 
is given by $i\phi_{\mbox{C}i}\rightarrow 0$.
Hence the colorless bubble is usually approximated to an ideal gas of 
quarks and gluons with the conserved color charge
$i\phi_{\mbox{C}i}\rightarrow \mu_{\mbox{C}i}/T\sim 0$.
On the other hand, the color singlet state 
is the ground symmetry projection for the $SU(N_c)$ representation.
This resemblance between the color singlet and colorless states
requires to search for the color saddle points around  
$\frac{\phi_{\mbox{C}i}}{\pi}\ll 1$ and to avoid 
the Cauchy integral which is used extensively 
in the literature in the large $N_c$ limit~\cite{Witten80}. 
%
%

Gorenstein {\em et. al.}~\cite{Gorenstein83a,Gorenstein82a,Gorenstein83b} 
have studied the gas of bags which correspond Hagedorn states using
the isobaric partition function.  
They measured the volume fluctuation 
for the hadronic bags by differentiating the micro-canonical ensemble 
with respect to the fixed bag volume.  
The micro-canonical ensemble only measures the mass
spectral density and does not include the actual volume fluctuation.  
The real deconfinement phase transition from the hadronic phase to the 
color quark and gluon bags did not take place in their approach.
The authors of Refs.~\cite{Gorenstein83a,Gorenstein82a,Gorenstein83b}
argue that the phase transition is only possible when 
Hagedorn bubble expands rapidly to form a big color singlet quark-gluon 
droplet which occupies the entire space.  
Recently, they have pointed out that the phenomenology 
of Hagedorn bubble internal symmetry decides the order of 
the phase transition for low chemical potential 
and high temperature~\cite{Gorenstein05rc}.
Unfortunately their model fails to predict the phase transition for 
deconfined quark-gluon plasma 
but instead a new phase of matter appears which retains 
the internal symmetry of Hagedorn bubble. 
They also were unable to explain how the order of the phase transition 
can be switched from a lower one to higher ones.
On the other hand, Auberson {\em et. al.}~\cite{Auberson86b} 
have shown that the phase transition to a deconfined QGP 
is granted when asymptotic volume fluctuation is taken into account 
correctly.  
They emphasize that the micro-canonical ensemble measures 
only the mass spectrum for a specific bag volume, 
which is not the actual volume fluctuation for the quark and gluon bag.  
Furthermore, it is well known that the bootstrap density of state 
can be obtained when the bag volume is fixed to $v=m/4B$.
Auberson {\em et. al.}~\cite{Auberson86b} have approximated 
the volume fluctuation by allowing the bag mass
variation with respect to the volume up to the second order 
in the distribution function.  
Furthermore, they have argued that by relaxing the internal symmetry
constraint for the color singlet state 
the resultant Gaussian-like volume fluctuation leads 
to a second order deconfined phase transition.  
Hence, the color singlet constraint imposed on the bag states 
is not a critical for the phase transition's existence.  
Therefore, the appearance of the deconfined phase transition depends 
essentially on the bubbles' volume fluctuations beside their internal 
structure constraints. 

Hagedorn bubbles likely appear for the highly compressed hadronic matter.
These Hagedorn bubbles coalesce and form hadronic bubbles foam.
This state of matter is not a deconfined phase. 
However, when this highly compressed foam of Hagedorn bubbles is heated,
the hadronic foam undergoes a smooth phase transition 
and forms a big quark-gluon droplet as the surfaces between bubbles 
dissolve.
%
At low baryonic density and high temperature, the situation is rather different.  
The formation of Hagedorn bubbles is unlikely to take place 
in the hadronic phase and whenever they appear because of the thermal 
fluctuations and they shall be suppressed by the gas pressure
of the external hadronic mass spectrum particles
in particular those relatively large Hagedorn bubbles.
However, when the temperature reaches the critical value, some explosive
hadronic bubbles can appear in the system. 
Whenever the internal pressure of the thermal fluctuated bubble approaches 
the external one for the gas of the mass spectrum hadrons, 
these bubbles grow up and expand rapidly forming the deconfined 
quark-gluon plasma at last.

The outline of the present paper is as follows. 
In section II, we derive the canonical ensemble for a quark-gluon bubble 
with internal color-flavor symmetry with massless and massive flavors.  
The micro-canonical ensemble is derived from the inverse Laplace transform 
of the grand-canonical one.  
In section III, we review the isobaric partition function for the gas of bags 
with excluded volume. We also summarize the conditions of the phase transition
in the isobaric partition function construction for the gas of bags. 
The small excluded volume for baryons and mesons and large excluded volume 
for Hagedorn bubbles are introduced in section IV.  
In section V, we present two models for the volume fluctuation 
in the isobaric partition function.  
The order and shape of the phase transition to quark-gluon plasma 
is analyzed in section VI.
Then we present our scenario
for the phase transition to quark-gluon droplets or plasma 
for low and intermediate chemical potentials and to foam of 
Hagedorn bubbles for large baryonic chemical potentials in section VII.
Finally, we give our conclusions in section VIII.

\section{Grand canonical and microcanonical ensembles for a quark and gluon bubble 
with internal color-flavor symmetry and massive constituent quarks}

In this section, we derive the (mixed-) grand microcanonical ensemble 
for an ideal gas of quarks and gluons which are confined inside 
a bubble of specific size 
which carry specific internal color-flavor symmetry. 
Contrary to previous 
calculations~\cite{Auberson86a,Gorenstein83b,Elze83a,Elze84a,Lang1981a,Skagerstam1984a,Tounsi98a}, 
we perform the derivation also for massive constituent quarks.
The microcanonical ensemble is calculated by taking 
the inverse Laplace transform of the mixed grand canonical ensemble. 
The volume fluctuation is introduced by going beyond 
the standard MIT bag model description with a sharp surface 
to bag models with an extended surface and bubbles with volume fluctuations. 
We assume that the volume fluctuation is measured 
by the bubble volume distribution function.
In this context, we propose that the volume distribution function is determined 
by a smeared volume and mass relation relaxing 
the corresponding constraint of the the standard bag model. 

\subsection{(Mixed-) Grand canonical ensemble}
The grand canonical ensemble for an ideal gas of quarks and gluons confined 
in a specific volume can be calculated 
by~\cite{Kapusta81a,Kapusta82b,Auberson86a,Gorenstein83b,Elze83a,Elze84a,Lang1981a,Skagerstam1984a,Tounsi98a}
\begin{eqnarray}
{\cal Z}_V(\beta,\vartheta)&=&\mbox{Tr} \hat{\cal P}_a 
e^{-\beta\hat{H}+i\vartheta\hat{N}},
\end{eqnarray}
where $\hat{H}$ is the Hamiltonian of the physical system, 
$\hat{N}$ 
is the conserved quantum number operator. 
The imaginary chemical potential $\vartheta$ can be written in term of fugacity 
$i\vartheta=\log\lambda$ in the mixed grand canonical ensemble where $\lambda=e^{\mu/T}$.
The operator $\hat{\cal P}_a$ selects those configurations 
that are allowed by specific constraints 
due to the internal 
symmetries~\cite{Redlich79a,Turko1981a,Auberson86a,Gorenstein83b,Elze83a}
of the system
\begin{eqnarray}
\hat{\cal P}_a\equiv \hat{\cal P}_j 
\cdot \hat{\cal P}_{\hat{\vec{P}}=0} \cdot \hat{\cal P}_{\vartheta}\cdots,
\label{project1}
\end{eqnarray}
where $\hat{\cal P}_j$ project the color component such as 
the color singlet state, $\hat{\cal P}_{\hat{\vec{P}}=0}$ transforms 
the system into the center of mass 
frame~\cite{Kapusta82b} 
while $\hat{\cal P}_{\vartheta}$ conserves the total number of particles 
in the fireball.
In the present model, we consider a quark-gluon gas with 
specific internal symmetry by introducing three projectors.
The color projector of the subspace of all states that transform under
the representation $j$ of the irreducible representation ${\cal G}(g)$
of a compact Lie group $\mbox{SU}(N_c)$ reads
\begin{eqnarray} 
\hat{\cal P}_j=
d_j\int_{SU({N_c})} d\mu(g) \chi_j(g), 
\label{project_j}
\end{eqnarray}
where 
\begin{eqnarray}
d\mu(g)=
\frac{1}{{N_c}!}
\left(\frac{1}{2\pi}\right)^{N_c}
\prod_{n>m}^{N_c}\left(2\sin\left(\frac{\theta_n-\theta_m}{2}\right)\right)^2 
2\pi\delta\left(\sum^{N_c}_{n=1} \theta_n\right)
d\theta_1d\theta_2\cdots d\theta_{N_c},
\label{Haar_ex}
\end{eqnarray}
is the normalized Haar measure on ${\cal G}(g)$.
However, for small color angles $\frac{\theta_n}{\pi}\ll 1$, the Haar measure 
can be approximated by
\begin{eqnarray}
d\mu(g)\approx
\frac{1}{N!}\left(\frac{1}{2\pi}\right)^{N_c}
\left(\prod^{N_c}_{n>m}\left(\theta_n-\theta_m\right)^2\right)
2\pi\delta\left(\sum^{N_c}_{n=1} \theta_n\right)
d\theta_1d\theta_2\cdots d\theta_{N_c}.
\label{Haar1}
\end{eqnarray} 
It is also convenience to introduce the set $\phi_i$ where
$\theta_i=\phi_i$ for $i=1,\cdots,N_c-1$ and 
$\phi_{N_c}=-\sum^{N_c-1}_{k=1}\phi_k$. 
We have introduced the set $\{\phi_i\}, i=1,\dots, N_c-1$ for the group representation $SU(N_c)$ 
while the set $\{\theta_i\}, i=1,\dots, N_c$ is introduced for the group $U(N_c)$.
In context of set $\phi_i$ the
Haar measure reads
\begin{eqnarray}
d\mu(g)\approx
\frac{1}{N_c!}\left(\frac{1}{2\pi}\right)^{N_c-1}
\left(\prod^{N_c}_{n>m}\left(\phi_n-\phi_m\right)^2\right)
d\phi_1d\phi_2\cdots d\phi_{N_c-1}.
\label{Haar1b}
\end{eqnarray}
The quantities $d_j$ and $\chi_j(g)$ are the dimension 
and the orthonormal basis $j$ of the representation, respectively.
The basis $\chi_{\mbox{singlet}}(g)=1$ with $d_j=1$ projects on 
the color singlet state for the quark-gluon bubble.
The zero momentum projector for the system enclosed in a volume ${\left<V\right>}$ reads
\begin{eqnarray}
\hat{\cal P}_{\hat{\vec{P}}=0} &=&
\int_{\left<V\right>} \frac{d^3 R}{\left<V\right>}
e^{i\vec{P}_G\cdot \vec{R}}
e^{i\vec{P}_Q\cdot \vec{R}}
e^{i\vec{P}_{\overline{Q}}\cdot \vec{R}},
\nonumber\\ 
&=&\int_{{\left<V\right>}/\beta^3} \frac{d^3
r}{{\left<V\right>}/\beta^3} 
e^{i\beta\vec{P}_G\cdot \vec{r}}
e^{i\beta\vec{P}_Q\cdot \vec{r}} 
e^{i\beta\vec{P}_{\overline{Q}}\cdot \vec{r}},
\label{project_mom}
\end{eqnarray}
where we have replaced $R=\beta r$.
This constraint means that we are working in the center of mass frame 
of the quark and gluon bubble~\cite{Kapusta82b}. 
However this projection correlates the quarks 
and gluons in a specific momentum configuration. 
This projection constrains the momenta of constituent quarks and gluon in a specific 
alignment 
\begin{eqnarray}
\sum_i \vec{p}_{Qi}+\sum_j \vec{p}_{\overline{Q}_j}+\sum_k \vec{p}_{Gk}=0~ \rightarrow
\int_{\left<V\right>} \frac{d^3 R}{\left<V\right>}
e^{i\left[
\vec{P}_G\cdot \vec{R}+\vec{P}_Q\cdot \vec{R}+\vec{P}_{\overline{Q}}\cdot \vec{R}\right]}.
\end{eqnarray} 
The thermal bath breaks the Lorenz invariance. 
Therefore, it is expected that the high temperature breaks badly this projection.
Breaking the momentum projection modifies the density of states significantly. 
We retain this projection in our calculations although
it is not a realistic one for the dense and hot medium 
as far it is related in somehow to flavor correlations. 
The argument of breaking this symmetry smoothly with temperature
is the beneath of the introduction of the phenomenological modification of the
density of states. 
We relate this modification to the color-flavor correlation.
However, in order to soften the equation of state for cold quark matter 
it is also possible to impose a higher order constraint in the momentum space 
such as 
\begin{eqnarray}
\sum_i \vec{p}_{Qi}=\sum_j \vec{p}_{\overline{Q}_j}=\sum_k \vec{p}_{Gk}=0~ \rightarrow
\int_{\left<V\right>} \frac{d^3 R_G}{\left<V\right>}
\int_{\left<V\right>} \frac{d^3 R_Q}{\left<V\right>}
\int_{\left<V\right>} \frac{d^3 R_{\overline{Q}}}{\left<V\right>}
e^{i
\left[
\vec{P}_G\cdot \vec{R}_G
+\vec{P}_Q\cdot \vec{R}_Q
+\vec{P}_{\overline{Q}}\cdot \vec{R}_{\overline{Q}}\right]}.
\end{eqnarray}
On the other hand, for the crystallized quark matter, it is convenience 
to introduce the orthogonal representation $O(N)$ for the rotational symmetries
in order to soften the equation of state. 
These aspects is not considered in the present work.  

The conserved charge, such as the baryon number $N_B$, is guaranteed by
\begin{eqnarray} 
\hat{\cal P}_{\vartheta}= \int^{\pi}_{-\pi}\frac{d\vartheta}{2\pi}
e^{i\vartheta\left[(N_Q-N_{\overline{Q}})-N_B\right]}. 
\label{project_mu}
\end{eqnarray} 
The grand canonical ensemble is calculated from the mixed grand canonical ensemble
(note that as we project on certain quantum states for the bubbles, we are not dealing with the grand canonical description
in a strict sense; however, the overall dependence of the system on the chemical potential is still kept, therefore the term mixed 
g.c.e.)
\begin{eqnarray}
{\cal Z}_V(\beta,\vartheta)=\int^{\pi}_{-\pi}\frac{d\vartheta}{2\pi}
e^{-i\vartheta N_B} \tilde{Z}_V(\beta,\vartheta).
\end{eqnarray}  
The Hilbert space of the gas of quarks and gluons has the structure 
of a tensor product of the three Fock spaces for gluons, quarks and anti-quarks. 
The mixed grand canonical ensemble for an ideal gas of quarks and gluons 
becomes
\begin{eqnarray}
\tilde{Z}_V(\beta,\vartheta)=
\hat{\cal P} 
\left[\mbox{Tr}_G\hat{U}_G (g) e^{-\beta\hat{H}_G}\right]
\left[\mbox{Tr}_q\hat{U}_Q (g) e^{-\beta\hat{H}_Q} e^{i\vartheta N_q}\right] 
\left[\mbox{Tr}_{\overline{Q}}\hat{U}_{\overline{Q}}(g) 
e^{-\beta\hat{H}_{\overline{Q}}}
e^{i\vartheta N_{\overline{Q}}} \right].
\label{ensemble_Haar1}
\end{eqnarray} 
The total ensemble is given by the product of ensembles for the constituents gluons,
quarks and antiquarks. The internal structure for each species is introduced by
the representation $U_i(g)$. The notation $\mbox{Tr}_i$ represents the traces over 
the energy states and the internal structure representation.   
Here, we have taken the operator 
$\hat{\cal P}=\hat{\cal P}_{(\hat{\vec{P}}=0)}\cdot\hat{\cal P}_j$,
where $\hat{\cal P}_{\hat{\vec{P}}=0}$ selects the zero momentum and 
$\hat{\cal P}_j$ projects on specific internal color-flavor symmetry.
The imaginary chemical potentials can be written in terms
of fugacities $\lambda_i=e^{i\vartheta_i}$.
The analytical continuation to the imaginary chemical potential by applying Wick rotation
to the conserved charge Fourier parameter
\begin{eqnarray}
i\vartheta_i\rightarrow \tilde{\vartheta}_i=\frac{\mu_i}{T}
\end{eqnarray}
introduces the real chemical potentials 
for the conserved charges in the mixed grand canonical ensemble.
The fugacities for Hagedorn bubble
$\lambda=\lambda(\lambda_B,\lambda_S,\lambda_I)$
are determined from the real chemical potentials $\mu_B, \mu_S$ 
and $\mu_Q$ for baryon, strange and isospin chemical potentials, respectively,
where $\lambda_B=e^{\mu_B/T}$, $\lambda_S=e^{\mu_S/T}$ and 
$\lambda_I=e^{\mu_I/T}$. 
The constituent quarks' fugacities are determined by 
$\lambda_Q=\lambda^{1/3}_B$ and
$\lambda_s=\lambda^{1/3}_B\lambda^{-1}_S$ 
for an isospin symmetric fireball.

The gluons and quarks satisfy the Bose-Einstein and Fermi-Dirac statistics, respectively.
Furthermore, the gluons are represented by the adjoint color $SU(N_c):N^2_c-1$ representation 
while the quarks by the fundamental color $SU(N_c):N_c$ representation. 
The three traces in the Hilbert spaces read~\cite{Auberson86a}
\begin{eqnarray} 
\mbox{Tr} \hat{U}_G (g) e^{-\beta\hat{H}_G}
&=&
\exp\left\{ -\sum_{\alpha} \mbox{Tr}_{\mbox{c}} \ln\left[1-{\bf R}_{\mbox{adj}}(g)
e^{-\beta {H_{G \alpha}}} \right] \right\}, \nonumber \\ 
&=&
\exp\left\{
-\left<V\right> d_{G}\int \frac{d^{3}p}{(2\pi)^3} \mbox{Tr}_{\mbox{c}}
\ln\left[1-{\bf R}_{\mbox{adj}}(g) e^{-\beta p} \right] \right\}, 
\label{Hilbert1a}
\end{eqnarray} 
for gluons and 
\begin{eqnarray} 
\mbox{Tr}_{\mbox{c}} \hat{U}_Q (g) e^{-\beta\hat{H}_Q} e^{i\vartheta N_Q}
&=&\exp\left\{ +\sum_{\alpha} \mbox{Tr}_{\mbox{c}}
\ln \left[ 1+{\bf R}_{\mbox{fund}}(g)
e^{-\beta {\hat{H}_{Q \alpha}}} e^{+i\vartheta} \right] \right\}, 
\nonumber \\
&=&
\exp\left\{ +\left<V\right> d_{Q}\int \frac{d^{3}p}{(2\pi)^3}
\mbox{Tr}_{\mbox{c}}
\ln\left[ 1+{\bf R}_{\mbox{fund}}(g) \lambda_Q
e^{-\beta E_Q(p)} \right] \right\}, 
\label{Hilbert1b}
\end{eqnarray} 
for quarks, and, finally, 
\begin{eqnarray}
\mbox{Tr}_{\mbox{c}}
\hat{U}_{\overline{Q}} (g) e^{ -\beta\hat{H}_{\overline{Q}} }
e^{-i\vartheta N_{\overline{Q}} } &=& 
\exp\left\{ +\sum_{\alpha} \mbox{Tr}_{\mbox{c}}
\ln\left[
1+{\bf R}^*_{\mbox{fund}}(g) e^{-\beta{\hat{H}}_{\overline{Q} \alpha}}
e^{-i\vartheta} \right]\right\} 
\nonumber \\ 
&=&
\exp\left\{ +\left<V\right> d_{Q}\int
\frac{d^{3}p}{(2\pi)^3} 
\mbox{Tr}_{\mbox{c}}
\ln\left[ 1+{\bf R}^*_{\mbox{fund}}(g)
\lambda^{-1}_Q e^{-\beta E_Q(p)} \right] \right\},
\label{Hilbert1c}
\end{eqnarray}
for antiquarks where 
$E_Q(p)=\sqrt{p^2+m^2_Q}$ is the constituent quark kinetic energy and
$d_{G}=2$ and $d_{Q}=2$
are the degeneracies for the gluon polarization states 
and quark spin states, respectively.
The trace $\mbox{Tr}_{\mbox{c}}$ runs over the color index.
The matrix ${\bf R}_{\mbox{adj}}(g)$ is the adjoint color group representation for 
gluons while 
the matrices ${\bf R}_{\mbox{fund}}(g)$ and ${\bf R}^*_{\mbox{fund}}(g)$ 
are the fundamental color group representations 
for quarks and antiquarks, respectively.
 
The flavor index is suppressed in order to simplify the calculations.
However, the specific color-flavor correlations can be studied
straightforward using the present formalism.
The fundamental representation
for the color compact Lie group $\mbox{SU}(N_c)$, respectively, read
\begin{eqnarray}
{\bf R}_{\mbox{fund}}(g^k)=
\left(
\begin{array}{cccc}
e^{ik\theta_1}&\cdots&0&0 \\
\vdots&\ddots &\vdots&\vdots \\
0&0&e^{ik\theta_{N_c-1}}&0 \\
0&0&0&e^{ik\theta_{N_c}}
\end{array}
\right)=
\left(
\begin{array}{cccc}
e^{ik\phi_1}&\cdots&0&0 \\
\vdots&\ddots &\vdots&\vdots \\
0&0&e^{ik\phi_{N_c-1}}&0 \\
0&0&0&e^{-ik\sum^{N_c-1}_i\phi_i}
\end{array}
\right),
\label{group_fund}
\end{eqnarray}
where the set $\{\theta_i\}$ is suitable for the unitary representation $U(N_c)$ 
while the set $\{\phi_i\}$ is suitable for the special unitary 
representation $SU(N_c)$.  
The trace for the adjoint representation reads
\begin{eqnarray}
\mbox{Tr}_{\mbox{c}}\left[{\bf R}_{\mbox{adj}}(g^k)\right]&=&
\mbox{Tr}_{\mbox{c}}\left[{\bf R}_{\mbox{fund}}(g^k)\right]
\mbox{Tr}_{\mbox{c}}\left[{\bf R^*}_{\mbox{fund}}(g^k)\right]
-1,
\nonumber\\
&=& \sum^{N_c}_{i=1} \sum^{N_c}_{j=1} \cos k(\theta_i-\theta_j)+(N_c-1),
~~(i\neq j).
\label{group_adj}
\end{eqnarray}
The sum of states is approximated by
\begin{eqnarray}
\sum_{\alpha} \equiv \left<V\right>
d_Q\int \frac{d^{3}p}{(2\pi)^3}.
\end{eqnarray}
The surface tension
\begin{eqnarray}
-c_1 \int \frac{d^{2}p}{(2\pi)^2}\int dS_{V}
\end{eqnarray}
is significant for small bubbles and could lead
to a bubble instability under specific circumstances
while it is negligible for large bubbles.    
However, the surface effect and curvature terms can be absorbed by introducing 
the so called ``quantum volume fluctuation''.
We will treat  the quantum volume fluctuation in detail in Sec.(\ref{volumsec}).
The mixed grand canonical ensemble for a gas of bubbles in the limit of large volumes becomes
\begin{eqnarray}
\tilde{{\cal Z}}_V(\beta,\vartheta)=
d_j\int_{SU(N_c)} d\mu(g) \chi_j(g)
\int_{{\left<V\right>}/\beta^3}
\frac{d^3 r}{{\left<V\right>}/\beta^3} 
\exp\left(\mbox{Tr}_{\mbox{c}}
\left[{\bf z}_{Q\overline{Q}G}(r,g)\right]\right),
\end{eqnarray}
where
\begin{eqnarray}
{\bf z}_{Q\overline{Q}G}(r,g)&=&
{\bf z}_{Q}(r,g)
+{\bf z}_{\overline{Q}}(r,g)
+{\bf z}_{G}(r,g),
\nonumber\\
z_{Q\overline{Q}G}(r,g)&=&
\mbox{Tr}_{\mbox{c}}
\left[{\bf z}_{Q\overline{Q}G}(r,g)\right]
=
z_{Q}(r,g)+z_{\overline{Q}}(r,g)+z_{G}(r,g).
\end{eqnarray}
Each integration in the multi-integration space 
is evaluated by making use of the saddle point approximation in the steepest descent method.
Using the Taylor expansion, we expand the function 
$\mbox{Tr}_{\mbox{c}}
\left[ {\bf z}_{Q\overline{Q}G}\left(r,g(\{\theta_i\})\right)\right]$
around the extremum points for the configuration radius $r_0$ and color angles $g_0(\{\theta_i\})$,
\begin{eqnarray}
z_{Q\overline{Q}G}(r,g)\approx
z_{Q\overline{Q}G}(r_0,g_0)
&+&
\frac{1}{2}\left.\frac{\partial^2
z_{Q\overline{Q}G}(r,g)}{\partial r^2}\right|_{r_0,g_0}(r-r_0)^2 \nonumber\\
&+& \frac{1}{2}\sum^{N_c}_i
\left. 
\frac{\partial^2
z_{Q\overline{Q}G}(r,g)
}{\partial \theta_i^2}
\right|_{r_0,g_0} 
(\theta_i-{\theta_i}_0)^2.
\end{eqnarray}
and
\begin{eqnarray}
z_{Q\overline{Q}G}(r,g)\approx
z_{Q\overline{Q}G}(r_0,g_0)
&+&
\frac{1}{2}\left.\frac{\partial^2
z_{Q\overline{Q}G}(r,g)}{\partial r^2}\right|_{r_0,g_0}(r-r_0)^2 \nonumber\\
&+& \frac{1}{2}\sum^{N_c-1}_i
\left.
\frac{\partial^2
z_{Q\overline{Q}G}(r,g)
}{\partial \phi_i^2}
\right|_{r_0,g_0}
(\phi_i-{\phi_i}_0)^2,
\end{eqnarray}
for the unitary representation $U(N_c)$ and 
the special unitary representation $SU(N_c)$, respectively.
The Hilbert space for a gas of quarks reads
\begin{eqnarray}
\mbox{Tr}_{\mbox{c}} \left[\hat{U}_Q (g)
e^{-\beta\hat{H}_Q} e^{i\vartheta N_Q} e^{i\beta p\cdot r}\right]
=\exp\left(\mbox{Tr}_{\mbox{c}}\left[{\bf z}_{Q}(r,g)\right]\right),
\end{eqnarray}
where the polar integration over the exponential term is evaluated as follows:
\begin{eqnarray}
{\bf z}_{Q}(r,g) &=&
+2{\left<V\right>}\int \frac{dpdx}{(2\pi)^2} p^2 \ln\left[
1+{\bf R}_{\mbox{fund}}(g) \lambda_Q e^{-\beta E_Q(p)}  e^{i\beta p r
x} \right],
\nonumber\\
&=&
+2{\left<V\right>}\int \frac{dpdx}{(2\pi)^2} p^2
\left\{-\sum^{\infty}_{n=1} (-1)^n \frac{1}{n}\left( {\bf R}_{\mbox{fund}}(g)
\lambda_Q e^{-\beta E_Q(p)}\right)^n e^{i n\beta p r x} \right\},
\nonumber\\
&=&
+2{\left<V\right>}\int \frac{dp}{(2\pi)^2} p^2
\left\{-2\sum^{\infty}_{n=1} (-1)^n \frac{1}{n}\left( {\bf R}_{\mbox{fund}}(g)
\lambda_Q e^{-\beta E_Q(p)}\right)^n 
\frac{\sin\left(n\beta p r\right)}{n\beta p r} \right\}.
\label{exponen1}
\end{eqnarray}
The function $\frac{\sin\left(n\beta pr\right)}{\left(n\beta p r\right)}$ 
has an extremum at $r$=0 and consequently Eq.(\ref{exponen1}) 
has an extremum (maximum) at $r$=0 in the configuration space. 
In order to evaluate the zero momentum projector operator, 
we make a Taylor expansion around the extremum $r=0$  
for the exponential up to the quadratic term
\begin{eqnarray}
{\bf z}_{Q(\overline{Q})}(r,g)\approx
\left.{\bf z}_{Q(\overline{Q})}(r,g)\right|_{r=0}+
\frac{1}{2}
\left.\frac{\partial^2 {\bf z}_{Q(\overline{Q})}(r,g)}
{\partial r^2}\right|_{r=0} r^2, 
\label{exponen2}
\end{eqnarray}
where
\begin{eqnarray} 
\left.{\bf z}_{Q}(r,g)\right|_{r=0}
&=& +2{\left<V\right>} \int 
\frac{dp}{(2\pi)^2} p^2
\left\{-2\sum^{\infty}_{n=1} (-1)^n 
\frac{1}{n}\left( {\bf R}_{\mbox{fund}}(g)
\lambda_Q e^{-\beta E_Q(p)}\right)^n\right\}, \nonumber\\ 
&=&
+2{\left<V\right>} \int \frac{d^3p}{(2\pi)^3} \ln\left[ 1+{\bf
R}_{\mbox{fund}}(g) \lambda_Q 
e^{-\beta E_Q(p)} \right]. 
\label{exponen3}
\end{eqnarray}
The quadratic term reads
\begin{eqnarray} 
\left.\frac{\partial^2
{\bf z}_{Q}(r,g)
}{\partial r^2}\right|_{r=0}
&=&+2{\left<V\right>} 
\int\frac{dp}{(2\pi)^2} p^2 \left\{-2\sum^{\infty}_{n=1} (-1)^n \frac{1}{n}\left(
{\bf R}_{\mbox{fund}}(g) \lambda_Q e^{-\beta E_Q(p)}\right)^n
\left(-\frac{1}{3}\beta^2 p^2 n^2 \right) \right\}, 
\nonumber\\ 
&=&+2{\left<V\right>} \int \frac{d^3p}{(2\pi)^3} \left\{ -\frac{1}{3}\beta^2 p^2
\frac{{\bf R}_{\mbox{fund}}(g) \lambda_Q e^{-\beta E_Q(p)}}
{\left[1+{\bf R}_{\mbox{fund}}(g) \lambda_Q e^{-\beta E_Q(p)}
\right]^2} \right\}. 
\label{exponen4}
\end{eqnarray} 
The above expansions are derived using the following relations 
\begin{eqnarray}
\sum^{\infty}_{n=1} (-1)^n \frac{1}{n} x^n=-\ln(1+x),
\label{series1}
\end{eqnarray}
and
\begin{eqnarray}
\sum^{\infty}_{n=1} (-1)^n n x^n= -x/(1+x)^2.
\label{series2}
\end{eqnarray}
The fundamental matrix 
${\bf R}_{\mbox{fund}}(g)|_{g_0(\{\theta_i=0\})}$
is approximated to a unit matrix 
near the extremum color angles $\{\theta_i=0\}$.
Hence, Eq.(\ref{exponen4}) becomes,
\begin{eqnarray}
\left[\frac{\partial^2 z_Q(r,g) }{\partial r^2} \right]_{r_0,g_0} 
&=&
\mbox{Tr}_{\mbox{c}}\left[\frac{\partial^2 {\bf z}_Q(r,g) }{\partial r^2} \right]_{r_0,g_0},
\nonumber\\
&=&-2\left({\left<V\right>}/\beta^3\right) {\cal D}_Q(m_Q\beta,\lambda_Q),
\label{exponenseq1}
\end{eqnarray} 
where 
\begin{eqnarray} 
{\cal D}_Q\left(m_Q\beta,\lambda_Q\right)&=& 
\frac{N_c}{3}
\int \frac{\beta^3 d^3p}{(2\pi)^3} \left\{
\beta^2 p^2 \frac{\lambda_Q e^{-\beta E_Q(p)}}
{\left[1+\lambda_Q e^{-\beta E_Q(p)} \right]^2} \right\},
\nonumber\\
&=&
\frac{N_c}{3}
\int^{\infty}_{m_Q\beta} \frac{d\epsilon}{2\pi^2}
\epsilon \left[\epsilon^2-{m^2_q\beta^2}\right]^{3/2} 
\left\{
\frac{\lambda_Q e^{-\epsilon}} {\left[1+\lambda_Q e^{-\epsilon}\right]^2}
\right\}. 
\label{expcoeff1} 
\end{eqnarray}
Hence for quark and antiquark it becomes
\begin{eqnarray}
{\cal D}_{Q\overline{Q}}\left(m_Q\beta,\lambda_Q\right)&=&
\frac{N_c}{3}
\int^{\infty}_{m_Q\beta} \frac{d\epsilon}{2\pi^2}
\epsilon \left[\epsilon^2-{m^2_q\beta^2}\right]^{3/2}
\left\{
\frac{\lambda_Q e^{-\epsilon}} {\left[1+\lambda_Q e^{-\epsilon}\right]^2}
+
\frac{\lambda^{-1}_Q e^{-\epsilon}} {\left[1+\lambda^{-1}_Q e^{-\epsilon}\right]^2}
\right\}.
\label{expcoeff1QQ}
\end{eqnarray}
For zero quark mass limit, it reduces to
\begin{eqnarray}
{\cal D}_{Q\overline{Q}}\left(0,\lambda_Q\right)&=&
\frac{N_c}{3}\left(
\frac{7\pi^2}{30}+\ln^2\lambda_Q\left[1
+\frac{1}{2\pi^2}\ln^2\lambda_Q\right]
\right).
\end{eqnarray}
On the other hand, the color state projection is calculated 
by evaluating the integral over the Haar measure.
The quark exponential term $Z_Q\left(r_0,g\right)$ around the point $r_0$ reads
\begin{eqnarray} 
Z_Q(r_0,g)&\equiv&
\exp\left( \mbox{Tr}_{\mbox{c}} \left[{\bf z}_Q(r_0,g)\right] \right) 
=\mbox{Tr}_{\mbox{c}} \hat{U}_Q (g) e^{-\beta\hat{H}_Q} e^{i\vartheta N_Q}, 
\nonumber\\ &=&
\exp\left\{ +2\left({\left<V\right>}/\beta^3\right) \int^{\infty}_{m_q\beta}
\frac{d\epsilon}{2\pi^2} \epsilon \sqrt{\epsilon^2-m_Q^2\beta^2}
\mbox{Tr}_{\mbox{c}}\ln\left[ 1+{\bf R}_{\mbox{fund}}(g) 
\lambda_Q e^{-\epsilon} \right]\right\}, 
\nonumber\\ &=&
\exp\left\{ +2\left({\left<V\right>}/\beta^3\right)
\int^{\infty}_{m_Q\beta} \frac{d\epsilon}{2\pi^2}
\frac{1}{3}(\epsilon^2-m_Q^2\beta^2)^{3/2} \mbox{Tr}_{\mbox{c}} \left( \frac{ {\bf
R}_{\mbox{fund}}(g)\lambda_Q e^{-\epsilon} } {\left[1+{\bf
R}_{\mbox{fund}}(g)\lambda_Q e^{-\epsilon}\right]} \right) \right\}.
\nonumber\\
\label{exponenq1}
\end{eqnarray} 
The exponential term for system of quarks and antiquark becomes
\begin{eqnarray}
Z_{Q\overline{Q}}(r_0,g)=\exp\left[z_{Q\overline{Q}}(r_0,g)\right]
\end{eqnarray}
where
\begin{eqnarray} 
z_{Q\overline{Q}}(r_0,g)=\mbox{Re}\left\{ +2\left({\left<V\right>}/\beta^3\right)
\int^{\infty}_{m_Q\beta} \frac{d\epsilon}{2\pi^2}
\frac{1}{3}(\epsilon^2-m_Q^2\beta^2)^{3/2} 
{\cal I}_{Q\overline{Q}}(\epsilon,g)
\right\}.
\label{exponenq1b}
\end{eqnarray} 
The color internal symmetry for the quark and antiquark is determined 
by
\begin{eqnarray}
{\cal I}_{Q\overline{Q}}(\epsilon,g)=
\mbox{Tr}_{\mbox{c}}
\left[
\frac{ 
{\bf R}_{\mbox{fund}}(g)\lambda_Q e^{-\epsilon} } 
{\left[1+{\bf R}_{\mbox{fund}}(g)\lambda_Q e^{-\epsilon}\right]}
+
\frac{{\bf R}^*_{\mbox{fund}}(g)\lambda^{-1}_Q e^{-\epsilon} } 
{\left[1+ {\bf R}^*_{\mbox{fund}}(g)
\lambda^{-1}_Q e^{-\epsilon}\right]}
\right]
\end{eqnarray}
and the straightforward calculation for the trace leads to
\begin{eqnarray}
\mbox{Re}{\cal I}_{Q\overline{Q}}(\epsilon,g)
&=&\sum^{N_c}_i
\left[
\frac{\lambda_Q e^{i\theta_i} e^{-\epsilon}}{1+\lambda_Q e^{i\theta_i} e^{-\epsilon}}
+
\frac{\lambda^{-1}_Q e^{-i\theta_i} e^{-\epsilon}}
{1+\lambda^{-1}_Q e^{-i\theta_i} e^{-\epsilon}}\right],
\nonumber\\
&=&\sum^{N_c}_i
\left[
\frac{ \lambda^{-1}_Q e^{\epsilon}\cos\theta_i+1 }
{\lambda^{-2}_Q e^{2\epsilon}
+2\lambda^{-1}_Q e^{\epsilon}\cos\theta_i+1}
+
\frac{ \lambda_Q e^{\epsilon}\cos\theta_i+1 }
{\lambda^{2}_Q e^{2\epsilon}
+2\lambda_Q e^{\epsilon}\cos\theta_i+1}
\right].
\end{eqnarray}
The variation with respect to $\theta_i$ in the $U(N_c)$ representation is given by
\begin{eqnarray}
\mbox{Re}\frac{\partial}{\partial \theta_i}
{\cal I}_{Q\overline{Q}}(\epsilon,g)=
&-&\left[
\frac{\lambda_Q e^{-\epsilon}\left(1- \lambda^2_Qe^{-2\epsilon}\right)}
{\left[
1+\lambda^2_Q e^{-2\epsilon}+2\lambda_Q e^{-\epsilon}\cos\theta_i
\right]^2}\right]\sin\theta_i
\nonumber\\
&-&
\left[\frac{\lambda^{-1}_Q e^{-\epsilon}\left(1- \lambda^{-2}_Qe^{-2\epsilon}\right)}
{\left[
1+\lambda^{-2}_Q e^{-2\epsilon}+2\lambda^{-2}_Q e^{-\epsilon}\cos\theta_i
\right]^2}\right]
\sin\theta_i.
\end{eqnarray}
In the limit $\frac{\phi_i}{\pi}\left(\frac{\theta_i}{\pi}\right)\ll 1$ and 
$\cos\theta_i\approx 1$, we have also for $SU(N_c)$ representation
\begin{eqnarray}
\mbox{Re}\frac{\partial}{\partial \phi_i}
{\cal I}_{Q\overline{Q}}(\epsilon,g)=
&-&
\left[
\frac{\lambda_Q e^{-\epsilon}\left(1- \lambda^2_Qe^{-2\epsilon}\right)}
{\left[
1+\lambda^2_Q e^{-2\epsilon}+2\lambda_Q e^{-\epsilon}\right]^2}
+
\frac{\lambda^{-1}_Q e^{-\epsilon}\left(1- \lambda^{-2}_Qe^{-2\epsilon}\right)}
{\left[
1+\lambda^{-2}_Q e^{-2\epsilon}+2\lambda^{-2}_Q e^{-\epsilon}\right]^2}\right]
\nonumber\\
&\times&
\left[
\sin(\phi_i) +\sin(\sum^{N_c-1}_k\phi_k)
\right].
\end{eqnarray}

On the other hand, the Taylor expansion 
of the gluonic exponential part around the extremum $(r_0=0)$ reads
\begin{eqnarray}
{\bf z}_{G}(r,g)\approx
\left.{\bf z}_{G}(r,g)\right|_{r=0}+
\frac{1}{2}
\left.\frac{\partial^2 {\bf z}_{G}(r,g)}
{\partial r^2}\right|_{r=0} r^2.
\label{exponen2g}
\end{eqnarray}
The first term is given by
\begin{eqnarray}
\left.{\bf z}_{G}(r,g)\right|_{r=0}
&=&
-2{\left<V\right>} \int \frac{d^3p}{(2\pi)^3}
\ln\left[ 1-{\bf R}_{\mbox{adj}}(g) e^{-\beta p} \right],
\nonumber\\
&=&
+2{\left<V\right>} \int \frac{d^3p}{(2\pi)^3} \frac{\beta p}{3}
\frac{ {\bf R}_{\mbox{adj}}(g) e^{-\beta p}}
{\left[
1-{\bf R}_{\mbox{adj}}(g) e^{-\beta p}
\right]}.
\label{exponen3g}
\end{eqnarray}
The second term reads
\begin{eqnarray}
\left.\frac{\partial^2 {\bf z}_{G}(r,g)}
{\partial r^2}\right|_{r=0}
=
-2{\left<V\right>} \int \frac{d^3p}{(2\pi)^3} 
\left(\frac{\beta^2 p^2}{3}\right)
\frac{ {\bf R}_{\mbox{adj}}(g) e^{-\beta p}}
{\left[
1-{\bf R}_{\mbox{adj}}(g) e^{-\beta p}
\right]^2}.
\label{expgdrr}
\end{eqnarray}

The straightforward calculations of the trace gives
\begin{eqnarray}
z_{G}(r_0,g)&=&\mbox{Tr}_{\mbox{c}}\left.{\bf z}_{G}(r,g)\right|_{r=0},
\nonumber\\
&=&
+2{\left<V\right>} \int \frac{d^3p}{(2\pi)^3}
\frac{\beta p}{3}
\mbox{Tr}_{\mbox{c}}\left[\sum^{\infty}_{k=1} {\bf R}_{\mbox{adj}}(g^k)
e^{-k \beta p}\right],
\nonumber\\
&=&
+2{\left<V\right>} \int \frac{d^3p}{(2\pi)^3}
\frac{\beta p}{3}
\left[\sum^{\infty}_{k=1}
\left( \sum^{N_c}_{i\neq j} \cos k(\theta_i-\theta_j) +(N_c-1)\right)
e^{-k \beta p}\right],
\nonumber\\
&=&
+
\sum^{N_c}_{i\neq j} 2\frac{\left<V\right>}{\beta^3}
\int\frac{\beta^3d^3p}{(2\pi)^3}
\frac{\beta p}{3}
\frac{e^{-\beta p}(\cos(\theta_i-\theta_j)-e^{-\beta p})}
{\left[1-2e^{-\beta p}\cos(\theta_i-\theta_j)+e^{-2\beta p}\right]}
\nonumber\\
&~&
+
(N_c-1) 2\frac{\left<V\right>}{\beta^3} \int\frac{\beta^3d^3p}{(2\pi)^3}
\frac{\beta p}{3}
\frac{e^{-\beta p}}{1-e^{-\beta p}}.
\end{eqnarray}
The first derivate reads
\begin{eqnarray}
\frac{\partial z_{G}(r_0,g)}{\partial \theta_n}
=+\sum^{N_c}_{j\neq n} f_1(\theta_n-\theta_j),
\end{eqnarray}
and 
\begin{eqnarray}
\frac{\partial z_{G}(r_0,g)}{\partial \phi_n}
=\sum^{N_c}_{j\neq n} f_1(\phi_n-\phi_j)
+\sum^{(N_c-1)}_{j=1} f_1(-\phi_{N_c}+\phi_j),
\end{eqnarray}
for $U(N_c)$ and $SU(N_c)$ representations, respectively, where
\begin{eqnarray}
f_1(\psi)=
-2\frac{\left<V\right>}{\beta^3}
\int \frac{\beta^3 d^3p}{(2\pi)^3}
\frac{e^{-\beta p}\sin\psi}
{\left[1-2e^{-\beta p}\cos(\psi)+e^{-2\beta p}\right]}.
\end{eqnarray}
The saddle points for the multi-integration 
over the color parameters are found 
by maximizing the exponential term for the quarks and gluons,
\begin{eqnarray}
\frac{\partial}{\partial \theta_i} 
\left[z_G(r_0,g)+z_{Q\overline{Q}}(r_0,g)\right]&=&0, ~i=1, \cdots, N_c
\nonumber\\
\frac{\partial}{\partial \phi_i}
\left[z_G(r_0,g)+z_{Q\overline{Q}}(r_0,g)\right]&=&0, ~i=1, \cdots, N_c-1
\end{eqnarray}
for $U(N_c)$ and $SU(N_c)$ representation, respectively.
The solution of this set of equations 
is $g_0\left(\theta_1\cdots \theta_{N_c}=0\right)$
$\left(g_0\left(\phi_1\cdots \phi_{N_c-1}=0\right)\right)$.
The sets $\left(\theta_1,\cdots\theta_{N_c}=0\right)$ and
$\left(\phi_1,\cdots\phi_{N_c-1}=0\right)$
are solutions for the saddle points in the $U(N_c)$ and $SU(N_c)$ 
representations.
These solutions are satisfactory under the assumption that the angles
$\frac{\theta_i}{\pi}\left(\frac{\phi_i}{\pi}\right)\ll 1$ 
are the dominated range for the saddle points in 
the multi-integration over the color parameters.
However, in general case this solution gives a good approximation for the integration over
the Haar measure but this approximation does not necessary work correctly for quark-gluon bubbles
in the extreme conditions in the entire $T-\mu_B$ plane.
Under certain conditions in particular in very high temperatures, 
one or more of the color parameters becomes $\frac{\phi_i}{\pi}\neq 0$
in particular at very high temperatures.
When such a case takes place, it is expected that another class of solution emerges 
and modifies the density of states.

In the limit $g_0\equiv g_0(\phi_i=0,i=1,\cdots ,N_c-1)$, we have
\begin{eqnarray}
\left.{\cal I}_{Q\overline{Q}}(\epsilon,g)\right|_{g_0}=
N_c\left[
\frac{\lambda_Q e^{-\epsilon}}{1+\lambda_Q e^{-\epsilon}}
+
\frac{\lambda_Q^{-1} e^{-\epsilon}}{1+\lambda_Q^{-1} e^{-\epsilon}}
\right]
\end{eqnarray}
and
\begin{eqnarray}
\mbox{Re}\frac{1}{2}
\left.\frac{\partial^2}{\partial \phi_i^2}
{\cal I}_{Q\overline{Q}}(\epsilon,g)\right|_{g_0}
&=&\mbox{Re}\left.\frac{\partial^2}{\partial \theta_i^2}
{\cal I}_{Q\overline{Q}}(\epsilon,g)\right|_{g_0},
\nonumber\\ 
&=&
-\left[\frac{\lambda_Q e^{-\epsilon}\left(1-\lambda_Q e^{-\epsilon}\right)}
{\left[1+\lambda_Q e^{-\epsilon}\right]^3}
+
\frac{\lambda^{-1}_Q e^{-\epsilon}\left(1-\lambda^{-1}_Q e^{-\epsilon}\right)}
{\left[1+\lambda^{-1}_Q e^{-\epsilon}\right]^3}\right].
\end{eqnarray}
We define the following terms 
\begin{eqnarray}
{{\cal A}_{Q\overline{Q}}}_{0}(m_Q\beta,\lambda_Q)
=
\int^{\infty}_{m_Q\beta}
\frac{d\epsilon}{2\pi^2} \frac{1}{3}(\epsilon^2-{m^2_q\beta^2})^{3/2}
\left[\left.{\cal I}_{Q\overline{Q}}(\epsilon,g)\right|_{g_0}\right]
\end{eqnarray}
and
\begin{eqnarray}
{{\cal B}_{Q\overline{Q}}}_{i}(m_Q\beta,\lambda_Q)
&=&
-\int^{\infty}_{m_Q\beta}
\frac{d\epsilon}{2\pi^2} \frac{1}{3}(\epsilon^2-{m^2_q\beta^2})^{3/2}
\left[\left.\frac{\partial^2}{\partial \theta_i^2}
{\cal I}_{Q\overline{Q}}(\epsilon,g)\right|_{g_0}\right]>0.
\end{eqnarray}
In the limit of massless quarks the above terms reduce to
\begin{eqnarray}
{{\cal A}_{Q\overline{Q}}}_{0}(m_Q=0,\lambda_Q)
&=&\frac{N_c}{3}\left(\frac{7\pi^2}{120}+
\frac{\ln^2\lambda_Q}{4}\left[1+\frac{\ln^2\lambda_Q}{2\pi^2}\right]\right),
\nonumber\\
{{\cal B}_{Q\overline{Q}}}_{0}(m_Q=0,\lambda_Q)
&=&\frac{1}{6\pi^2}\left(3\ln^2\lambda_Q+\pi^2\right).
\end{eqnarray}
At the color saddle points $g_0\left(\{\theta_i=0\}\right)$, 
the gluonic terms reduce to
\begin{eqnarray}
z_{G}(r_0,g_0)&=&
+[N_c(N_c-1)] 2\frac{\left<V\right>}{\beta^3}
\int\frac{\beta^3 p^2d p}{2\pi^2} \frac{\beta p}{3}
\frac{1}{\left[e^{\beta p}-1\right]}
\nonumber\\
&~&
+(N_c-1) 2\frac{\left<V\right>}{\beta^3}
\int\frac{\beta^3 p^2dp}{2\pi^2} \frac{\beta p}{3}
\frac{1}{\left[e^{\beta p}-1\right]},\nonumber\\
&=& 2\frac{\left<V\right>}{\beta^3} {\cal A}_G,
\end{eqnarray}
where 
\begin{eqnarray} 
{\cal A}_G=(N^2_c-1)\frac{\pi^2}{90}.
\end{eqnarray}
The second derivative of the gluonic part with respect to $r$ in the limit $g_0$
reads
\begin{eqnarray}
\left.\frac{\partial^2 {\bf z}_{G}(r,g_0)}
{\partial r^2}\right|_{r=0}
=
-2\frac{\left<V\right>}{\beta^3} {\cal D}_G,
\end{eqnarray}
where 
\begin{eqnarray}
{\cal D}_G=4(N^2_c-1)\frac{\pi^2}{90}.
\end{eqnarray}
Furthermore, the second derivative for the gluonic part reads
\begin{eqnarray}
\frac{\partial^2 z_{G}(r_0,g)}{\partial \theta_m\partial \theta_n}
=
+\sum^{N_c}_{j\neq n}f_2(\theta_n-\theta_j)\delta_{nm}
-\left.f_2(\theta_n-\theta_m)\right|_{m\neq n},
\end{eqnarray}
and
\begin{eqnarray}
\frac{\partial^2 z_{G}(r_0,g)}{\partial \phi_m\partial \phi_n}
=
&+&\sum^{N_c}_{j\neq n}f_2(\phi_n-\phi_j)\delta_{nm}
-\left.f_2(\phi_n-\phi_m)\right|_{m\neq n}
\nonumber\\
&+&\sum^{N_c-1}_{j} f_2(-\phi_{N_c}+\phi_j)
+f_2(\phi_n-\phi_{N_c})
+f_2(\phi_m-\phi_{N_c}),
\end{eqnarray}
for $U(N_c)$ and $SU(N_c)$, respectively, where
\begin{eqnarray}
f_2(\psi)=
-2\frac{\left<V\right>}{\beta^3}
\int \frac{\beta^3 d^3p}{(2\pi)^3}
e^{-\beta p}
\frac{
\left[
\left(1+e^{-2\beta p}\right)\cos\psi-2e^{-\beta p}
\right]
}
{\left[1-2e^{-\beta p}\cos(\psi)+e^{-2\beta p}\right]^2}.
\end{eqnarray}
In the limit of the color saddle points 
$g_0\left(\phi_i=0, i=1, \cdots, N_c-1\right)$ we have 
\begin{eqnarray}
f_2(0)&=&
-2\frac{\left<V\right>}{\beta^3}
\int \frac{\beta^3 p^2 dp}{2\pi^2}
\frac{e^{-\beta p}}
{\left[1-e^{-\beta p}\right]^2}
\nonumber\\
&=&
-2\frac{\left<V\right>}{\beta^3}\frac{{\cal B}_G}{2N_c},
\end{eqnarray}
where
\begin{eqnarray}
{\cal B}_G=\frac{N_c}{3}.
\end{eqnarray}

The resulting quadratic expansion for quarks and antiquarks reads
\begin{eqnarray} 
z_{Q\overline{Q}}(r,g)\approx
&+&2\frac{{\left<V\right>}}{\beta^3} 
\left[ {\cal A}_{Q\overline{Q}}(m_Q\beta,\lambda_Q) 
\right.
\nonumber\\
&-&
\left.
\frac{1}{2}
\sum^{N_c}_{i=1}
{{\cal B}_{Q\overline{Q}}(m_Q\beta,\lambda_Q)} \theta^2_i
-\frac{1}{2} {\cal D}_{Q\overline{Q}} (m_Q\beta,\lambda_Q) r^2\right]. 
\end{eqnarray} 
The same procedure can be done for gluons. 
The expansion for the gluon term around the extrema 
$\left(r_0,g_0\right)$ reads
\begin{eqnarray}
z_{G}(r,g)\approx
2\left(\frac{\left<V\right>}{\beta^3}\right)\left[ {\cal A}_{G}
-\frac{1}{2}{\cal B}_{G}\sum^{N_c}_{i=1}\theta^2_i -\frac{1}{2}
{\cal D}_{G} r^2\right], 
\end{eqnarray} 
where 
\begin{eqnarray} 
{\cal A}_{G}=\frac{\pi^2}{90} \left(N^2-1\right), 
{\cal D}_{G}=4{\cal A}_{G},
{\cal B}_{G}=N_c/3.
\end{eqnarray} 

The mixed grand-canonical ensemble is calculated via 
\begin{eqnarray} 
\tilde{Z}_{V}(\beta,\theta)\equiv
\tilde{\sigma}_{V}\left(\beta,\lambda\right)= 
d_j \int_{SU(N_c)}
d\mu(g)\chi(g)\int_{{\left<V\right>}/\beta^3} 
\frac{d^3 r}{{\left<V\right>}/\beta^3} 
\exp\left(\mbox{Tr}_{\mbox{c}}\left[ {\bf z}_{Q\overline{Q}G} \right] \right), 
\label{canonicalsig1}
\end{eqnarray} 
where
\begin{eqnarray} 
z_{Q\overline{Q}G}&=& z_{G}+\sum_Q z_{Q\overline{Q}}\nonumber\\
&=&
2({\left<V\right>}/\beta^3)\left[ 
\left({\cal A}_{G}+
\sum_Q\left({\cal A}_{Q\overline{Q}}(m_Q\beta,\lambda_Q)
\right)\right)\right.
\nonumber\\ 
&-&
\left.
\frac{1}{2} 
\left(
{\cal D}_{G}+
\sum_Q {\cal D}_{Q\overline{Q}}(m_Q\beta,\lambda_Q)
\right) r^2  
-  \frac{1}{2} \left({\cal B}_{G}
+\sum_Q 
{\cal B}_{Q\overline{Q}}(m_Q\beta,\lambda_Q)
\right) 
\sum^{N_c}_{i=1}\theta^2_i \right].
\label{qqgcanonical1}
\end{eqnarray} 
Each integration of this function is evaluated using 
the saddle point approximation.
The color singlet state for gluons and flavorless quarks blob is integrated 
over the Haar measure with the help of the following relation,
\begin{eqnarray} 
\int_{SU(N_c)} d\mu(g) \chi_{\mbox{singlet}}(g)
\delta\left(\sum^{N_c}_{i=1}\theta_i=0\right)
\exp\left(-\frac{C}{2}\sum^{N_c}_{i=1} \theta^2_i \right)\approx
\frac{C^{(1-N^2)/2}}{  (2\pi)^{({N_c}-1)/2}\sqrt{N_c} } \prod^{N_c-1}_{j=1} j!,
\label{singlet1}
\end{eqnarray} 
where 
\begin{eqnarray}
C&=&\frac{\left<V\right>}{\beta^3} C^{\star}, \nonumber\\
C^{\star}&=&2\left[{\cal B}_{G}
+ \sum_Q \left({\cal B}_{Q\overline{Q}}(m_Q\beta,\lambda_)
\right)  \right].  
\end{eqnarray} 

In order to fit the phenomenological prediction about the existence of the critical point,
we shall generalize this relation for $N_c=3$ to arbitrary color symmetry configurations 
with the phenomenological parameter $\gamma$,
\begin{eqnarray}
\int_{SU(N_c)} d\mu(g) \chi_{\mbox{Sym}(\gamma)}(g)
\delta\left(\sum^{N_c}_{i=1}\theta_i=0\right)
\exp\left(-\frac{C}{2}\sum^{N_c}_{i=1} \theta^2_i \right) \propto
\frac{2}{2\pi\sqrt{3}} \left[\frac{\left<V\right>}{\beta^3} 
C^{\star}\right]^{-4\left(\gamma-1/2\right)},
\label{symmetry1}
\end{eqnarray}
where $\gamma=1/2$ and $\gamma=3/2$ for colored and color singlet states 
respectively. The above result is motivated by using the general result 
for Gaussian integration
\begin{eqnarray}
\int d\theta \theta^{\gamma} e^{-C \theta^2}\sim C^{-(\gamma+1)/2},
C\propto \frac{\left<V\right>}{\beta^3},
\end{eqnarray}
and
\begin{eqnarray}
\int d^{n}\theta \theta_{lk}^{\gamma} e^{-C f(\theta)}=
\left(\frac{2\pi}{C}\right)^{\frac{n}{2}}
\left[
\mbox{det}
\frac{\partial^2 f(\theta_0)}{\partial\theta_i\partial\theta_j}
\right]^{-\frac{1}{2}} 
\left[\frac{C}{2}
\mbox{det}
\frac{\partial^2 f(\theta_0)}{\partial\theta_l\partial\theta_k}
\right]^{-\frac{\gamma}{2}}
e^{-C f(\theta_0)}, 
\end{eqnarray}
where $i,j=1,\cdots, N_c-1$ and $l,k$ are the exponents of the 
pre-exponential.
The parameter $\gamma$ depends on the underlying overall symmetry 
of the system and increases when the symmetry configuration becomes more complicated.
Furthermore, we assume that this phenomenological parameter $\gamma$ also absorbs 
the breaking of the Lorenz symmetry invariance of the momentum projection
due to the thermal excitations
\begin{eqnarray}
\int dr r^{\gamma} e^{- D r^2}\sim D^{-(\gamma+1)/2}, 
D\propto \frac{\left<V\right>}{\beta^3}.
\end{eqnarray}
The momentum alignment, $\vec{P}_Q+\vec{P}_{\overline{Q}}+\vec{P}_G=0$, 
which is introduced by Kapusta\cite{Kapusta82b} in order to reproduce the bootstrap 
density of
states is no longer preserved in the extreme temperature and diluted matter.
The phenomenological parameter $\gamma$ increases as the color-flavor correlation 
of the system increases and the color configuration symmetry becomes more intricate.
We will use this parameter $\gamma$ to study the impact of these color-flavor correlation 
on the phase transition to quark-gluon plasma. 

The zero momentum projection is calculated using
\begin{eqnarray}
\int_{{\left<V\right>}/\beta^3} \frac{d^3
r}{{\left<V\right>}/\beta^3}\exp\left(-\frac{r^2}{2}D\right)=
\frac{1}{{\left<V\right>}/\beta^3} \left(\frac{2\pi}{D}\right)^{3/2},
\end{eqnarray} 
where 
\begin{eqnarray} 
D&=& \frac{\left<V\right>}{\beta^3} D^{\star},
\nonumber \\
D^{\star}&=&2 \left(
{\cal D}_{G}+\sum_{Q}{\cal D}_{Q\overline{Q}}(m_Q\beta,\lambda_Q) 
\right).
\label{gaussrad1}
\end{eqnarray} 
The resulting grand-canonical ensemble becomes
\begin{eqnarray}
{\cal Z}\left(\beta,\left<V\right>\right)
\sim\frac{2\sqrt{2\pi}}{\sqrt{3}} 
{\left(D^{\star}\right)}^{-3/2} 
{\left(C^{\star}\right)}^{-4[\gamma-1/2]}
{\left( \frac{\left<V\right>}{\beta^3} \right)}^{-[4\gamma+1/2]}\exp\left(
\frac{1}{3}\frac{\left<V\right>}{\beta^3} d^*(\beta)
\right),
\label{result_g_micro1}
\end{eqnarray}
where
\begin{eqnarray}
d^{\star}(\beta)= 3\cdot 2
\left({\cal A}_{G}+
\sum_{Q}{\cal A}_{Q\overline{Q}}(m_Q\beta,\lambda_{Q})
\right).
\label{saddlelaplace1}
\end{eqnarray}
%
%
%

\subsection{Microcanonical ensemble}
The asymptotic density of states for the microcanonical ensemble for
large fireball energy $W$ is calculated by taking the inverse Laplace transform 
for the grand canonical ensemble. 
The inverse Laplace transform for large $W\rightarrow \infty$ 
is evaluated using the steepest descent method,
\begin{eqnarray} 
{\cal Z}\left(W,\left<V\right>\right)&=& 
\frac{1}{2\pi i}
\int^{\beta_0+i\infty}_{\beta_0-i\infty}
d\beta e^{\beta W} {\cal Z}\left(\beta,\left<V\right>\right)
\nonumber\\
&=&
\frac{1}{2\pi i} 
\int^{\beta_0+i\infty}_{\beta_0-i\infty}
d\beta g(\beta) 
e^{ \beta W+\frac{1}{3}\frac{\left<V\right>}{\beta^3} d^{\star}(\beta)}, 
\nonumber\\
&\sim&
\frac{1}{2\pi}
\sqrt{\frac{2\pi}{\left<V\right>f''(\beta_{\mbox{S}})}} g(\beta_{\mbox{S}}) 
e^{\left<V\right>f(\beta_{\mbox{S}})}.
\label{laplacecanon1}
\end{eqnarray}
where $\beta_{\mbox{S}}$ is the saddle point. 
The pre-exponential function in Eq.(\ref{result_g_micro1}) reads,
\begin{eqnarray}
g(\beta)=\frac{2\sqrt{2\pi}}{\sqrt{3}}
{\left(D^{\star}\right)}^{-3/2}
{\left(C^{\star}\right)}^{-4[\gamma-1/2]}
{\left(\left<V\right>\right)}^{-[4\gamma+1/2]} {\beta}^{3[4\gamma+1/2]},
\end{eqnarray}
while the exponential function reads,
\begin{eqnarray} 
f(\beta)=\beta\frac{W}{\left<V\right>}+ 
\frac{1}{3\beta^3} d^{\star}(\beta)=
\beta x + \frac{1}{3\beta^3} d^{\star}(\beta).
\end{eqnarray} 
In order to separate the volume fluctuation from the mass fluctuation,
we define the energy density parameter $x=W/\left<V\right>$.
The second derivative reads
\begin{eqnarray}
f''(\beta)&=&\frac{\partial^2 f(\beta)}{\partial \beta^2},\nonumber\\
&=& 4\frac{d^{\star}(\beta)}{\beta^5}-2\frac{{d^{\star}}'(\beta)}{\beta^4}
+\frac{1}{3}\frac{{d^{\star}}''(\beta)}{\beta^3}
\label{saddleff2}
\end{eqnarray}
Eq.(\ref{saddleff2}) is evaluated with the help of Eq.(\ref{saddlelaplace1}).
The gluon terms read
\begin{eqnarray}
\frac{\partial^2}{\partial \beta^2} {\cal A}_{G}
=\frac{\partial}{\partial \beta}{\cal A}_{G}=0.
\end{eqnarray}
The quark and antiquark terms read
\begin{eqnarray}
\frac{\partial}{\partial \beta}
{\cal A}_{Q\overline{Q}}(m_Q\beta,\lambda_{Q})&=&
-m^2_q\beta\int^{\infty}_{m_Q\beta}
\frac{d\epsilon}{2\pi^2} (\epsilon^2-{m^2_q\beta^2})^{1/2}
\left[
\frac{
\lambda_Q e^{-\epsilon} } {\left[1+\lambda_Q e^{-\epsilon}\right]}
+
\frac{
\lambda^{-1}_Q e^{-\epsilon} } {\left[1+\lambda^{-1}_Q e^{-\epsilon}\right]}
\right],
\end{eqnarray}
and 
\begin{eqnarray}
\frac{\partial^2}{\partial \beta^2}
{\cal A}_{Q\overline{Q}}(m_Q\beta,\lambda_{Q})&=&
-m^2_q\int^{\infty}_{m_Q\beta}
\frac{d\epsilon}{2\pi^2} 
\frac{(\epsilon^2-2{m^2_q\beta^2})}
{(\epsilon^2-{m^2_q\beta^2})^{1/2}} 
\left[
\frac{
\lambda_Q e^{-\epsilon} } {\left[1+\lambda_Q e^{-\epsilon}\right]}
+
\frac{
\lambda^{-1}_Q e^{-\epsilon} } {\left[1+\lambda^{-1}_Q e^{-\epsilon}\right]}
\right].
\end{eqnarray}

The value of the saddle point $\beta_{\mbox{S}}$ 
given in Eq.(\ref{laplacecanon1})
is determined by maximizing the function  $f(\beta)$,
\begin{eqnarray} 
f'({\beta})|_{\beta=\beta_{\mbox{S}}}=0.  
\label{refzrs1}
\end{eqnarray} 
This point has a nonlinear solution for massive flavors
\begin{eqnarray} 
\left.
\beta=\left[\frac{d^{\star}(\beta)}{x}\right]^{1/4}
\left[ 1-\frac{\beta}{3}
\frac{ {d^{\star}}'(\beta) }{d^{\star} (\beta)} \right]^{1/4}
\right|_{\beta_{\mbox{S}}}.
\label{refzrs2} 
\end{eqnarray} 
For massless flavors, the saddle point has 
a trivial linear solution~\cite{Gorenstein83b,Auberson86a}.
Finite constituent quark masses do not seem to effect
the order of the phase transition although 
it modifies the nonlinear solution for the micro-canonical ensemble.
The nonlinear solution of Eq.(\ref{refzrs2}) is obtained numerically.
The micro-canonical ensemble with massive flavors will be studied in 
our forthcoming work in particular to study quark and gluon bubbles with 
strangeness. 
The incorporation of strangeness in the canonical ensemble is particularly 
important for high temperatures~\cite{Derreth1985a}. 

The remainder of the article will focus on quark and gluon bubbles 
(i.e. fireballs and hereinafter we abbreviate them as FB) with massless quarks.
For massless flavors, 
the function $d^{\star}(\beta)=d^{\star}(0)$ becomes independent on $\beta$ 
and the saddle point solution for the steepest descent method
is known analytically.
In this case the inverse Laplace transform becomes 
\begin{eqnarray} 
\frac{1}{2\pi i} \int^{\beta_0+i\infty}_{\beta_0-i\infty}
d\beta \beta^{\mbox{n}} e^{{V}\left[\beta x+ 
d^{\star}(0)/3\beta^3\right]} \sim \frac{1}{2\sqrt{2\pi}}
\frac{1}{ V^{1/2} x^{1/2} }
\left[d^{\star}(0)/x\right]^{(1+2\mbox{n})/8} 
e^{\frac{4}{3}{ V }\left[d^{\star}(0)x^3\right]^{1/4}} .
\label{laplaceinve0}
\end{eqnarray} 
The density of states 
${\sigma(W,V)}_{\mbox{asym}}\approx {\cal Z}(W,V)$
with respect to the bag energy $W$ with massless flavors reads 
\begin{eqnarray} 
{ \sigma(W,V) }_{\mbox{asym}}= 
A {V}^{-4\gamma-1} x^{-3\gamma-1}
\exp\left(\frac{4}{3} {V}\left[d^*(0) x^3\right]^{1/4}\right), 
\label{statedensity1}
\end{eqnarray} 
where $x=(m-BV)/V$ and 
\begin{eqnarray}
u&=&d^*(0),\nonumber\\
 &=&3\cdot 2\cdot\left[ (N^2_c-1)\frac{\pi^2}{90}
+\sum_Q \frac{N_c}{3}\left(
\frac{7\pi^2}{120}
+\frac{\ln^2\lambda_Q}{4}\left[1+\frac{\ln^2\lambda_Q}{2\pi^2}\right]
\right)
\right]. 
\label{u_dens1}
\end{eqnarray}
The pre-exponential coefficient reads,
\begin{eqnarray}
A&=&
\frac{1}{\sqrt{3}}
{\left(D^{\star}\right)}^{-3/2}
{\left(C^{\star}\right)}^{-4(\gamma-1/2)}
\left[d^{\star}(0)\right]^{3\gamma+\frac{1}{2}},
\nonumber\\
&=&
\frac{3}{8} {\left(C^{\star}\right)}^{-4(\gamma-1/2)}
\left[d^{\star}(0)\right]^{3\gamma-1},
\label{a_dens1}
\end{eqnarray}
where
\begin{eqnarray}
D^{\star}&=&\frac{4}{3} d^{*}(0),\nonumber\\
C^{\star}&=&
2\cdot\left[ \frac{N_c}{3}
+\sum_Q \frac{1}{6\pi^2}\left(
3\ln^2\lambda_Q+\pi^2\right)\right].
\end{eqnarray}
We have introduced the variable $x$ in order to simplify the equation
when we study the isobaric partition function below.
In the MIT-bag like models, the energy for the quark and gluon bubble $W$ 
is fixed by $W=m-B V$.
However, the ansatz for color-flavor correlations alters the quark and gluon 
bubble internal symmetry and have a strong impact on the phase transition 
diagram, as will show below.
%
\subsection{Bubble volume fluctuation\label{volumsec}}

The quantity $\sigma(P,\left<V\right>)$ with $\sigma({P,\left<V\right>})d^4P$ 
describes the density of states of noninteracting quarks, anti-quark and gluons 
confined in a bag with the total 4-momentum in 
the interval $(P,P+dP)$ and with the specific volume ${\left<V\right>}$.
The canonical ensemble measures only the mass spectral density.
Therefore, the density of states given by Eq.(\ref{statedensity1}) 
does not measure the bubble volume fluctuation.
In order to measure the bubble volume fluctuation we need to find the wavefunction 
for the quarks and gluons blob bound state. 
The density of single particles levels can be calculated using the multiple reflection
expansion~\cite{Balian70a,Balian70b,Balian70c}. 
The volume fluctuation is considered by smoothing the sharp boundary 
of the MIT bag due to extended potential profile. 
The density of states can then be calculated for a bag with an extended surface profile.

Let the system comprises a gas of bubbles with various species.
Each species type is labeled by the index $[I]$ and is 
classified by the bubble internal structure 
of constituent quarks and gluons. 
The square bracket is to distinguish the species type index from 
the quantum numbers of the bubble's constituent particles. 
The eigenstate for a constituent particle $i$ in the bubble of species type $[I]$ 
is given by the energy eigenstate $E_{[I]n_i}$ and the wavefunction $\Psi_{[I]n_i}(X_i)$. 
The label $n_i$ indicated the energy quantum number which can be occupied 
by the particle $i$.
In the standard MIT bag model, the interaction between the particles is neglected
and the confined constituent particle wavefunction for a bag 
with a sharp boundary is given by 
$\Psi_{[I]n_i}(X)=N_{[I]n_i} j_l\left(\frac{x_{{n_i}l}}{R_{[I]}} r\right) 
Y_l\left(\hat{r}\right)$
where the Bessel function$j_l$ 
vanishes at the bag's surface $r=R_{[I]}$, i.e. $j_l\left(x_{n_il}\right)=0$.
In this example the bubble's species type is specified by the size 
of the bag radius $R_{[I]}$. 

The level density of states for a bubble of species $[I]$ with 
a sharp surface boundary can be calculated from
\begin{eqnarray}
\sigma_{[I]}(E)&=&
\sum^{\mbox{States}}_{n_i} 
{\cal D}_{[I]n_i}\delta\left(E-E_{[I]n_i}\right),
\end{eqnarray}
where ${\cal D}_{[I]n_i}$ is the constituent particles 
degeneracy factor of the energy level $n_i$ 
while $E_{[I]n_i}$ is the energy eigenstate and $[I]$ 
is the bubble's species quantum number.
The sum over $n_i$ runs over all the energy states.
However, the level density of states for the bag with an extended surface 
and properly adjusted potential reads
\begin{eqnarray}
\rho_{[I]}(E,r)&=&\sum_{n_i} {\cal D}_{[I]n_i} 
\delta\left(E-E_{[I]n_i}\right)\left|\Psi_{[I]n_i}(r)\right|^2,
\end{eqnarray}
where $\Psi_{[I]n_i}(r)$ is the bubble's constituent particle wave-function.
After integrating the configuration space distribution function, the density of states becomes
\begin{eqnarray}
\sigma_{[I]}(E)&=&\int d^3r \rho_{[I]}(E,r).
\end{eqnarray}
When the system consists bubbles with several species, the density of states becomes 
\begin{eqnarray}
\rho(E,r)&=&\sum^{\mbox{Types}}_{[I]} \rho_{[I]}(E,r),
\end{eqnarray}
where each species is labeled by type $[I]$.
The (mixed-)grand canonical ensemble for constituent particles 
in the bubble of type $[I]$ with specific internal color-flavor structure,  reads,
\begin{eqnarray}
{\cal Z}_{[I]}&=&\mbox{Tr} e^{-\beta \hat{H}_{[I]}}
=\mbox{Tr} \left<[I]\right|e^{-\beta \hat{H}}\left|[I]\right>
\nonumber\\
&=& 
\sum_{n_1, n_2 \cdots} 
\left<[I]|n_1,n_2\cdots\right>\left<n_1,n_2\cdots\right|
e^{-\beta 
\hat{H}}\left|n_1,n_2\cdots\right>\left<n_1,n_2\cdots|[I]\right>
\nonumber\\
&=&
\int d^3X \Psi^*_{[I]}(X)\Psi_{[I]}(X) 
\prod_i \sum_{n_i}
e^{-\beta E_{[I]n_i}}
=
\int d^3X \left|\Psi_{[I]}(X)\right|^2 \prod_i \sum_{n_i} e^{-\beta 
E_{[I]n_i}},
\label{ensembledensity1}
\end{eqnarray}
where the sum runs over the occupation number 
$n_i$ for one-particle state while the product $\prod_i$ is 
the number of states which can fill the one particle state.
The coordinate $X$ is the cluster's configuration space while 
\begin{eqnarray}
\left<[I]|n_1,n_2\cdots\right>=\Psi_{[I]}(X)\prod^{\mbox{constituents}}_{i} \phi_{n_i}, 
\end{eqnarray}
is the cluster wavefunction. 
The function $\Psi_{[I]}(X)$ is bubble wavefunction 
in the configuration space while $\phi_{n_i}$ 
are the constituent particle's Fock states.

The number of ``one particle'' states reads
\begin{eqnarray}
\left. \prod_{i=1}^{\mbox{states}} 
\sum_{n_i} e^{-\beta E_{[I]n_i}}\right|_{[I]}
&=&
\exp\left[ \sum_{i}^{\mbox{states}} 
\ln \sum_{n_i} e^{-\beta E_{[I]n_i}}\right],
\nonumber\\
&=&
\exp\left[ {V_X} \int \frac{d^3p}{(2\pi)^3} 
\ln \sum_{n} e^{-\beta E_{[I]n}}\right].
\label{ensemblesum}
\end{eqnarray}
The resulting ensemble for quark and gluon bubble of type $[I]$ with the extended surface becomes
\begin{eqnarray}
{\cal Z}_{[I]}&=&
\left[\int d^3 X |\Psi_{[I]} (X)|^2
\exp\left( {V_X} \int \frac{d^3p}{(2\pi)^3}
\ln \sum_{n} e^{-\beta E_{[I]n}}\right)\right],
\nonumber\\
&=&
\exp\left(
\left<V_{[I]}\right>\int \frac{d^3p}{(2\pi)^3}
\ln \sum_{n} e^{-\beta E_{[I]n}} \right).
\label{ensemble_wave}
\end{eqnarray}
where $V_X=\frac{4\pi}{3}\pi X^3$.
The density of states for single-particle levels has been 
studied in~\cite{Balian70a,Balian70b,Balian70c}
extensively using the multiple reflection expansion method.
When we go beyond the standard bag model with a sharp surface to
a smooth one, it is possible to express the bag's volume fluctuation as follows
\begin{eqnarray}
\delta V= V_X-\left<V_{[I]}\right>,
\end{eqnarray}
for bubbles with a conserved number density.
Therefore, the volume distribution function 
for bubbles with smooth boundaries can be found by the following replacement  
\begin{eqnarray}
f\left(\left<V_{[I]}\right>\right)=
\int dV_X P_{[I]}(V_X-\left<V_{[I]}\right>) f\left(V_X\right),
\end{eqnarray}
which reduces to
\begin{eqnarray}
{\mbox{P}}_{[I]}\left(V_X-\left<V_{[I]}\right>\right)
&\approx& \delta\left(V_X-\left<V_{[I]}\right>\right),
\end{eqnarray}
for bubbles with sharp surfaces. 
Hence the density of states for a bag with a smoothed boundary 
can be written as follows
\begin{eqnarray}
\rho(W,v)=\mbox{P}\left(x,v\right)\sigma(x,v)_{\mbox{asym}}, ~\mbox{and}~ x=\frac{W}{v},
\label{density_approx}
\end{eqnarray} 
where ${\mbox{P}}_{[I]}\left(V_X-\left<V_{[I]}\right>\right)\rightarrow \mbox{P}\left(x,v\right)$
and $V_X\rightarrow v$.
For the standard MIT bag model, we have $\left<V_{[I]}\right>=\frac{m}{4B}$
and 
\begin{eqnarray}
{\mbox{P}}_{[I]}\left(V_X-\left<V_{[I]}\right>\right)\sim
\delta\left(V_X-m/4B\right)\equiv \delta\left(\frac{W}{V_X}-3B\right),
\label{wave_volume}
\end{eqnarray}
where $m$ and $W$ are the mass and energy, respectively, 
while $V_X$ and $B$ are the volume and bag constant, respectively 
for the specified bag. 
In order to fit the phenomenology, the spectral density is the average 
density for quark and gluon bubbles
with smoothed boundaries and their volumes are close 
to $\left<V_{[I]}\right>$. 
This density of states is given by
\begin{eqnarray}
\sigma\left(\varepsilon,\left<V_{[I]}\right>\right)=
\int dV_X f_{\Delta_V} 
\left(V_X-\left<V_{[I]}\right>\right)\sigma(\varepsilon,V_X),
\end{eqnarray}
where the function $f_{\Delta_V}$ is a smoothing function normalized to
\begin{eqnarray}
\int dV_X f_{\Delta_V}\left(V_X-\left<V_{[I]}\right>\right)=1,
\end{eqnarray}
and $\varepsilon=E/V_X$ is the bubble's energy density. 
%
In the phenomenological calculations, 
the Gaussian smoothing function is given 
by~\cite{Balian70a,Balian70b,Balian70c} 
\begin{eqnarray}
\delta(x)\rightarrow f_{\Delta_V}(x)=
\sqrt{\frac{\Delta/E}{\pi}} 
e^{-\frac{\Delta}{E} x^2} L^{\alpha}_n\left[\frac{\Delta}{E} x^2\right].
\label{Smootheq}
\end{eqnarray}
The $L^{\alpha}_n[x]$ denotes the associated Laguerre polynomials, $\Delta$ 
is the smoothing parameter,
and $E$ is the bubble's energy (or mass) while $\left<V_{[I]}\right>\propto E$ 
in MIT bag model.
The smoothing function is analogous to the volume distribution function 
defined in Eq.(\ref{wave_volume}),
\begin{eqnarray}
{\mbox{P}}_{[I]}\left(V_X-\left<V_{[I]}\right>\right)&=&
f_{\Delta_V}\left(V_X-\left<V_{[I]}\right>\right),
\nonumber\\
&=&f_{\Delta_V}(V-V_{0}).
\end{eqnarray}
In general, the cluster's wave-function is very complicated 
and in principle is not known in detail.
However, the standard MIT bag model can be extended by smoothing 
the sharp boundary condition.
The bubble with an extended surface boundary is approximated 
by replacing the delta function with a smoothing function 
as given in Eq.(\ref{Smootheq}),
\begin{eqnarray}
\delta(V-V_0)
\rightarrow
f_{\Delta_V}\left(V-V_0\right).
\end{eqnarray}
This can be done e.g. using the multiple reflection
expansion~\cite{Balian70a,Balian70b,Balian70c} for calculating
the density of states and the Strutinsky smoothing
method~\cite{Strutinsky67a,Strutinsky67b}.
The volume fluctuation to lowest order for the Gaussian smoothing function arrives then at
\begin{eqnarray}
\delta(V-V_0)
\rightarrow
\sqrt{\frac{\Delta/E}{\pi}} e^{-\frac{\Delta}{E} \left(V-V_0\right)^2}, 
\sqrt{\frac{\Delta/E}{\pi}}  \left[\frac{\Delta}{E} V^2\right] e^{-\frac{\Delta}{E} \left(V-V_0\right)^2}, \cdots
\label{deltasmooth1}
\end{eqnarray}
for $n=0, 1$ and $\cdots$, respectively.
Nevertheless, the order of the volume fluctuation plays a crucial role 
in the order of phase transition at the extreme conditions. 
The parameter $\Delta$ is chosen small as possible, but nevertheless sufficient large 
to taken into account the extended surface of the bag effectively. 
The high excitation of the smoothing function $f_{\Delta_V}(x)$ 
due to the thermal excitation presumably leads to the bag instability.  

Note that our model differs from the Strutinsky smoothed 
density~\cite{Strutinsky67a,Strutinsky67b}.
We soften the density of states by allowing the volume fluctuation 
with a specific energy density instead of smoothing 
the energy spectral function with respect to the bag energy. 
Our ansatz can be justified by the multiple reflection method 
where the potential is deformed by $\delta r$ 
and the particle wavelength is varied by $L=L_0+\delta L$. 
A more detailed analysis will be considered in forthcoming work 
to study the volume fluctuation due to potential deformation 
and smoothing the bag boundary 
by considering a bag with an extended surface using the multiple 
reflection method~\cite{Balian70a,Balian70b,Balian70c}.  
The Strutinsky smoothed density of states $\overline{\sigma}(E)$
is obtained from the density of states, $\sigma(E)$, 
by the Gaussian smoothing
\begin{eqnarray}
\overline{\sigma}(E)=\int dE' f_{\Delta_E}(E'-E)\sigma(E'),
\end{eqnarray}
where $f_{\Delta}(x)$ is given by Eq.(\ref{Smootheq}) 
but instead $\Delta_V=\Delta/E$ 
we have $\Delta_E=\Delta/V$.
However, it is possible to derive the smoothed density for the bubble's volume 
from  the Strutinsky smoothed density by substituting 
$E=\varepsilon V$ and $dE=\varepsilon dV$.


\section{The gas of bags with van der Waals repulsion}

The partition function for $N$ relativistic noninteracting particles 
reads~\cite{Gorenstein98a,Gorenstein82a}
\begin{eqnarray} 
Z(T,V,\lambda_B,\lambda_S)&=&
\sum^{\infty}_{n_B=-\infty} \lambda^{n_B}_B
\sum^{\infty}_{n_S=-\infty} \lambda^{n_S}_S
Z(T,V,n_B,n_S) \nonumber\\
&=&\sum^{\infty}_{N=0}\frac{1}{N!} 
\int \prod^{N}_{i=1}
\left[
\left(V-\sum^{N}_{j=1}{v_{\mbox{VdW}}}_j\right) 
\frac{d^3 p_i}{(2\pi)^3} dm_i d {v_{\mbox{VdW}}}_i  
\tau(m_i,{v_{\mbox{VdW}}}_i,\cdots)\right]
\nonumber\\ 
&\cdot& \theta(V-\sum^{N}_{j=1}  {v_{\mbox{VdW}}}_j)
\int d^{4}P e^{-\frac{P_0}{T}} \delta^{4}(P-\sum^{N}_{j=1}p_j)
\label{ensemble_particle}
\end{eqnarray} 
where
\begin{eqnarray}
\tau(m_i,{v_{\mbox{VdW}}}_i,\cdots)&\equiv&
\tau(m_i,T, {v_{\mbox{VdW}}}_i,\lambda_B,\lambda_S,\cdots)
\nonumber\\
&=&
\sum^{\infty}_{n_B=-\infty}\lambda^{n_B}
\tau(m_i,T,{v_{\mbox{VdW}}}_i,n_B,n_S,\cdots)
\label{dens_particles}
\end{eqnarray}
is the particles density of states with the fugacities 
$\lambda_B=e^{\frac{\mu_B}{T}}$ and $\lambda_S=e^{\frac{\mu_S}{T}}$  
and $n_B, n_S$ are baryonic and strangeness densities and are the Fourier modes 
with $i\theta_B=\mu_B/T$ and $i\theta_S=\mu_S/T$, respectively.
The density of states given by Eq.(\ref{dens_particles}) consists 
the known hadronic mass spectrum particles and the continuous
density of states for Hagedorn bubbles
\begin{eqnarray}
\tau(m_i,T, {v_{\mbox{VdW}}}_i,\lambda_B,\lambda_S,\cdots)=&~&
\sum^{\mbox{Baryons}}_i D_{\mbox{FD}i}\left(m,T,v,\lambda_i\right)\delta(m-m_i)\delta(v-{v_{\mbox{VdW}}}_H) 
\nonumber\\
&+& 
\sum^{\mbox{Mesons}}_i D_{\mbox{BE}i}\left(m,T,v,\lambda_i\right)\delta(m-m_i)\delta(v-{v_{\mbox{VdW}}}_H)  
\nonumber\\
&+&
\sum^{\mbox{Types}}_I \sum_{m,v} \rho_I(m,T,v,\lambda_I),
\end{eqnarray}
where the sum runs over the baryons and mesons mass spectrum and they satisfy 
Fermi-Dirac and Bose-Einstein statistics respectively. 
The terms 
$D_{\mbox{FD}i}\left(m,T,v,\lambda_i\right)$
and
$D_{\mbox{BE}i}\left(m,T,v,\lambda_i\right)$ are 
the degeneracies for the Fermi and Bose particles, respectively. 
The continuous densities of states run over the various species 
of Hagedorn bubbles
and for large masses they obey Maxwell-Boltzeman statistics.
Each species is labeled by the type $[I]$ and characterized by specific quantum numbers 
such as mesonic fireball, baryonic fireball, $\cdots$, etc. 
The excluded volume is taken as the hard core Van der Waals repulsion volume ${v_{\mbox{VdW}}}_i=4 v_i$ for a system 
of identical particles with size $v_i$.
The grand partition function for a gas of identical particles of mass $m_i$ 
in a volume $V$ with specific quantum statistics reads
\begin{eqnarray}
Z_{\mbox{Stats}}(T,V;m_i,\lambda_i)= 
\exp\left[V\varphi_{\mbox{Stats}}(T;m_i,\lambda_i)\right], 
\end{eqnarray} 
where $m_i$ and $\lambda_i$ are hadronic particle mass and fugacity, respectively, while
${\cal D}_i$ is the degeneracy factor stemming from the internal degrees of freedom.
The volume $V$ is the hadronic particle's Van der Waals excluded volume.
The subscripts $\mbox{Stats}\equiv \mbox{MB},\mbox{FD}$ or $\mbox{BE}$ 
correspond to Maxwell-Boltzmann, Fermi-Dirac and Bose-Einstein statistics, respectively.
The function $\varphi_{\mbox{Stats}}(T;m_i,\lambda_i)$ reads
\begin{eqnarray} 
\varphi_{\mbox{MB}}(T;m_i,\lambda_i)= 
{\cal D}_i\lambda_i 
\int\frac{d^{3}k}{(2\pi)^{3}}e^{-\sqrt{k^{2}+m_i^{2}}/T},
\label{quant_stats_mb}
\end{eqnarray} 
\begin{eqnarray} 
\varphi_{\mbox{FD}}(T;m_i,\lambda_i)=
{\cal D}_i\int\frac{d^{3}k}{(2\pi)^{3}} 
\ln\left[1+\lambda_i e^{-\sqrt{k^{2}+m_i^{2}}/T}\right], 
\label{quant_stats_fd}
\end{eqnarray} 
and
\begin{eqnarray}
\varphi_{\mbox{BE}}(T;m_i,\lambda_i)= -{\cal D}_i\int\frac{d^{3}k}{(2\pi)^{3}} 
\ln\left[1-\lambda_i e^{-\sqrt{k^{2}+m_i^{2}}/T}\right].
\label{quant_stats_be}
\end{eqnarray}  
After integration by parts, they become
\begin{eqnarray} 
\varphi_{\mbox{Stats}}(T;m_{i},\lambda_{i})=
\frac{1}{T} p^{\mbox{ideal}}_i =
\frac{1}{2\pi^{2}}\frac{1}{3}\frac{1}{T} {\cal D}_{i}
\int dk \frac{k^{4}}{\sqrt{k^{2}+m_{i}^{2}}}
f_{\mbox{Stats}}(T;m_{i},\lambda_{i}),
\label{stats_quant}
\end{eqnarray} 
where the quantum statistics distribution functions read
\begin{eqnarray} 
f_{\mbox{MB}}(T;m_{i},\lambda_{i})=
\lambda_{i}\exp\left(-\sqrt{k^{2}+m_{i}^{2}}/T\right), 
\label{f_mb}
\end{eqnarray}
\begin{eqnarray} 
f_{\mbox{FD}}(T;,m_{i},\lambda_{i})=
\left[\lambda_{i}^{-1}
\exp\left(\sqrt{k^{2}+m_{i}^{2}}/T\right)+1\right]^{-1},
\label{f_fd}
\end{eqnarray} 
and 
\begin{eqnarray} 
f_{\mbox{BE}}(T;m_{i},\lambda_{i})=
\left[\lambda_{i}^{-1}
\exp\left(\sqrt{k^{2}+m_{i}^{2}}/T\right)-1\right]^{-1}.
\label{f_de}
\end{eqnarray} 
for Maxwell-Boltzmann, Fermi-Dirac and Bose-Einstein statistics, respectively.
The pressure and energy density for each particle  
are determined, respectively, by 
\begin{eqnarray}
p_i=T\varphi_{\mbox{Stats}}(T;m_{i},\lambda_{i}),
\label{ensemble_pr}
\end{eqnarray}
\begin{eqnarray}
\epsilon_i= T^2 \frac{\partial}{\partial T} 
\varphi_{\mbox{Stats}}(T;m_{i},\lambda_{i}),
\label{ensemble_en}
\end{eqnarray}
The baryonic density for each particle reads 
\begin{eqnarray}
n_{Bi}(T;m_{i},\lambda_{i})
&=&T\frac{\partial \lambda_{i}}{\partial \mu_B} 
\frac{\partial}{\partial\lambda_{i}}
\varphi_{\mbox{Stats}}(T;m_{i},\lambda_{i}), 
\nonumber\\
&=& 
\frac{\partial\ln\lambda_{i}}{\partial\ln\lambda_B}
{\cal D}_{i}\int\frac{d^{3}k}{(2\pi)^{3}} 
f_{\mbox{Stats}}(T;m_{i},\lambda_{i}),
\nonumber\\
&\equiv& 
{\cal D}_{i}\int\frac{d^{3}k}{(2\pi)^{3}} 
f_{\mbox{Stats}}(T;m_{i},\lambda_{i}).
\label{ensemble_nn}
\end{eqnarray}
The above quantities can also be calculated for the  anti-particles be simply replacing 
$\lambda_i\rightarrow \lambda^{-1}_i$.
The grand canonical ensemble  for a gas of non-interacting 
multi-particle species obeys the relation~\cite{Gorenstein98a}, 
\begin{eqnarray}
Z(T,V;m_{1},\lambda_{1},\cdots,m_{n},\lambda_{n})=\prod_{i=1}^{n}Z(T,V;m_{i},\lambda_{i}).  
\end{eqnarray}
The isobaric partition function is calculated 
by taking the Laplace transformation of the grand partition 
function~\cite{Gorenstein1981d,Gorenstein98a,Auberson86b} 
\begin{eqnarray}
\hat{Z}(T,s;m_{i},{v_{\mbox{VdW}}}_H,\lambda_i)&\equiv&\int_{v_{\mbox{VdW}}}^{\infty}dV
\exp\left(-sV\right)Z(T,V;m_i,{v_{\mbox{VdW}}}_H,\lambda_i)
\nonumber\\
&=& 
1/\left[s-
\varphi_{{\cal S}_i}(T;m_{i},\lambda^{\star}_{i})\right], 
\label{isobaric_z1}
\end{eqnarray} 
where ${v_{\mbox{VdW}}}_H$  is the hadron's Van der Waals excluded volume. 
The effective fugacity becomes
\begin{eqnarray}
\lambda^{\star}_i&=&\exp\left(-{v_{\mbox{VdW}}}_H s\right) \lambda_i, 
\nonumber\\
\overline{\lambda}^{\star}_i&=&\exp\left(-{v_{\mbox{VdW}}}_H s\right) \lambda^{-1}_i,
\end{eqnarray}
for quark and antiquark, respectively.
In order to simplify the notation for different 
statistics cases, we use the ``Maxwell-Boltzmann''-like notation  
\begin{eqnarray}
\exp\left(-{v_{\mbox{VdW}}}_i s\right)
\varphi_{{\cal S}_i}(T;m_{i},\lambda_{i})
&~& \rightarrow~
\varphi_{{\cal S}_i}(T;m_{i},\lambda^{\star}_{i}) 
\nonumber\\
\sum\mspace{-23.0mu}\int
\exp\left(-{v_{\mbox{VdW}}}_i s\right)\varphi_{{\cal S}_i}(T;m_{i},\lambda_{i})
&~& \rightarrow
\sum\mspace{-23.0mu}\int
\varphi_{{\cal S}_i}(T;m_{i},\lambda^{\star}_{i}).
\end{eqnarray}
We have assumed the Van der Waals excluded volume ${v_{\mbox{VdW}}}_H$ is the same 
for all the known mass spectrum hadrons.  
The isobaric partition function with the isobaric ensemble
$\left(T,s\right)$ of a system is characterized by 
the external pressure $p=Ts$ rather than the fixed volume $V$. 
The point $s=s^*$ is the extreme right singularity point
in the limit of infinite external volume $V\rightarrow \infty$.
The isobaric partition function with multi-particle species 
reads~\cite{Gorenstein82a,Gorenstein83a,Gorenstein98a,Auberson86b} 
\begin{eqnarray} 
\hat{Z}(T,s;m_{1},{v_{\mbox{VdW}}}_1,\cdots,m_{n},{v_{\mbox{VdW}}}_n)= 
1/\left[s -\sum\mspace{-23.0mu}\int
\exp\left(-{v_{\mbox{VdW}}}_i s\right)\varphi_{{\cal S}_i}(T;m_{i},\lambda_{i})\right],
\label{isobaric_zs}
\end{eqnarray}
where
\begin{eqnarray}
\sum\mspace{-23.0mu}\int
\exp\left(-{v_{\mbox{VdW}}}_i s\right)\varphi_{{\cal S}_i}(T;m_{i},\lambda_{i})
=f_{H}(T,\lambda;s)+f_{Q}(T,\lambda;s).
\label{isobaric_den}
\end{eqnarray}
The first term $f_{H}(T,\lambda;s)$ denotes the ensemble for the known hadronic 
mass spectrum including their resonances and antiparticles 
in a hot and dense medium,
\begin{eqnarray}
f_{H}(T,\lambda;s) = 
\sum_{i=1}^{n}
\left[
\varphi_{{\cal S}_i}\left(T;m_{i},\lambda^{\star}_{i}\right)
+
\varphi_{{\cal S}_i}\left(T;m_{i},\overline{\lambda}^{\star}_{i}\right)
\right],
 ~(\mbox{fermions}), (\mbox{bosons}).
\label{isobaric_spectrum}
\end{eqnarray}
The masses of these particles are taken as listed 
in the particle data group book~\cite{databook2004}.  
In our numerical calculations, 
we have included the spectrum of all 76 nonstrange mesons 
and all 64 nonstrange baryons and their antiparticles 
as well as Hagedorn bubbles for the highly excited hadronic states. 
The effect of strangeness will be considered elsewhere. 
These Hagedorn bubbles naturally appear as fireballs in the heavy ions 
collision at high temperature.
The second term in Eq.(\ref{isobaric_den}) corresponds to the ensemble 
of Hagedorn bubbles which exist in a color singlet 
state~\cite{Gorenstein82a,Gorenstein83b,Tounsi98a,Gorenstein98a}. 
The isobaric ensemble for Hagedorn bubbles becomes
\begin{eqnarray}
f_{Q}(T,\lambda;s)&=&
\int_{V_0}^{\infty}\int_{m_0}^{\infty} dv dm
e^{-4v s} \rho(T,m,v,\lambda_Q)\varphi_{Q}(T;m)
\nonumber\\
&\equiv& 
\int_{V_0}^{\infty}\int_{m_0}^{\infty} dv dm
e^{-4v s} \rho(m,v)\varphi_{Q}(T;m), ~(\mbox{Maxwell-Boltzmann}),
\label{isobaric_bubble}
\end{eqnarray}
where $\rho(m,v)$ measures the mass spectral density {\em and} 
the volume fluctuation. 
The asymptotic behavior $m\gg T$ of the function $\varphi_{Q}(T;m)$ 
reads
\begin{eqnarray}
\varphi_{Q}(T;m)&=&\int \frac{d^3 k}{(2\pi)^3}
\exp\left(-\sqrt{k^2+m^2}/T\right), 
\nonumber\\
&=& \left[\frac{m^{2}T}{2\pi^{2}}\right]K_{2}(m/T) 
\nonumber\\
&\approx&
\left(\frac{mT}{2\pi}\right)^{3/2}e^{-m/T}.
\label{kinetic_app}
\end{eqnarray}
Furthermore, in addition to the bubble's quantum ground state, 
it is possible to take into account the higher quantum excitations 
of Hagedorn bubbles such as states with higher angular quantum 
momenta as well as mesonic and baryonic dominated 
bubbles (fermionic, bosonic fireballs), $\dots$ etc.
The density of states for the standard MIT bag model 
(with a sharp surface) reads
\begin{eqnarray}
\rho(m,v)=\delta\left(m-4Bv\right)
{\sigma(W,v)}_{\mbox{asym}},
\label{mit_sharp1}
\end{eqnarray}
where the gas of quarks and gluons is projected on a color singlet state 
and is confined in a spherical cavity with a specific volume $v=m/4B$. 
The subscript $\mbox{asym}$ denotes the asymptotic density
of states for the excited hadronic states.

The isobaric ensemble Eq.(\ref{isobaric_bubble}) for Hagedorn bubbles 
with sharp surfaces and density of states 
given by Eq.(\ref{mit_sharp1}) and Eq.(\ref{statedensity1}) 
can then be written as follows
\begin{eqnarray}
f_{Q}(T,\lambda:s)&=&
\int^{\infty}_{V_0}\int^{\infty}_{m_0}
\delta\left(m-4Bv\right){\sigma(W,v)}_{\mbox{asym}}
\left(\frac{mT}{2\pi}\right)^{3/2}e^{-m/T},
\nonumber\\
&=&
\int^{\infty}_{V_0} {\sigma(3Bv,v)}_{\mbox{asym}}
\left(\frac{4BvT}{2\pi}\right)^{3/2}e^{-4Bv/T}
\nonumber\\
&=& C \int_{V_0}^{\infty} dv v^{-(4\gamma-3/2)-1}e^{-4v(s-s_0)},
\label{bubble_eval_1}
\end{eqnarray}
where $W=m-Bv$ and the prefactor
\begin{eqnarray}
C=A \left(3B\right)^{-3\gamma-1}
\left(\frac{4BT}{2\pi}\right)^{3/2}.
\end{eqnarray}
Note, that we have introduced here the parameter $\gamma$ which describes
the internal color-flavor configuration of the bubble (see section IIA).
The quark and gluon bubble's internal isobaric pressure is given by
\begin{eqnarray}
s_0=\left(\frac{1}{3}u^{1/4}(3B)^{3/4}T-B\right)/T.
\label{bubble_eval_pr}
\end{eqnarray}
The pressure stems from the isobaric pressure at $P=Ts$.
The external pressure $p_Q=T f_{Q}(T,\lambda:s)$ 
for Hagedorn bubbles is written as follows 
\begin{eqnarray}
f_{Q}(T,\lambda:s)=
C \left(\frac{z_0}{V_0}\right)^{4\gamma-3/2}\int^{\infty}_{z_0}
dz z^{-(4\gamma-3/2)-1} e^{-z}
\label{bubble_eval_int}
\end{eqnarray}
where
\begin{eqnarray}
z_0=4V_0\left(s-s_0\right).
\label{z0singular}
\end{eqnarray}
In order to analyze the phase transition, it is useful 
to introduce the function
\begin{eqnarray}
\int^{\infty}_{z} dz
z^{-n-1} e^{-z}=\Gamma(-n,z),
\label{exp_integ}
\end{eqnarray}
with
\begin{eqnarray}
\Gamma(-n,z)=\frac{{(-1)}^n}{n!} \left[ E_1(z)-e^{-z}\sum^{n-1}_{j=0} 
\frac{{(-1)}^j j! }{ z^{j+1} } \right].
\end{eqnarray}
The exponential integral function $E_1(z)$ can be defined as
\begin{eqnarray}
E_1(z)=-\left[\gamma_{E}+ \ln(z) 
+ \sum^{\infty}_{n=1} \frac{ {(-1)}^n }{n n!} z^n \right],
\end{eqnarray}
where
$\gamma_E=0.5772$ is Euler's constant. 
It reads 
\begin{eqnarray}
\Gamma(0,z)=E_1(z),
\end{eqnarray}
for $n=0$.
Close to a phase transition when the external isobaric pressure $s$ 
reaches the hadronic bubble's internal isobaric pressure $s_0$ and 
with the definition of Eq.(\ref{z0singular}), we find
\begin{eqnarray}
\lim_{z\rightarrow 0} 
z^n \Gamma(-n,z)&=& 
1/n,~~~~~~\mbox{for}~~~~~~ n>0 
\nonumber \\ 
&=&\lim_{z\rightarrow 0}
\left(-\ln(z)\right),~~~~~~\mbox{for}~~~~~~ n=0.
\label{exp_asymp}
\end{eqnarray}
Note, that the expression is logarithmically divergent for $n\le 0$.
It might be of interest for the reader that the integration 
of the delta function over the volume instead of over the mass 
will lead to the mass spectral density of the bootstrap model
\begin{eqnarray}
\sigma(W,\left<V\right>)=
\int dv \rho(W,v) \delta\left(v-\left<V\right>\right).
\label{bootstrap_dens}
\end{eqnarray}

On the other hand, it is wonderful to note here 
that it is possible to start from Hagedorn bootstrap 
mass spectral density~\cite{Hagedorn65a,Frautschi71} 
to write the mass and volume density of states as follows
\begin{eqnarray}
\rho(m,v)=\left[ c m^{-a} e^{b m}\right]\delta\left(m-4Bv\right)
\label{bootstrap_dens_in}
\end{eqnarray}
where $B$ is the bag constant for Hagedorn bubbles. 
The order of the phase transition for the bootstrap mass spectral 
density with phenomenological input parameters $a$ and $b$ 
can be analyzed by using the isobaric partition function
\begin{eqnarray}
f_Q(T,\lambda;s)=
C \int_{V_0}^{\infty} dv v^{-(a-5/2)-1}e^{-4v(s-s_0)},
\label{hagedorn_isobaric}
\end{eqnarray}
where
\begin{eqnarray}
C=c\left(\frac{4BT}{2\pi}\right)^{3/2}
\left(4B\right)^{-a}.
\end{eqnarray}
The bubble's internal pressure is calculated by
\begin{eqnarray}
s_0=B\frac{bT-1}{T}.
\end{eqnarray}
Hence, the internal MIT bag color-flavor structure parametrized 
with the parameter $\gamma$ is related to the bootstrap model by 
\begin{eqnarray}
4\gamma-3/2 \propto a-5/2~~~~~~~~~~~~
\left(\gamma\sim\frac{a-1}{4}\right).
\end{eqnarray}
The case $a>5/2$ is of particular interest in the bootstrap model 
as it will lead (see below) to the phase transition. 
It corresponds to $\gamma> 3/8$ in the bag model. 
Values of $a>7/2$ corresponds to a first order phase transition. 
A higher order phase transition occurs in the range $7/2\ge a>5/2$.

The shape and order of the phase transition for this simple model 
will be discussed below.
However, we stress again that the density of states can be generalized 
to consider the effect of the bag model with a smoothed boundary 
where quarks and gluons are confined by the interacting potential 
in finite volume but have an extended surface. 
Furthermore, the above model can be extended, in principle, 
to take into account higher quantum excitations 
for Hagedorn bubbles in a straightforward fashion.  

\section{Excluded volume with small and large volume components}
The isobaric partition function for multi-particle species 
with a small and large volume components reads~\cite{Gorenstein99a}
\begin{eqnarray} 
\hat{Z}= 
1/
\left[\left[s\right] -\sum\mspace{-23.0mu}
\int \exp\left(-\left[{v_{\mbox{VdW}}} s\right]_i\right)
\varphi_{{\cal S}_i}(T;m_{i},\lambda_{i})\right],
\label{isobaric_small_large}
\end{eqnarray} 
where 
\begin{eqnarray}
\hat{Z}\equiv 
\hat{Z}(T,s;m_{1},{v_{\mbox{VdW}}}_1,\cdots,m_{n},{v_{\mbox{VdW}}}_n),
\end{eqnarray}
and
\begin{eqnarray} 
\sum\mspace{-23.0mu}\int
\exp\left(-\left[{v_{\mbox{VdW}}} s\right]_i\right)
\varphi_{i}(m_{i},\lambda_{i})
=f_{H}(T,\lambda;\{s\})+f_{Q}(T,\lambda;\{s\}), 
\end{eqnarray} 
and the set $\{s\}\equiv \{\xi_H,\xi_Q\}$, the parameter 
$[s]=\xi_H+\xi_Q$, the set
$\left\{\left[v_{\mbox{VdW}} s\right]\right\}\equiv 
\left\{\left[v_{\mbox{VdW}} s\right]_H,\left[v_{\mbox{VdW}} s\right]_Q\right\}$,
the parameter
$[{v_{\mbox{VdW}}} s]_H={v_{\mbox{VdW}}}_H \xi_H+{v^*_{\mbox{VdW}}}_H \xi_Q$ 
and the parameter
$[{v_{\mbox{VdW}}} s]_Q={v_{\mbox{VdW}}}_Q \xi_Q+{v^*_{\mbox{VdW}}}_Q \xi_H$.  
The excluded volume effects for small and large components 
(i.e. for the known mass spectrum hadrons and Hagedorn bubbles)
with the asymptotic approximation $v_Q\gg v_H$, respectively, read
\begin{eqnarray} 
{v_{\mbox{VdW}}}_H&=&4
v_H=\frac{16}{3}\pi r^3_H \nonumber\\ 
{v_{\mbox{VdW}}}_Q&=&4 v_Q=\frac{16}{3}\pi r^3_Q
\nonumber\\ 
{v^*_{\mbox{VdW}}}_H&=&
\frac{\left(v^{1/3}_H+v^{1/3}_Q\right)^3}{\left(v_H+v_Q\right)} v_H 
\approx v_H 
\nonumber\\ 
{v^*_{\mbox{VdW}}}_Q&=&
\frac{\left(v^{1/3}_H+v^{1/3}_Q\right)^3}{\left(v_H+v_Q\right)} v_Q 
\approx v_Q 
\nonumber\\
\lim_{v_Q\gg v_H}&~&
\left[\frac{\left(v^{1/3}_H+v^{1/3}_Q\right)^3}{\left(v_H+v_Q\right)}\right]
\approx 1.
\label{excl_vol_large_small}
\end{eqnarray}
The isobaric pressure for the hadronic mass spectrum reads
\begin{eqnarray}
\xi_H&=&f_{H}(T,\lambda;\{s\})
\nonumber\\
&=& \left\{
\sum_{i=1}^{n}
\exp\left(-\left[{v_{\mbox{VdW}}}_H \xi_H
+{v^*_{\mbox{VdW}}}_H \xi_{Q}\right]\right)
\varphi_{\mbox{Stats}}(T;m_{i},\lambda_{i})\right\},
\nonumber\\
&=& 
\sum_{i=1}^{n}
\left[
\varphi_{\mbox{Stats}}(T;m_{i},\lambda^{\star}_{i})
+
\varphi_{\mbox{Stats}}(T;m_{i},\overline{\lambda}^{\star}_{i})
\right],  
\label{spectrum_large_small}
\end{eqnarray}
where 
\begin{eqnarray}
\lambda^{\star}_i&=&e^{-v_H(4\xi_H+\xi_{Q})}\lambda_i,
\nonumber\\
\overline{\lambda}^{\star}_i&=&e^{-v_H(4\xi_H+\xi_{Q})}\lambda^{-1}_i,
\end{eqnarray}
are the
effective excluded volume fugacity for the hadrons and their anti-particles.
For the gas of Hagedorn bubbles, it reads
\begin{eqnarray} 
{\xi_Q}&=&{f_{Q}}(T,\lambda,\{s\})
\nonumber\\
&=&
\int_{V_0}^{\infty}\int_{m_0}^{\infty} dv 
dm\exp\left(-\left[{v_{\mbox{VdW}}}_Q \xi_Q
+{v^*_{\mbox{VdW}}}_Q \xi_{H}\right]\right) 
\rho(m,v)\varphi_{Q}(T;m), 
\nonumber\\ 
&=&\int_{V_0}^{\infty}dv 
\exp\left(-\left[{v_{\mbox{VdW}}}_Q \xi_Q
+{v^*_{\mbox{VdW}}}_Q \xi_{H}\right]\right)
\left[\int_{m_0}^{\infty} dm\rho(m,v)\varphi_{Q}(T;m)\right],
\nonumber\\ 
&=& 
\int_{V_0}^{\infty}dv 
\exp\left(-\left[{v_{\mbox{VdW}}}_Q \xi_Q
+{v^*_{\mbox{VdW}}}_Q \xi_{H}\right]\right) 
{\cal I}(v),
\nonumber\\
&\approx&
\int_{V_0}^{\infty}dv
\left. e^{-v(4\xi_Q+\xi_H)} {\cal I}(v)\right|_{x=\overline{x}}.
\label{fireb1} 
\end{eqnarray}
where
\begin{eqnarray}
x=\frac{m}{v} -B.
\end{eqnarray}
The last line in the above equation is a good approximation 
for sufficient large bubbles with $V_0\ge m_0/B$. 
The values $M_0$ and $V_0$ are the initial bag mass and volume, respectively. 
The fireball initial mass is taken just above the highest mass of 
the known hadronic mass spectrum particles (i.e. hadrons) listed 
in the particle data group book~\cite{databook2004} 
$M_0\equiv M_Q\approx 2.0 \mbox{GeV}$ and 
the initial volume is the phenomenological volume $V_0=M_Q/4B$ 
where $B$ is the MIT bag constant. 
The integration over mass density is evaluated as follows
\begin{eqnarray}
{\cal I}(v)&=&\int^{\infty}_{M_0} dm \rho(m,v)\varphi_Q(T;m)=
\int^{\infty}_{M_0} dm \rho(m,v)\varphi_Q(T;m)
\nonumber\\
&=&
\int^{\infty}_{M_0} dm \rho(m,v)
\left(\frac{mT}{2\pi}\right)^{3/2} e^{-m/T}
\nonumber\\
&\rightarrow&
\int^{\infty}_{x_0} dx f(x,v)e^{v h(x)}
\approx\left[f(x,v)
e^{v h(x)}\sqrt{\frac{2\pi}{-v h''(x)}} \right]_{x=\overline{x}}.
\label{fireb1b}
\end{eqnarray}
where the $\overline{x}$ value is determined by
\begin{eqnarray}
\frac{\partial h(x)}{\partial x}|_{x=\overline{x}}=0.
\end{eqnarray}
The values of $f(x)$ and $h(x)$ are model dependents and are determined in the Sec. V.

\subsection{One component excluded volume approximation}
It is worth to note that in the classical one component excluded VdW 
volume approximation, the isobaric partition function reduces to 
\begin{eqnarray}
\hat{Z}=\frac{1}{\left[s-
\sum_{i} 
\left(
\varphi_{\mbox{Stats}}(\beta;m_i,\lambda^{\star}_i)
+
\varphi_{\mbox{Stats}}(\beta;m_i,\overline{\lambda}^{\star}_i)
\right)
+\int_{V_0}^{\infty} dv e^{-v s} {\cal I}(v)
\right]},
\end{eqnarray}
where $\lambda^{\star}_i= e^{-v_H s}\lambda_i$ and 
$\overline{\lambda}^{\star}_i= e^{-v_H s}\lambda^{-1}_i$.
The isobaric pressure for Hagedorn bubbles reads
\begin{eqnarray}
f_Q&=&\int_{V_0}^{\infty} dv e^{-v s} {\cal I}(v),
\nonumber\\
&=&
\left.\int_{V_0}^{\infty} dv e^{-v \left[s-h(x)\right]} 
f(x,v)\sqrt{\frac{2\pi}{-v h''(x)}}\right|_{x=\overline{x}}.
\end{eqnarray}
Furthermore, in the original VdW approximation we have simply 
$v_{\mbox{VdW}}=v_0$ instead of $v_{\mbox{VdW}}=4 v_0$. 
In Eq.(\ref{Goren_isobaric_x}) the isobaric pressure 
for Hagedorn bubbles in a specific model is given by
\begin{eqnarray}
f_Q= C(\overline{x})\int^{\infty}_{V_0} dv v^{-4\gamma+1} e^{-v[s-h(\overline{x})]}.
\end{eqnarray}
The baryon density for the hadronic matter is calculated by
\begin{eqnarray}
n^{HG}_B= T\frac{\partial s} {\partial \mu_B} = \frac{n_{\cal N}}{n_{\cal D}},
\end{eqnarray}
where 
\begin{eqnarray}
n_{\cal N}&=&\
\sum^{\mbox{Baryons}}_i [n^{\star}_i-\overline{n}^{\star}_i]
\nonumber\\
&+&
\left[\frac{T}{C(\overline{x})}\frac{\partial C(\overline{x})}{\partial \mu_B}\right] f_Q
+
\left[T\frac{\partial h(\overline{x})}{\partial \mu_B}\right]
C(\overline{x})\int^{\infty}_{V_0} dv v^{-4\gamma+2} e^{-v[s-h(\overline{x})]},
\end{eqnarray}
and
\begin{eqnarray}
n_{\cal D}&=&
\left(1+ v_H \sum_i^{\mbox{Baryons}} 
\left[n^{\star}_i+\overline{n}^{\star}_i\right]
+ v_H \sum_i^{\mbox{Mesons}}
\left[n^{\star}_i+\overline{n}^{\star}_i\right] 
\right.
\nonumber\\
&+&
\left.
C(\overline{x})
\int^{\infty}_{V_0} dv v^{-4\gamma+2} e^{ -v[s-h(\overline{x})] }
\right).
\end{eqnarray}
The density for each particle reads,
\begin{eqnarray}
n^{\star}_i&=&n(T;m_{i},\lambda^{\star}_{i}),\nonumber\\
     &=&{\cal D}_i \int \frac{d^3 k}{(2\pi)^3} f_{\mbox{Stats}}(T;m_i,\lambda^{\star}_i).
\end{eqnarray}
The density of anti-particle is given by 
$\overline{n}^{\star}=n(T;m_{i},\overline{\lambda}^{\star}_{i})$.
%
%
\section{Applications with two models for the volume fluctuation}

We consider the isobaric function for Hagedorn bubbles formed
in the hadronic phase in the context of two different models.
The proper choice for the bubble's volume fluctuation plays the crucial rule 
to determine the order of the phase transition for bubbles 
with specific internal color symmetries.
The first model assumes the maximal volume fluctuation
for Hagedorn bubbles in the ground state.
Asymptotically, it is equivalent to the density of states which is 
considered extensively in the 
literature~\cite{Gorenstein83a,Gorenstein82a,
Gorenstein83b,Letessier88a,Tounsi98a}.
Nevertheless, this choice does not lead to 
a real deconfinement phase transition.
The second one assumes Gaussian-like volume fluctuation~\cite{Auberson86b}. 
It leads to a second order phase transition to the deconfined quark-gluon 
plasma. 
In fact the order of the phase transition depends strongly 
on a proper choice for the volume fluctuation.
The volume fluctuation is expected to be enhanced 
for low densities and high temperatures while it is supposed 
to be suppressed for high densities and low temperatures.
\subsection{Volume variation {\em {\`a} la} Gorenstein 
{\em et. al.}~\cite{Gorenstein83a,Gorenstein82a,Gorenstein83b}}
 
The density of states as function of mass and volume 
was introduced by differentiating the mass spectral density 
(e.g. Eq.(\ref{statedensity1}))
derived from the micro-canonical ensemble with respect to $v$. 
Nevertheless, Gorenstein 
{\em et. al.}~\cite{Gorenstein83a,Gorenstein82a,Gorenstein83b} 
assumed that the density of states 
for a bag of unspecified volume less than $v$ is determined by: 
\begin{eqnarray}
\sigma(W,\left<V\right>)=
\int_{\left<V\right>} dv \frac{\partial}{\partial v} \sigma(W,v)
\equiv \int_{\left<V\right>} dv \rho(W,v).
\label{Goreneq_1}
\end{eqnarray}
The volume fluctuation asymptotically behaves 
as the original mass spectral density
\begin{eqnarray}
\rho(W,v)=\frac{\partial}{\partial v} \sigma(W,v)
\approx \mbox{P}(W/v,v) \sigma(W,v)
\label{Goren_dens_1}
\end{eqnarray}
where
\begin{eqnarray}
\mbox{P}(x=W/v,v)\sim 
\frac{1}{3} u^{1/4} v^{-3/4} W^{3/4}\equiv \frac{1}{3} u^{1/4} x^{3/4},
\label{Goren_dens_v}
\end{eqnarray}
This term is derived directly from the Eq.(\ref{Goren_dens_1}) 
after the terms arrangement to take the form of Eq.(\ref{density_approx}). 
The density $\sigma(W,v)$ is given by Eq.(\ref{statedensity1}).
This volume fluctuation gives the maximal one for $0\le v\le \infty$ 
and it might be appropriate for hot and diluted matter.
However, this case is not expected to be correct 
for highly compressed matter since 
the volume fluctuation is expected 
to be suppressed in that regime.
In this case, the integral which appears 
in Eqs.(\ref{fireb1}) and (\ref{fireb1b}) becomes
\begin{eqnarray} 
{\cal I}(v)=C(\overline{x}) 
v^{-4\gamma+1}e^{v h(\overline{x})}, 
\label{Goren_isobaric_1}
\end{eqnarray} 
where 
\begin{eqnarray} 
h(x)=\frac{4}{3}
u^{1/4} x^{3/4} - (x+B)/T, 
\end{eqnarray} 
where $u$ is given by Eq.(\ref{u_dens1})
and 
$\overline{x}$ is the maximum
point for the function $h(x)$ and satisfies $\frac{\partial}{\partial x}
h(x)|_{x=\overline{x}}=0$.  
It should be noted that 
\begin{eqnarray}
\overline{x}= u T^{4}, 
\end{eqnarray} 
and 
\begin{eqnarray}
h(\overline{x})=\frac{1}{T}P_{QGP}= 
\frac{1}{T}\left[\frac{1}{3} u T^4-B\right]. 
\end{eqnarray} 
We have also 
\begin{eqnarray}
C(\overline{x})&=&\frac{1}{3} u^{1/4} A {\overline{x}^{-(3\gamma+1/4)}}
\left( \frac{(\overline{x}+B)T}{2\pi}\right)^{3/2}
\sqrt{\frac{2\pi}{-h''(\overline{x})}},
\nonumber\\
&=&2\frac{\sqrt{2\pi}}{3} u^{1/8} A {\overline{x}^{-3(\gamma-1/8)}}
\left( \frac{(\overline{x}+B)T}{2\pi}\right)^{3/2},
\end{eqnarray}
where $A$ is given by Eq.(\ref{a_dens1}).
The isobaric pressure for Hagedorn bubbles becomes
\begin{eqnarray} 
{f_Q}(v)&=&\xi_Q=\int^{\infty}_{V_0} dv
e^{-v\left(4\xi_Q+\xi_H\right)} {\cal I}(v), 
\nonumber\\ 
\xi_Q&=& C(\overline{x}) \int^{\infty}_{V_0} dv v^{-4\gamma+1}
e^{-v\left(4\xi_Q+\xi_H-h(\overline{x})\right)}, 
\nonumber\\ 
&=& C(\overline{x}) \left(\frac{z_0}{V_0}\right)^{4\gamma-2}
\int^{\infty}_{z_0} dz z^{-(4\gamma-2)-1} e^{-z}, 
\label{Goren_isobaric_x}
\end{eqnarray} 
where 
\begin{eqnarray}
z_0=V_0\left(4\xi_Q+\xi_H-h(\overline{x})\right). 
\label{Goren_singularity}
\end{eqnarray} 
Hagedorn bubble ensemble is solved self-consistently with 
the isobaric ensemble for the gas 
of the known hadronic mass spectrum particles 
\begin{eqnarray} 
\xi_H=f_H(T,\lambda;\xi_H,\xi_Q).  
\end{eqnarray}

%
%
\subsection{Volume variation {\em {\`a} la} Auberson {\em et. al.}~\cite{Auberson86b}}

The basic objects are glueballs as described within the simplest version of the
MIT bag model. In this spherical cavity approximation the energy for each
glueball state is given in terms of the cavity volume
\begin{eqnarray} 
m_i=\frac{y_i}{v^{1/3}}+Bv, 
\end{eqnarray} 
where the bag constant energy $B$ density simulating the confining forces 
is the only free parameter. 
The $y_i$ are pure numbers determined by the various modes of the
eight (Abelian) gauge fields filling the cavity and subject to appropriate
boundary conditions.  The mass $m_i$ and volume $v_{i}$ of the standard
glueball (static bags) are obtained by minimizing the expression with respect
to the volume $v$. 
It is natural to image that the ground state 
Hagedorn bubble's volume fluctuates in the equilibrium.
The mass expression is expanded 
around its minimum $m_i$ up to second order 
\begin{eqnarray} 
m\approx
m_0+\frac{8}{3}\frac{B^2}{m_0}(v-v_0)^2. 
\label{emass_vol_aub}
\end{eqnarray} 
The bubble's volume, $v$, dependence of the density of states 
is unknown, since the dynamics of the new degree of freedom $v$ 
is not known.  
However, the mass dependence is given by the mass spectral density 
$\sigma(W,v)_{\mbox{asym}}$ and $W=m-Bv$. 
On the other hand, the function 
\begin{eqnarray} 
\varphi(m)&\sim& e^{-m/T}, \nonumber\\ 
&\sim& e^{-m_0/T}\cdot
e^{-\frac{1}{T}\frac{8B^2}{3m_0}(v-v_0)^2}, 
\label{auberson_mass}
\end{eqnarray} 
asymptotically measures the classical volume fluctuation for the bag with mass $m_0$.
The density of states is extracted from the statistical ensemble
\begin{eqnarray}
\rho(m_0,v)\varphi_Q(m_0)&\sim& \rho(m,v) e^{-m/T},
\nonumber\\
&\sim&
\left[\sigma(m_0-Bv,v)_{\mbox{asym}}
e^{-\frac{1}{T}\frac{8B^2}{3m_0} \left(v-\frac{m_0}{4B}\right)^2}
\right]e^{-m_0/T}.  
\end{eqnarray} 
It is worth to remind the reader the microcanonical ensemble 
$\sigma(m_0-Bv_0,v_0)_{\mbox{asym}}$ is derived for 
a bag with specific volume $v_0$ and energy $W_0=m_0-Bv_0$. 
The bag's volume and mass 
are related by a strict constraint $m_0=4Bv_0$ 
for the standard MIT bag with a sharp boundary. 
The aim of the Auberson {\em et. al.}~\cite{Auberson86b} scenario is simply 
to soften the bag volume/mass constraint by allowing 
a small volume fluctuation $m_0\approx 4B\left(v_0+\delta v\right)$. 
Therefore, we can write 
\begin{eqnarray}
\rho(m,v)= 
\frac{3}{4}\mbox{P}(x,v)\sigma(x,v)_{\mbox{asym}}, ~\mbox{and}~ x=\frac{W}{v}
\label{aub_dens_m_v}
\end{eqnarray}
where 
\begin{eqnarray}
\mbox{P}(x,v)&=&\sqrt{\frac{8B^2}{2\pi T m}}
e^{-\frac{1}{T}\frac{8B^2}{3m} \left(v-\frac{m}{4B}\right)^2},
\nonumber\\
&=&
\sqrt{\frac{8B^2/T}{2\pi(x+B)}}
e^{-\frac{1}{T}\frac{8B^2}{3(x+B)} v\left(1-\frac{x+B}{4B}\right)^2},
\label{aub_v_fluct}
\end{eqnarray}
where the bag energy density $x=W/v$ is introduced. 
The above equation can be written as follows
\begin{eqnarray}
\mbox{P}(x,v)=\sqrt{{a}/\pi} \exp\left[-{a}(v-v_0)^2\right]
\sim \delta\left(v-v_0\right)
\end{eqnarray}
where $v_0=\frac{m}{4B}$ and ${a}=\frac{1}{T}\frac{8B^2}{3m}$. 
When the bag is deformed due to the highly thermal excitations, the smoothing 
function given by Eq.(\ref{deltasmooth1}) becomes appropriate to fit the phenomenology.

Hence with this choice we can write 
\begin{eqnarray} 
\rho(m,v)\varphi_Q(m)=
\frac{3}{4} A x^{-3\gamma-1}\sqrt{\frac{8B^2/T}{2\pi(x+B)}}
\left(\frac{x+B}{2\pi/T}\right)^{3/2} v^{-4\gamma} e^{v h(x)} 
\end{eqnarray}
where 
\begin{eqnarray} 
h(x)=\frac{4}{3} u^{1/4} x^{3/4} -
\frac{(3B-x)^2}{6(x+B)T} -(x+B)/T, 
\end{eqnarray} 
with $x\equiv W/v=m/v-B$.
In this approximation the density of states has a narrow Gaussian distribution 
function for high temperatures and a wide one for low temperatures. 
Although this choice is suitable to describe the volume fluctuation
for moderate density and temperature, it leads, unfortunately, to incorrect 
asymptotic behavior for both high and low temperatures.
It is expected to have a very strong volume fluctuation for high temperature 
and diluted matter while the overlap effect suppresses the volume fluctuation 
for low temperature.
As done in the previous section, the integration of the density of states over
the mass is evaluated using the saddle point approximation 
(e.g. Eqs. (\ref{fireb1}) and (\ref{fireb1b})).
It is given by 
\begin{eqnarray} 
{\cal I}(v)=C(\overline{x}) v^{-4\gamma+1/2}e^{v h(\overline{x})},
\label{aub_isobaric_1}
\end{eqnarray} 
where 
\begin{eqnarray} 
C(\overline{x})
=\frac{3}{4}A \overline{x}^{-3\gamma-1} 2\sqrt{2} B 
\frac{(\overline{x}+B)T}{(2\pi)^2}
\sqrt{\frac{2\pi}{-h''(\overline{x})}}.  
\end{eqnarray} 
The saddle point $\overline{x}$ is the maximum point for the function 
$h_2(x)$ and satisfies
$\frac{\partial}{\partial x} h(x)|_{x=\overline{x}}=0$.  
The isobaric singularity point
for the fireball becomes 
\begin{eqnarray} 
\xi_Q&=& C(\overline{x})
\int^{\infty}_{V_0} dv e^{-v\left(4\xi_Q+\xi_H-h(\overline{x})\right)}
v^{-(4\gamma-3/2)-1}, \nonumber\\ 
&=& C(\overline{x})\left(\frac{z_0}{V_0}\right)^{4\gamma-3/2}
\int^{\infty}_{z_0} dz z^{-(4\gamma-3/2)-1} e^{-z}, 
\label{aub_isobaric_x}
\end{eqnarray} 
where 
\begin{eqnarray}
z_0=V_0\left(4\xi_Q+\xi_H-h(\overline{x})\right).  
\label{aub_singularity}
\end{eqnarray} 

\subsection{Densities for the hadronic mass spectrum particle
gas and the Hagedorn bubble gas}
%
The pressure for a gas comprising the mass spectrum of all known hadrons and 
fireballs reads
\begin{eqnarray}
p=T(\xi_H+\xi_Q),
\end{eqnarray}
Here the fireballs are simply the Hagedorn states.
The total isobaric pressure is calculated from the extreme right singularity as
given Eqs.(\ref{isobaric_zs}) and (\ref{isobaric_den}) 
and demonstrated 
by Eq.(\ref{isobaric_small_large}) for small and large volume components.
The isobaric pressures for the gas 
of the hadronic mass spectrum particles 
and the gas of the fireballs are given, respectively, by 
\begin{eqnarray}
\xi_H=
\sum_i \varphi_{S_i}(T;m_i,\lambda^{\star}_i),
\nonumber\\
\xi_Q=\int^{\infty}_{V_0} dv e^{-v\left(4\xi_Q+\xi_H\right)}{\cal I}(v),
\end{eqnarray}
where $\lambda^{\star}_i=e^{-v_H\left(4\xi_H+\xi_Q\right)}\lambda_i$ and
$\lambda_i=\exp\left(B_i\mu_B/T+S_i\mu_S/T+\cdots\right)$.
The baryonic density for the hadronic gas can be calculated by differentiating 
the total pressure with respect to the baryonic chemical potential
\begin{eqnarray}
n_B&=&n^{\mbox{Spectrum}}_B+n^{\mbox{Fireballs}}_B
\nonumber\\
&=& T\frac{\partial}{\partial \mu_B}\xi_H
  + T\frac{\partial}{\partial \mu_B}\xi_Q.
\end{eqnarray}
%

\section{The order of the phase transition to Quark-Gluon 
droplets or plasma state}

The order of the phase transition from the hadronic phase 
to quark-gluon droplets or plasma state is determined 
by scrutinizing the properties of the isobaric pressure 
for Hagedorn bubbles when integrating over the volume.
The volume fluctuation in the model is calculated 
after rescaling the mass and volume $(W,v)$ variables  
in the (grand-) micro-canonical density
to mass density and volume $(x=W/v,v)$ variables. 
This mass density/volume scaling is justified by the assumption 
that $v\propto W$ and $v=\frac{m}{4B}$ for the MIT bag mode.  

Although Hagedorn bubble's internal color-flavor symmetry 
is important to determine the order and shape 
of the phase transition, it is not the only the criteria.  
The bubble volume fluctuation after mass/volume scaling is, indeed, 
crucial to fix the order of the phase transition, in particular
for hot and diluted hadronic matter.
It is reasonable to expect that the volume fluctuation varies differently 
for compressed matter where the bubbles start to overlap each other,
and for diluted and hot matter where the bag's surface is expected to dissociate 
spontaneously near the critical temperature.
The hadronic phase comprises of all the known hadronic states 
(mass spectra of baryons, mesons and their resonances)
as well as the highly excited hadronic fireballs,
i.e. Hagedorn bubbles with higher internal color-flavor symmetry.  
Each hadronic fireball is approximated as an ideal gas of quarks and gluons 
moving freely inside the bag within a specific color-flavor quantum state.

However, our numerical calculations show that Hagedorn bubbles appear 
only in highly dense matter for large baryo-chemical potential. 
In contrast they are unlikely to appear in diluted matter 
for low baryo-chemical potential even when the system is approaching 
the critical temperature. 
This observation apparently contradicts to the common thought 
that Hagedorn bubbles always show up below $T_c$ for dilute 
and hot hadronic matter. 
Usually Hagedorn states are supposed to develop below $T_c$.
However this apparent contradiction is resolved easily by noting that 
the bubble's internal pressure is quartic temperature dependent
\begin{eqnarray}
p\sim \frac{1}{3} u T^4 - B.
\end{eqnarray}   
This pressure grows up and exceeds the external hadronic pressure quickly 
with respect to the temperature. Below the critical temperature, 
these Hagedorn bubbles are suppressed by the external hadronic 
pressure of the mass spectrum gas.
Furthermore, in the numerical calculations we have taken a relatively small
Van der Waals repulsive volume for the mass spectrum particles 
(e.g. ${v_{\mbox{VdW}}}_H=4\times 0.0654 \mbox{fm}^3$) 
while a relatively large one for Hagedorn bubbles 
(e.g. ${v_{\mbox{VdW}}}_Q=4\times 4.190 \mbox{fm}^3$). 
This {\em small-large} excluded volume scenario suppresses 
these Hagedorn bubbles with relatively large excluded volumes strongly. 
The numerical calculations also show that 
smaller values for the initial Hagedorn bubble's excluded volume 
enhance the appearance of Hagedorn states significantly 
at temperatures below $T_c$ for diluted hadronic matter.
The small excluded volume for the mass spectrum particles and
the large excluded volume for Hagedorn states effect concerns 
us in the present work in order 
to demonstrate clearly the different scenarios 
for the phase transition diagrams. 
These relatively large Hagedorn bubbles can 
in principle appear in the system 
due to the high thermal excitations of the vacuum.
Whenever they appear in dilute hot matter, however, 
they are mechanically unstable. 
The bubbles with an internal pressure less than the external one 
collapse and disappear rapidly.
However, when the internal pressure of Hagedorn bubbles 
reach the external one from below near the critical temperature, 
they will eventually expand and explode because 
the bubble's overlap affect is negligible in this regime.
The resulting exploding bubbles expand rapidly forming quark-gluon droplets 
by merging with each other eventually filling the whole space.  
This big quark-gluon droplet loses its internal structure as it expands 
and undergoes a true deconfinement phase transition.
On the other hand, Hagedorn bubbles can exist in the compressed 
and cold hadronic phase. 
Their appearance at large baryonic chemical potential is essential to soften
the equation of state being a measure of the highly excited mass spectrum 
of compressed hadronic matter.
These bubbles expand slowly in dense matter because of the overlap 
with other bubbles. 
As the system is compressed further at low temperature, many bubbles 
are likely to merge into each other to form denser Hagedorn bubbles 
with more complicated internal color-flavor structure.
As this effect increases with baryon density, 
the bubble volume fluctuation will decrease correspondingly.
Therefore, it is expected that the hadronic system undergoes 
a phase transition to foam of highly dense Hagedorn bubbles 
at large baryonic chemical potential and low temperatures. 
Furthermore, when the foam is heated due to e.g. compression, 
the internal surfaces dissolve the bubbles merge to form bigger droplets.
At some point the quark-gluon droplet surface collapses
and the system undergoes a phase transition 
to fully deconfined quark-gluon plasma.

The integration over volume for Hagedorn bubble's isobaric pressure 
reads
\begin{eqnarray} 
\xi_Q\sim \Xi(-\alpha,z_0)=C(x_0) \int^{\infty}_{V_0} dv
v^{-\alpha} e^{-v(4\xi_Q+\xi_H-s_0)}, 
\end{eqnarray} 
where 
$z_0=V_0(4\xi_Q+\xi_H-s_0)$.
The exponent $\alpha$ is some phenomenological parameter depending 
on both the bubble's internal color symmetry
and the volume fluctuation of the bubble itself.  
The asymptotic behavior near
the phase transition point acts as 
\begin{eqnarray} 
\lim_{z_0\rightarrow
0^+}\Xi(-\alpha,z_0) &\sim& 
\lim_{z_0\rightarrow 0^+}
z^{\alpha-1}_0\Gamma(-\alpha+1,z_0)
\rightarrow~~\mbox{finite for} ~~~\alpha >1 \nonumber\\ 
&\sim& 
\lim_{z_0\rightarrow 0^+}
\ln(z_0)\rightarrow~~\mbox{diverges for}~~~\alpha=1, \nonumber\\ 
&\rightarrow&~~\mbox{diverges for}~~~~\alpha\le 1, 
\end{eqnarray}
where $z_0=V_0(4\xi_Q+\xi_H-s_0)$. 
The factor $4$ which appears in front of $\xi_Q$ comes from the hard core 
Van der Waals repulsion for large and small components 
for Hagedorn bubbles and the mass spectrum particles, respectively, 
as discussed in Sec. IV. 
Therefore, the convergence or divergence of the isobaric pressure near 
the point of the phase transition can be summarized as follows 
\begin{eqnarray} 
\lim_{z_0\rightarrow 0^+}
\xi_Q &\rightarrow& \mbox{diverges} ~~~~~~~ (\alpha-1\le 0),     \nonumber\\
\lim_{z_0\rightarrow 0^+} \xi_Q &\rightarrow& \mbox{finite} ~~~~~~~
(\alpha-1>0).  
\end{eqnarray} 
Note, that the phase transition does not exist for Hagedorn bubbles 
characterized by volume structure $\alpha\le 1$.  
Hagedorn bubble's external pressure diverges and subsequently the 
internal pressure is always less than the external one
and consequently these bubbles collapse and are strongly suppressed 
in the hadronic phase.

The phase transition occurs only for bubbles with a volume parameter 
$\alpha>1$.  
Nonetheless, one can expect a rapid and smooth phase transition for bubbles 
with internal structure $\alpha=1^+$.
In this case the  bubble external pressure is large but finite 
near the phase transition.  
Hence, it is possible that expanding Hagedorn bubbles emerge and
form big quark-gluon droplet which occupies most of the available space. 
The instability of Hadronic bubbles increases when the hadronic 
external pressure increases to large values just below the phase transition.
The phase transition will take place when Hagedorn bubble's
internal pressure becomes equal to the external pressure of the hadronic gas
\begin{eqnarray}
4\xi_Q + \xi_H=s_0.
\end{eqnarray}
Despite the quark-gluon droplet forming rapidly and expanding
quickly, not the whole hadronic gas undergoes a phase transition.  
In these circumstances, it would be difficult to distinguish between the hadronic phase
and the quark-gluon plasma and the system undergoes a smooth but rapid cross-over phase
transition.  
The order of the phase transition is determined by the discontinuity 
of the n-th derivative of the isobaric pressure 
at the surface of Hagedorn bubble just below the phase transition line.
If the first derivative is discontinuous, 
then the system undergoes a first order phase transition.  
When the first derivative is continuous, then the second derivative 
should be inspected.  
If the first derivative is continuous while the second derivative
is discontinuous, then the system undergoes a second order phase transition.
Furthermore, when the first and second derivatives are both continuous while
the third derivative is discontinuous then a third order phase transition
takes place and so on. 
Consequently, the $n^{th}$ order phase transition is determined by 
the discontinuity of the $n^{th}$ derivative of the isobaric pressure 
at the surface of Hagedorn bubble.
The first derivative 
\begin{eqnarray} 
(4\xi_Q+\xi_H-s_0)'&\propto&  \xi'_Q/\Xi(-\alpha+1,z_0) \propto
1/z_0^{\alpha-2}\Gamma(-\alpha+2,z_0), 
\nonumber \\ 
&\propto& z_0^{-(\alpha-1)+1}, z_0=V_0\left(4\xi_Q+\xi_H-s_0\right),
\end{eqnarray} 
at the surface of Hagedorn bubble is continuous only with the volume 
parameter 
$-(\alpha-1)+1>0$.  
The prime notation in $X'$ indicate the partial derivative with respect to the
thermodynamical ensemble such as the temperature and  chemical potential
(e.g. $X'\equiv \frac{\partial X}{\partial T}$, $\frac{\partial X}{\partial \mu_B}$) .

The continuity of the second and third derivatives 
\begin{eqnarray} 
(4\xi_Q+\xi_H-s_0)''&\propto&
\Xi(-\alpha+2,z_0)/\Xi^3(-\alpha+1,z_0), \nonumber \\ 
&\propto&
z_0^{-2(\alpha-1)+1}, z_0=V_0\left(4\xi_Q+\xi_H-s_0\right), 
\end{eqnarray} 
and 
\begin{eqnarray} 
(4\xi_Q+\xi_H-s_0)'''
&\propto&
\left[\Xi(-\alpha+3,z_0)\Xi(-\alpha+1,z_0)-
3\Xi^2(-\alpha+2,z_0)\right]/\Xi^5(-\alpha+1,z_0),
\nonumber \\ 
&\propto& z_0^{-2(\alpha-1)+1}, z_0=V_0\left(4\xi_Q+\xi_H-s_0\right), 
\end{eqnarray} 
appears only for 
$-2(\alpha-1)+1>0$ 
and 
$-3(\alpha-1)+1>0$, 
respectively.
The discontinuity of the $n^{th}$ derivatives 
\begin{eqnarray}
(4\xi_Q+\xi_H-s_0)^{(n)} &\propto& \lim_{z_0\rightarrow 0} \ln(z_0) 
\rightarrow \mbox{diverges}, z_0=V_0\left(4\xi_Q+\xi_H-s_0\right),
\end{eqnarray} 
are determined by $-n(\alpha-1)+1\le 0$.  
The $n^{th}$ derivative is continuous
\begin{eqnarray} 
(4\xi_Q+\xi_H-s_0)^{(n)} &\propto& \lim_{z_0\rightarrow 0}
z_0^{-n(\alpha-1)+1}\sim 0, 
\end{eqnarray} 
for $-n(\alpha-1)+1>0$.  
Hence the condition for the n-th order phase transition reads 
\begin{eqnarray} 
\alpha > 1 + 1/n.  
\end{eqnarray} 
The $1^{st}$, $2^{nd}$, $n^{th}$ order phase transitions 
are given by $\alpha>2$, $2\ge\alpha>1+\frac{1}{2}$ and 
$\left[1+\frac{1}{n-1}\right]\ge\alpha>\left[1+\frac{1}{n}\right]$, 
respectively.  
However, a cross-over phase transition takes place only 
when all derivatives of $n^{th}$ order are becoming equal.  
This corresponds to the volume parameter $\alpha=1^+$.
The external pressure of Hagedorn bubble diverges smoothly 
as $\alpha$ approaches $1^+$ 
and the system subsequently undergoes a phase transition.

The order of the phase transition depends on both 
the bubble volume fluctuation and its internal symmetry.
We demonstrate this dependence by considering 
the two different models for the volume fluctuation
as presented in section III.
The volume fluctuation structure as given by 
Gorenstein {\cal et. al.}~\cite{Gorenstein83a,Gorenstein82a,Gorenstein83b}
(see Eq.(\ref{Goren_dens_1}))
with specific internal color symmetry 
is handled by the redefinition
\begin{eqnarray} 
\alpha=4\gamma -1.  
\end{eqnarray} 
In this approach, the ground state Hagedorn bubble 
expands freely to infinity, 
i.e. the maximal volume fluctuation for 
the bubble in the ground state.
We have find then that the phase transition is not possible 
for colored bag of gluons and flavorless quarks with 
structure $\gamma=\frac{1}{2}$ as $\alpha=1$.
The first and second order phase transitions are possible 
for bags with color structures characterized by $\gamma>\frac{3}{4}$ 
and $\gamma>\frac{5}{8}$, respectively. 
Furthermore, the $n^{th}$ order phase transition takes 
place at $\gamma>\frac{2n+1}{4n}$.
The colorless bag of gluons and flavorless quarks
has $\gamma=\frac{3}{2}$ undergoes a first order phase transition.

On the other hand, the Gaussian volume fluctuation
given by Eq.(\ref{aub_dens_m_v}) and
suggested by Auberson {\cal et. al.}~\cite{Auberson86b}
corresponds to
\begin{eqnarray} 
\alpha=4 \gamma -1/2.
\end{eqnarray}
for Hagedorn bubbles with specific internal color-flavor symmetry.
Hence, the bubbles with color-flavor structure $\gamma=1/2$ 
undergo a third order phase transition 
to the real deconfined quark-gluon plasma as $(\alpha=3/2)$.  
The second order phase transition occurs 
for bubbles with $\frac{5}{8}\ge \gamma>\frac{1}{2}$
while those with $\gamma>\frac{5}{8}$ have 
a first order phase transition. 
Hence, the Hagedorn bubbles in the color singlet state 
and internal color-flavor structure $\gamma=3/2$ ($\alpha=11/2$) 
undergoes a first order phase transition.

Generally speaking, the volume fluctuation basically depends on 
the quantum wave-function of the quark-gluon cluster and varies 
with respect to temperature and chemical potential. 
However, for the bag model with a deformed boundary, 
the bag volume is expected to fluctuate around 
the mean volume $(v-v_0)$ uniformly which can be approximated by a Gaussian 
for some range of chemical potentials.
Nonetheless, this approximation is not necessarily
correct for the entire $\mu-T$ phase diagram.
The Gaussian approximation fails to measure the real volume fluctuation 
for Hagedorn bubbles embedded in the compressed matter in particular 
when these bubbles overlap with each other. 
In this regime, the bubbles squeeze into each other and 
there is a little room for further expansion. 
Hence, the volume fluctuation is likely damped 
for highly compressed and cold matter.
Moreover, the Gaussian approximation also fails for hot and dilute
baryonic matter regime as the hadronic bag surface dissociates 
spontaneously near the critical temperature and Hagedorn bubbles expand 
rapidly and resulting in a first or even higher order phase transition 
to quark-gluon droplets or real deconfined quark-gluon plasma.

\section{Results and discussions}

\begin{figure} 
\includegraphics{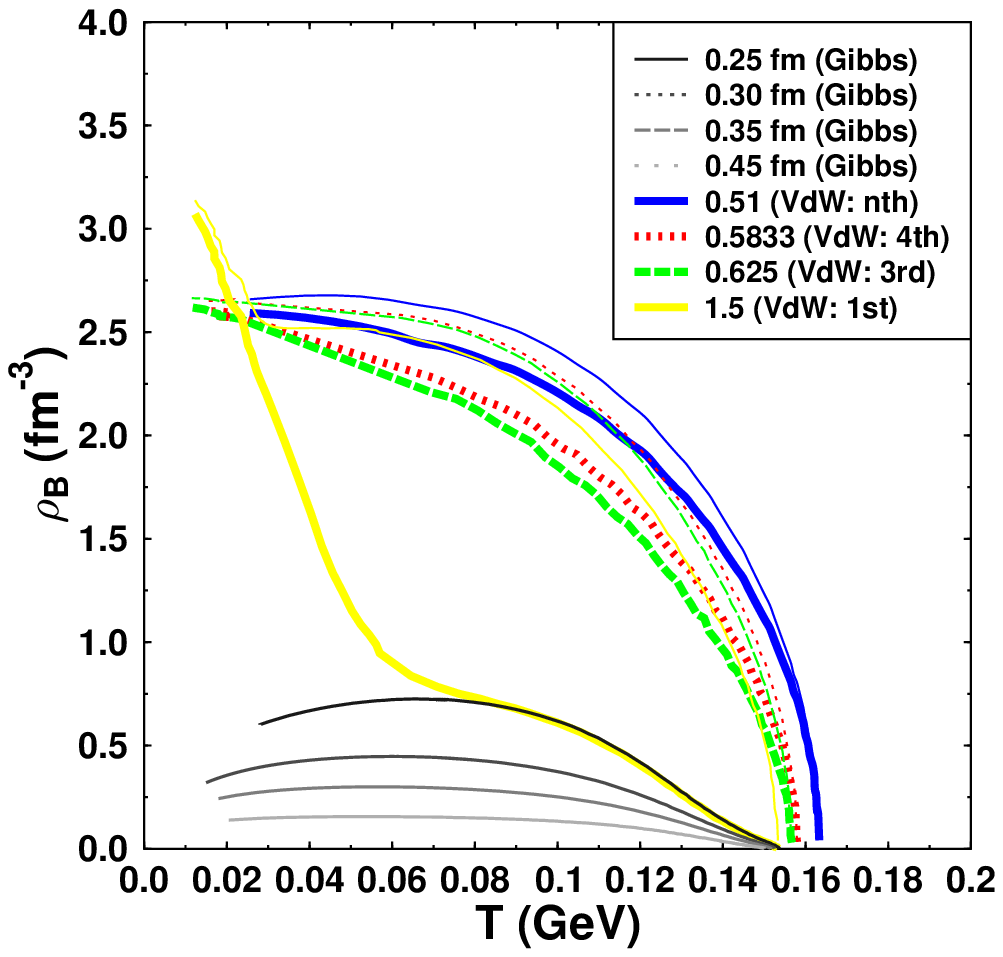}
\caption{\label{spectrum_rho_t} 
The phase transition diagram in the baryonic density and temperature ($\rho_B-T$)  plane. 
The baryonic density is calculated for the hadronic phase 
below the phase transition line while for QGP is calculated above the phase transition line. 
The low thin lines show the phase diagram calculated using the Gibbs construction 
with various hadronic excluded volume. 
In Gibbs construction, Hagedorn bubbles (i.e. hadronic bubbles)
are not included in the hadronic phase.
The thick lines show $\rho^{HG}_B$ for the hadronic gas including 
Hagedorn bubbles with initial radii $r_{\mbox{FB}}=1.0$ fm.
The Hagedorn bubbles' internal color-flavor structure 
is given by Eq.(\ref{Goren_dens_1}). 
The bag constant is taken to be $B^{1/4}=210$ MeV.
The above thin lines show $\rho^{QGP}_B$ for the QGP phase above 
the phase transition line for hadronic matter consisting
Hagedorn bubbles with various internal color structure.} 
\end{figure}

\begin{figure} 
\includegraphics{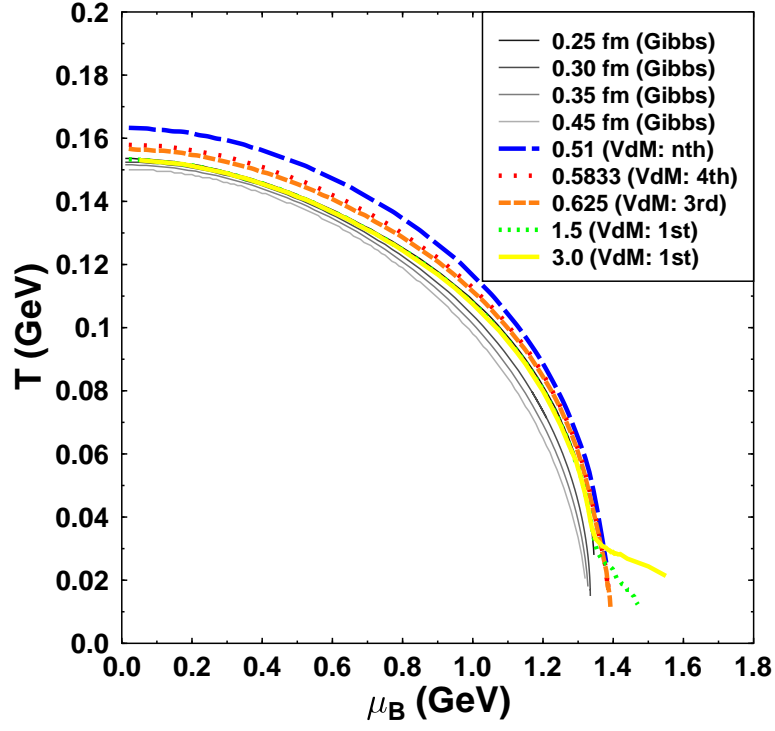} 
\caption{\label{spectrum_mu_t} 
Same as Fig.~\ref{spectrum_mu_t} but in the baryonic chemical potential 
and temperature ($\mu_B-T$) plane.
At low temperatures, the hadronic matter consisting 
both the hadronic mass spectrum particles and Hagedorn bubbles shift 
the phase transition line to larger chemical potentials than hadronic matter 
consisting only the hadronic mass spectrum particles.
The bag constant for Hagedorn bubbles is taken to be $B^{1/4}=210$ MeV.
The critical temperature $T_c(\mu_B=0)$ is found sensitive to $B$.} 
\end{figure}

\begin{figure}
\includegraphics{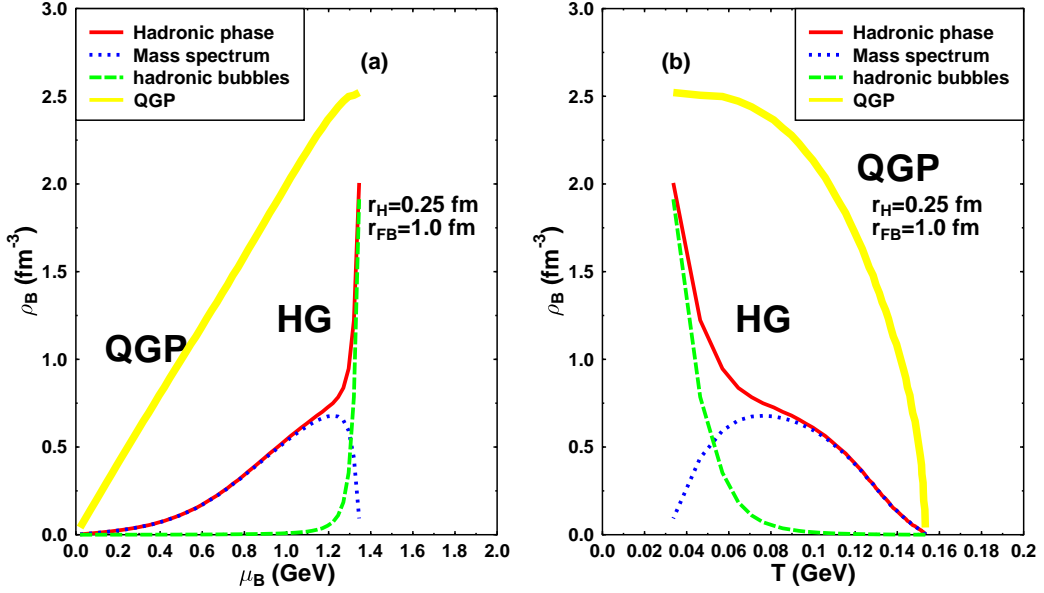}
\caption{\label{spectrum_d_m_t}
The baryonic densities for both Hagedorn bubbles (i.e. hadronic bubbles) 
and the hadronic mass spectra particles 
in the hadronic phase below the phase transition line.
The above thick line shows the baryonic density $\rho^{QGP}_B$ for the QGP phase
above the phase transition line for hadronic matter consisting 
hadronic mass spectrum particles and
Hagedorn states with initial radius $r_{\mbox{FB}}=1.0$ fm
for Hagedorn bubble's volume fluctuation.
The excluded volume for the hadronic mass spectrum 
particles is taken to be $r_H=0.25$ fm. 
At high temperatures and low chemical potentials, 
the hadronic mass spectrum particles is dominated 
while Hagedorn bubbles become the dominant
for large chemical potentials and low temperatures.
a) versus baryonic chemical potential $\mu_B$.
b) versus temperature.}
\end{figure}

\begin{figure} 
\includegraphics{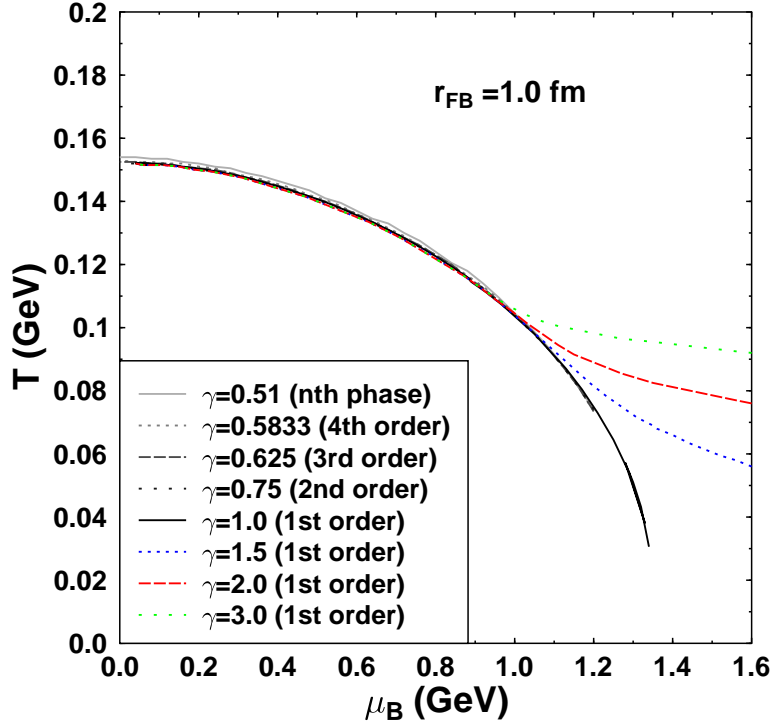} 
\caption{\label{Goren_phase1} 
The phase transition diagram to QGP
in the baryonic chemical potential and temperature ($\mu_B-T$) plane
for the hadronic phase consisting all 
the known hadronic mass spectrum particles as well as Hagedorn bubbles 
(i.e. hadronic bubbles) 
with various color-flavor internal symmetry and a specific volume fluctuation 
as in the model (I) given in the text.
The phase transition diagram is displayed for various values of the color-flavor structures 
as determined by the pre-exponential power 
$\gamma(\alpha)$ factor 0.5, 1.00, 1.50, 2.00 and 3.00. 
The hadronic phase is chosen to be an ideal gas of the hadronic mass spectrum particles 
with the excluded volume of $r_H$=0.30 fm 
as well as Hagedorn bubbles with an initial radii of $r_{\mbox{FB}}$=1.0 fm.
The density of states for Hagedorn bubbles is given by Eq.(\ref{Goren_dens_1}).
The small excluded volume component for the hadronic mass spectrum particles 
and the large excluded volume component for the Hagedorn states 
approximation is considered in the numerical calculations.} 
\end{figure}

\begin{figure} 
\includegraphics{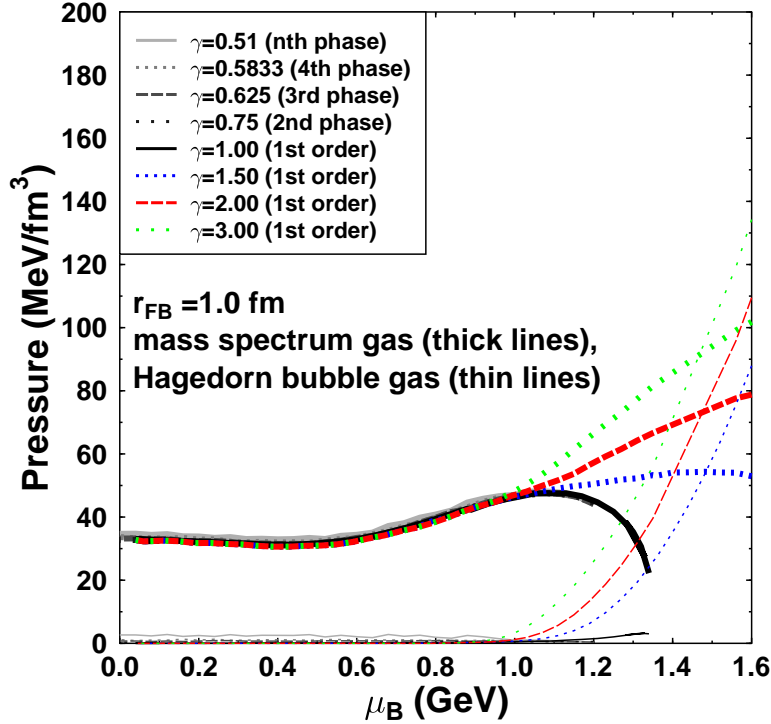} 
\caption{\label{Goren_pres1}
The pressure for the gas of hadronic mass spectrum particles and 
the pressure for the gas of Hagedorn states (i.e. hadronic bubbles)
versus the baryonic chemical potential $\mu_B$ at temperatures 
just below the phase transition line. 
These pressures are displayed for Hagedorn bubbles with various values 
of internal color-flavor structure $\alpha=4\gamma-1$ 
factor of 0.5, 1.00, 1.50, 2.00 and 3.00 
where the density of states is given by Eq.(\ref{Goren_dens_1}).
It is shown that the gas pressure for Hagedorn bubbles becomes larger than that 
for the hadronic mass spectrum particles for large baryonic chemical potentials.
The small excluded volume component for the mass spectrum particles and the large
excluded volume component for Hagedorn states approximation
is considered in the numerical calculations.} 
\end{figure}

\begin{figure}
\includegraphics{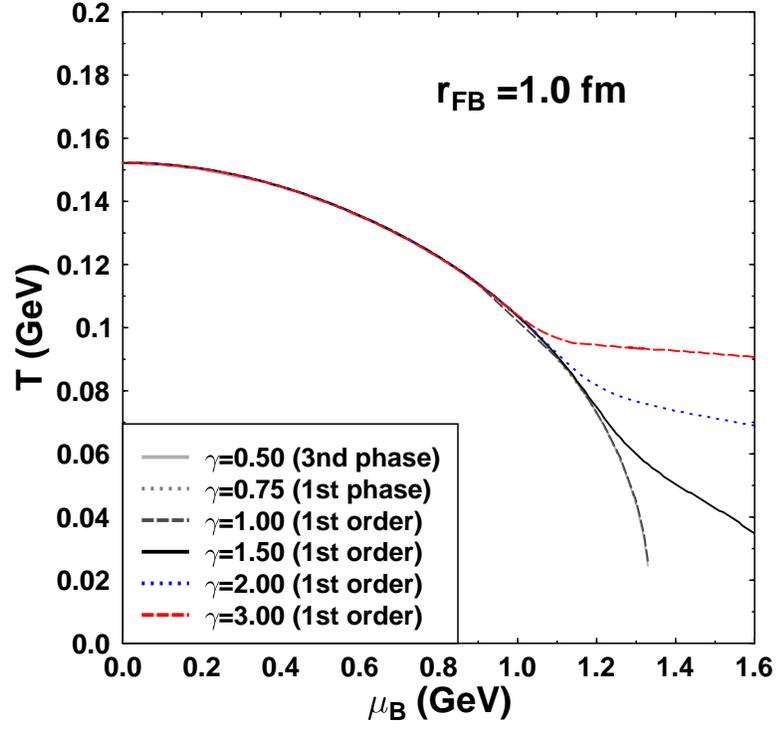}
\caption{\label{Aub_phase1}
Same as Fig.~\ref{Goren_phase1} but for density of states including the volume fluctuation
consistently as given in model (II) (see Eq.(\ref{aub_dens_m_v})).}
\end{figure}

\begin{figure} 
\includegraphics{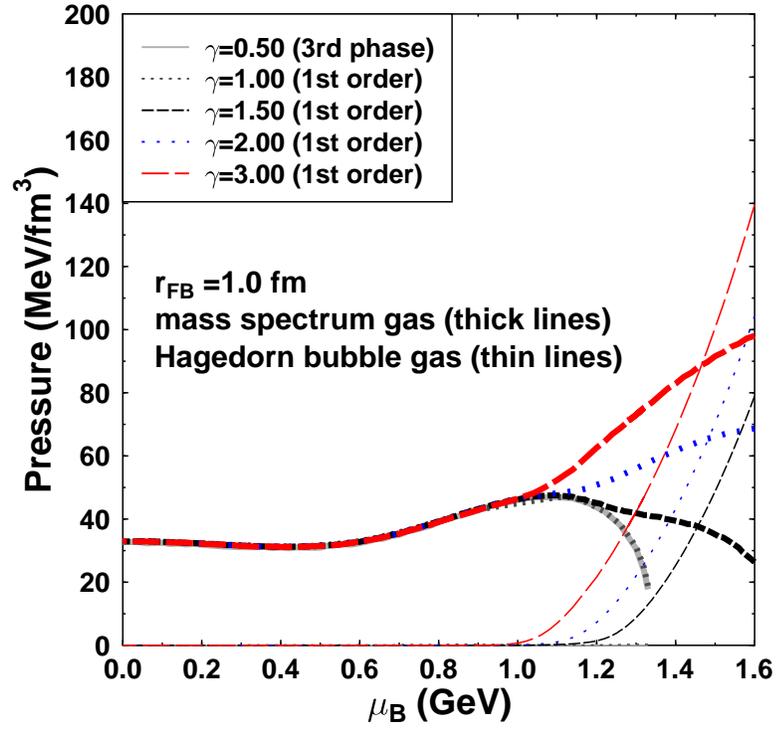} 
\caption{\label{Aub_pres1}
The same as Fig.~\ref{Goren_pres1} but for the density Eq.(\ref{aub_dens_m_v}), i.e.  model (II).} 
\end{figure}

In the following, the hadronic matter is treated as Van der Waals gas 
consisting of all the known particles as given 
in the particle data group book~\cite{databook2004}. 
Hereinafter, we call these particles the mass spectrum particles.
The highly excited hadronic states are also taken into account 
and they are taken as Van der Waals gas of Hagedorn bubbles. 
The density of states for Hagedorn bubbles is derived from 
Laplace inverse of the mixed canonical ensemble of blob
of quarks and gluons with 
specific internal color-flavor structure. 
These states are similar to Hagedorn states 
in the bootstrap model~\cite{Hagedorn65a,Frautschi71} 
where the density of states for the hadronic matter 
is modified significantly for the highly compressed 
or heated matter because the appearance 
of exotic hadronic (i.e. {\em Hagedorn}) bubbles with large masses 
$m_H>2.0$ GeV in the hadronic phase.
The rich mass spectrum of Hagedorn states is essential for softening 
the equation of state in particular for a large baryonic chemical potential.
In order to demonstrate the importance of Hagedorn bubbles 
in the hadronic phase, we have studied the quark-gluon plasma 
phase transition with and without Hagedorn bubbles.  
The effect of the initial volume fluctuation of Hagedorn bubbles 
in the phase transition is also considered.   
The density of states for Hagedorn bubbles is considered for the quark 
and gluon bags with various internal color-flavor symmetries range 
from colored bubbles 
(or colorless bubbles with the minimum color-flavor correlation) 
to singlet ones with strong color-flavor correlations specified 
by the parameter 
$\gamma_{\mbox{color}}\le\gamma\le\gamma_{\mbox{singlet}}\ll 
\gamma_{\mbox{strong}}$ 
and in the context of two approaches for the bubble's volume fluctuation. 
The first approach is based on Gorenstein 
{\em et. al.}~\cite{Gorenstein83a,Gorenstein82a,Gorenstein83b} 
ansatz for the maximum volume fluctuation while second one is based 
on Auberson {\em et. al.}~\cite{Auberson86b} ansatz for Gaussian 
like volume fluctuation.
The bag constant for the hadronic pressure is taken $B^{1/4}= 210 \mbox{MeV}$ 
and the initial bag volume starts at 
$\frac{4}{3}\pi \left(1.0 \mbox{fm}\right)^3$.
In Figs.~\ref{spectrum_rho_t}, \ref{spectrum_mu_t} 
and \ref{spectrum_d_m_t}, 
the one exclude volume component approximation is considered 
in the numerical calculations while the excluded volume 
with small and large volume components approximation is considered in 
Figs.~\ref{Goren_phase1}, \ref{Goren_pres1}, \ref{Aub_phase1} 
and \ref{Aub_pres1}.

We display the baryonic density and temperature $(\rho_B-T)$ 
phase transition diagram in Fig.~\ref{spectrum_rho_t}. 
The thin lines depict the phase transition diagram for hadronic 
matter consisting of Van der Waals gas of the known particles without 
Hagedorn bubbles with various excluded volumes.
In this approximation, the phase transition to quark-gluon plasma 
is calculated using the standard Gibbs construction scheme 
where the pressures and the chemical potentials are equal in both phases.
The quark-gluon plasma is treated as an ideal gas of quarks and gluons.
The excluded volume for the mass spectrum particles
is taken ${v_{\mbox{VdW}}}_H=4v_H$ where $V_H=\frac{4}{3}\pi r^3_H$ 
and the phase transition diagrams are displayed for 
excluded volume radii $r_{\mbox{H}}$=0.25, 0.35, 0.40 and 0.45 fm.   
The number of particles per volume is reduced as the size of the excluded volume increases. 
The proper Van der Waals excluded volume is chosen to fit the phenomenology 
and it is taken to be rather small 
${r_{\mbox{VdW}}}_H=4^{1/3} r_H$ $\approx$ 0.40-0.65 fm 
($r_H\approx$ 0.25-0.45 fm).
The classical Van der Waals excluded volume corresponds 
the phenomenological nucleon radius
$r_N\approx$ 0.40-0.65 fm.
On the other hand, the thick lines display the baryonic density for 
the hadronic matter consisting 
all the known hadronic spectrum particles 
in the particle data group book~\cite{databook2004}
as well as the highly excited Hagedorn bubbles. 
The excluded volume for the known hadronic particles 
is taken to be $r_H=0.25\mbox{fm}$, 
i.e. ${r_{\mbox{VdW}}}=0.40\mbox{fm}$.
The density of states for Hagedorn bubbles is taken as given in 
Eqs(\ref{statedensity1}), (\ref{Goren_dens_1}) and (\ref{Goren_dens_v}).
The initial volume fluctuation, $V_0$, for Hagedorn bubbles 
is taken $r_{\mbox{FB}}$=1.0 and the bubble's Van der Waals 
excluded volume $V_{\mbox{VdW}Q}=4V_0$. 
These Hagedorn bubbles are also supposed to be the fireballs (FB) 
could appear in the hadronic phase.
The bag constant for the Hagedorn bubble is taken 
$B^{1/4}=210 \mbox{MeV}$.
The internal symmetry for the color-flavor correlation in  
Hagedorn bag is chosen to be one for the massless flavorless 
and color singlet state, $\gamma$=1.5 with a smoothed 
volume fluctuation as given in Eq.(\ref{Goren_dens_v}).  
When Hagedorn bubbles are included in the calculation, 
the phase transition diagram is found self-consistently 
by searching the second singularity in the isobaric pressure.
The first singularity is the extreme right singularity point  
in the limit of infinite external volume given by 
Eqs.(\ref{isobaric_zs}) and (\ref{isobaric_den}),
while the second singularity is for the isobaric pressure 
of Hagedorn bubbles.
This second singularity is analogous to the condition for 
the phase transition determined by Hagedorn bubble's
mechanical instability arises when the bubble's internal pressure 
reaches the external one from below.
Due to this pressure instability, the bubbles start to expand rapidly 
and fill the entire space, 
consequently, quarks and gluons move freely and 
the phase transition to the quark-gluon plasma phase is reached.
Fig.~\ref{spectrum_rho_t} shows that although 
the Van der Waals gas of 
the known hadronic mass spectrum particles  
without Hagedorn bubbles gives a reasonable  
phase transition diagram for low and intermediate densities, it fails  
to predict the phase transition for large baryonic density 
and low temperature.
The finite size effect for the hadronic phase and quark-gluon plasma 
is also found important 
here~\cite{Rafelski_1980,Rafelski2002a,Pratt03a}.
Including Van der Waals gas for color singlet bubbles 
$\gamma=\frac{3}{2}$ and bubbles with 
more complicated internal color-flavor structure $\gamma=3.0$ 
shifts the phase transition line to higher baryonic densities.
The phase transition diagram exhibits a long ``tail'' 
at low temperatures and high densities. 
Furthermore, it is found that increasing the bubble's initial volume 
suppresses Hagedorn bubbles population. 
This supports the intuitive idea that Hagedorn bubble spectrum 
is the continuation for the excited mass spectrum for 
the known hadronic particles found 
in the particle data group book~\cite{databook2004} 
where Hagedorn bubble mass spectrum starts at the end 
of the hadronic mass spectrum particles. 

The above thin lines show the baryonic densities for 
the quark-gluon plasma phase 
$\rho^{QGP}_B$ above the phase transition 
point for hadronic matter consisting both the spectrum of known hadrons 
and Hagedorn bubbles with various internal color-flavor structures.
It is shown that $\rho^{QGP}_B$ above the phase transition point exceeds 
significantly that for the hadronic gas $\rho^{HG}_B$ 
consisting the known hadronic particles 
and Hagedorn bubbles with internal structure $\gamma=1.5$ below the phase 
transition point. 
This measurable discrepancy in the baryonic density between two phases 
indicates a first order phase transition and a discontinuity 
in the baryonic density at the point of phase transition. 
However, the case is rather different for hadronic matter consisting 
Hagedorn bubbles with internal color-flavor structure $\gamma=0.51$ 0.5833 
and 0.625. 
It is found that the split in baryonic density between two phases 
is tiny for $\gamma=0.51$ and small for 0.625.
This could indicate a continuity for the baryonic density 
at the point of the phase transition and a discontinuity for higher 
derivatives of the thermodynamical grand potential density.
Indeed the baryonic density split above and below 
the phase transition line sheds some 
information about the order of the phase transition and the 
discontinuity of the isobaric pressure and its first n-th derivatives.
Indeed, it is supposed that the higher order phase transition reduces 
the split size at the point of the phase transition significantly where 
the isobaric pressure is continuous while its $n^{th}$ derivative is 
supposed to be discontinuous. 
Furthermore, the numerical calculations show that this density split decreases
significantly as the internal structure decreases from $\gamma=1.5$ to $0.51$. 
This evidently indicates that the hadronic system consisting 
Hagedorn bubbles with internal structure $\gamma=1.5$ undergoes 
a first order phase transition 
while the system consisting bubbles with internal structure 
$\gamma=0.51$ undergoes a higher order phase transition 
and the $n^{th}$ higher derivatives of the 
thermodynamical grand potential density is continuous.

%
The phase transition diagram in the baryonic chemical potential 
and temperature  $(T-\mu_B)$ plane is displayed in 
Fig.~\ref{spectrum_mu_t}. 
The phase transition line for the hadronic gas of only the known hadronic 
mass spectrum particles without any Hagedorn state ends at 
the chemical potential $\mu_B\approx 1300$ MeV for low temperatures. 
However, the size effect for the excluded volume 
in $T-\mu_B$ plane is not apparent as for the ($\rho_B-T$) 
phase transition diagram shown in Fig.~\ref{spectrum_rho_t}. 
When Hagedorn bubbles are included, the phase transition diagram is 
shifted to larger baryonic chemical potentials for low temperatures. 
The phase transition diagram is modified to have a long tail 
at large baryonic chemical potentials and low temperatures. 
When Hagedorn bubbles appear in highly dense matter 
and low temperatures, 
the system prefers to stay in the hadronic phase which is dominated 
by highly compressed Hagedorn bubbles. 
Furthermore, the system experiences a smooth phase transition 
from hadronic gas dominated 
by the hadronic mass spectrum particles
to another one dominated by foam of dense Hagedorn bubbles.
The baryonic densities for both the mass spectrum particles
$\rho^{HG:spectrum}_B$
and Hagedorn bubbles $\rho^{HG:FB}_B$ in the hadronic phase below the phase 
transition line to QGP are displayed in Fig.~\ref{spectrum_d_m_t}.
We have also displayed the baryonic density $\rho^{QGP}_B$ for 
QGP above the line of the phase transition.
It is shown that the hadronic gas of the hadronic mass spectrum particles 
dominates the hadronic phase for low and intermediate baryonic chemical potentials 
with temperatures higher than 60 MeV while Hagedorn bubbles 
become the dominant one in the hadronic phase for the chemical potential 
$\mu_B$ exceeds 1200 MeV and the temperature falls below 60 MeV.

As outlined in the previous sections, the density of states for 
Hagedorn bubbles is given by the microcanonical ensemble 
for blobs of quarks and gluons with specific color-flavor symmetry.
The micro-canonical ensemble is derived from the inverse Laplace transform 
of the mixed grand canonical ensemble of quarks and gluons confined in a cavity.
The micro-canonical ensemble measures the mass spectral density 
Eqs.(\ref{laplacecanon1}) and (\ref{statedensity1}) for Hagedorn 
bubble with specific volume for a cavity with sharp surface. 
The volume fluctuation can be calculated by finding a solution 
for a bag model with a deformed boundary and 
this solution is, in general, not known.
The bag model with a smoothed boundary can be mimicked 
by using several assumptions to fit the phenomenology. 
For the nuclear shell model, the density of states is modified 
by using the ansatz of 
Strutinsky~\cite{Strutinsky67a,Strutinsky67b,Balian70a,Balian70b,Balian70c} 
in order to fit the nuclear data and is based on smearing 
the delta function $\delta(E-E_0)$ respecting energy conservation. 
  
Here, we are going to study different scenarios for incorporating volume 
fluctuation and the phase transition impact on extending our previous
discussions to more realistic systems. 
We emphasize in the following that the bubble's color-flavor internal 
symmetry and its volume fluctuation plays a vital role 
in determining the order of the phase transition at low baryonic chemical 
potential and high temperature and the shape of the phase transition line 
at large chemical potential as well. 
In order to emphasize the volume fluctuation's role 
in the phase transition we studied two models 
with different ansatz for the volume fluctuation.  
%

In the model of 
Gorenstein {\em et. al.}~\cite{Gorenstein83a,Gorenstein82a,Gorenstein83b}
hereafter denoted as model (I), 
the volume fluctuation is measured by 
differentiating the microcanonical ensemble
with respect to the bag's volume (see Eq.(\ref{Goreneq_1})). 
This approximation leads to a smoothed volume fluctuation 
as given in Eq.(\ref{Goren_dens_v}) 
which is independent on the bubble's volume 
and depends only on the bubble energy density. 
It is asymptotically equivalent to replacing the volume delta function 
$\delta(v-v_0)$ given in Eq.(\ref{mit_sharp1}) in the standard MIT bag  
by the power law $x^{3/4}$ where $x$ is the energy density. 
This variation is volume independent  
and subsequently it behaves the maximum volume fluctuation.
We display the phase transition diagram in Fig.~\ref{Goren_phase1} 
with various values for Hagedorn bubble's color-flavor 
correlation parameter, 
$\gamma$= 0.51, 0.5833, 0.625, 0.75, 1.0, 1.5, 2.0 and 3.0. 
The excluded volume for the mass spectrum particles is fixed 
to $r_{\mbox{H}}=0.30\mbox{fm}$ and Hagedorn bubbles' 
initial volume fluctuation starts from $r_{\mbox{FB}}=1.0 \mbox{fm}$.
The $\gamma=0.5$ case corresponds to colored bubbles 
while $\gamma=1.5$ corresponds to color singlet bubbles. 
For values of $\gamma>1.5$ Hagedorn bubbles have more 
complicated color-flavor structures but still remain in a color 
singlet state, while for $\gamma<1.5$ the bubbles have 
color-flavor structure with color nonsinglet components. 
The cases with $\gamma=$~0.51, 0.5833, 0.625, 0.75 or 1.0
do not alter the shape of the phase transition line 
for large baryonic chemical potential. 
The phase transition line ends at a baryonic chemical potential of 
$\mu_B\approx$~1350~MeV for zero temperature.  
The order of the phase transition changes from 1st order to 
$2^{nd}$, $3^{rd}$, $4^{th}$ and $n^{th}$ order for 
$\gamma=$~0.75, 0.625, 0.5833 and 0.51, respectively.
For $\gamma=$~1.5, 2.0 and 3.0, the phase transition diagram 
changes significantly and the phase transition line is shifted 
to larger baryonic chemical potential for low temperature.
The shift of the phase transition line increases drastically 
with $\gamma$. However, in the one component excluded volume 
approximation (see Fig.\ref{spectrum_mu_t}) this shift 
in the phase transition diagram to larger baryonic chemical potential 
at low temperature is less pronounced for $\gamma=1.5$ than 
that for the small and large excluded volume components approximation 
displayed in Fig.\ref{Goren_phase1}. 
The reason is that the bubbles 
in the one excluded volume component approximation
are effectively more suppressed than that for the small and large 
excluded volume components approximation. 
The system prefers to remain in the hadronic phase dominated 
by Hagedorn bubbles for large baryonic chemical potential rather 
than undergoes a phase transition 
to real deconfined quark-gluon plasma. 
When the medium becomes sufficiently hot, these Hagedorn bubbles 
undergo a phase transition to the quark-gluon plasma.    

We display in Fig.~\ref{Goren_pres1} the pressures for the gas of all known
nonstrange particles and the gas of Hagedorn bubbles versus 
the baryonic chemical potential $\mu_B$ at temperatures 
just below the phase transition diagram.
The sum of both pressures gives the total pressure for the hadronic phase. 
Most of the thermodynamical quantities such as particle multiplicities 
are derived from the isobaric pressure. 
The thermodynamical grand potential density is also given
in terms of the isobaric pressure.  
In the dilute hadronic matter, the pressure for Hagedorn bubble gas  
is suppressed and Hagedorn states unlikely appear in the hadronic phase. 
The gas of the hadronic mass spectrum particles 
is the dominant in the hadronic system for baryonic
chemical potential up to $\mu_B\approx$~1200~MeV. 
In this regime the colored bubbles with $\gamma=0.5$ are strongly suppressed.
If these bubbles appear because of the high thermal excitations, 
their gas pressure diverges and whenever they appear they explode 
and occupy the entire space and mix with 
the hadronic mass spectrum particles instantly. 
This mechanism leads to the so called crossover phase transition. 
The bubbles with color nonsinglet components 
and exponents $\gamma\le\frac{3}{4}$ 
may appear slightly just below the phase transition line 
and the system undergoes a higher order phase transition to quark-gluon plasma. 
These bubbles are also suppressed in the hadronic phase, their appearance 
just below the phase transition diagram allows a higher order phase transition.  
On the other hand, the bubbles with the net color singlet state 
do not appear for small and intermediate baryonic chemical potentials  
$\mu_B\le$~1000~MeV. 
When the baryonic chemical potential exceeds $\mu_B\ge 1000$~MeV, 
these Hagedorn bubbles appear and dominate the hadronic phase. 
For the small and intermediate baryonic chemical potential, 
Hagedorn states appear just below the phase transition line. 
Fig.~\ref{Goren_pres1} shows also that with increasing $\gamma$,
the pressure for Hagedorn bubbles increases and the system prefers 
to remain in the hadronic phase for dense and cold nuclear matter.
However, at high temperatures, the system undergoes 
a phase transition to quark-gluon plasma.

\begin{figure}
\includegraphics{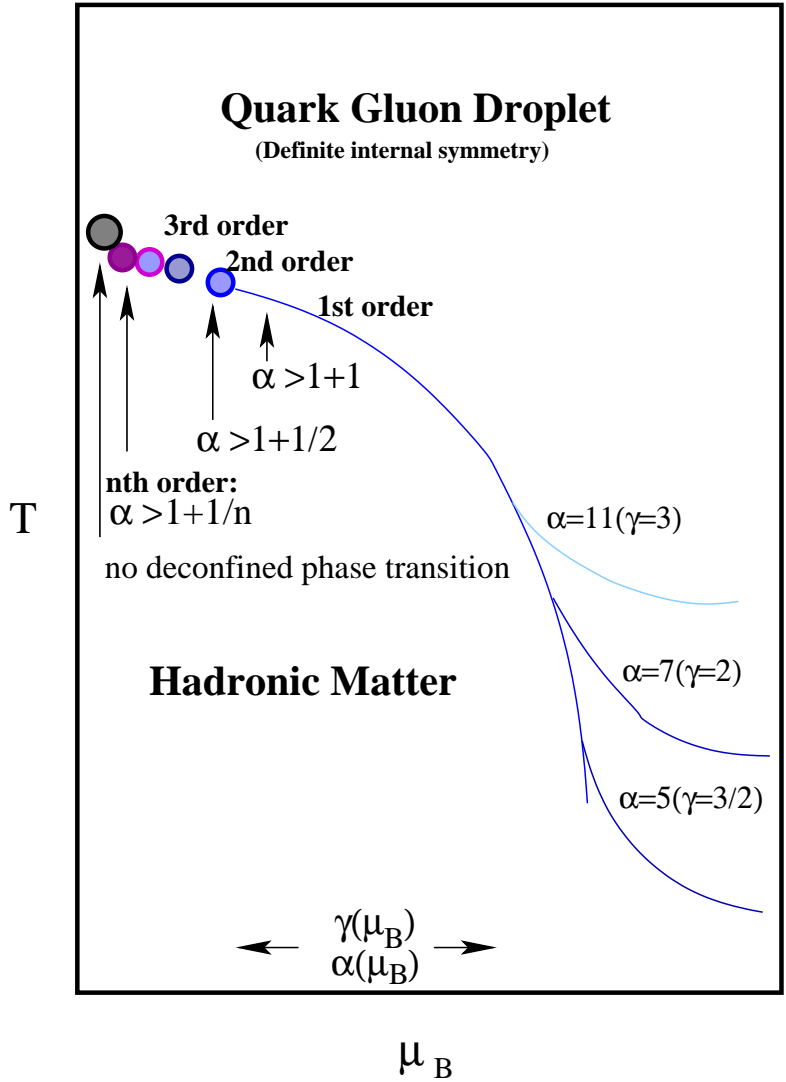}
\caption{\label{phase_sketch_goren}
The sketch for the phase transition diagram outlining the shape and order 
of the phase transition between the hadronic and quark-gluon plasma phases.  
The hadronic gas consists all of the known hadronic mass spectrum particles 
and Hagedorn bubbles with various values for color-flavor correlations 
as determined by the pre-exponential power factor 
$\alpha=4\gamma-1$.  
The sketch is for model (I) with the density of states given by  
Eq.(\ref{Goren_dens_1})  (e.g. Hagedorn bubbles with strong volume fluctuation).
The strength of the color-flavor correlation factor 
$\gamma\equiv\gamma(\mu)$ (e.g. $\alpha(\mu)$)   
changes the order of the phase transition for small baryonic chemical potentials 
and high temperatures while it modifies the shape of the phase transition 
at large chemical potentials and low temperatures.
The phenomenological assumption is that the exponent $\gamma$ is modified 
self-consistently with respect 
to the baryonic chemical potential  $\mu_B$. 
The variation of the phenomenological exponent $\gamma$ in the medium causes 
the existence of critical point in the phase transition diagram.}
\end{figure}

In order to summarize the scenario for the order and shape of 
the phase transition in model (I), 
the corresponding phase transition diagram is sketched in 
Fig.~\ref{phase_sketch_goren}. 
Let us discuss the influence of the parameter $\alpha=4\gamma-1$ 
on changes to the phase transition diagram by varying 
it with the baryochemical potential.
There is no phase transition for hadronic matter 
with $\alpha=1$ (e.g. bubbles structure $\gamma=0.5$) 
and the system undergoes a crossover phase transition 
to the quark-gluon plasma for dilute and hot matter. 
For slightly larger baryonic chemical potential, 
the system undergoes n-th order phase transition 
to quark-gluon plasma 
for $\alpha<2$ $\left(\gamma<\frac{3}{4}\right)$
as Hagedorn bubbles appear near the phase transition 
due to the thermal excitations.  
The first order phase transition takes 
place only for bubbles with $\alpha>2$. 
Hadronic matter consisting bubbles with $\alpha=2$ 
undergoes a second order phase transition while the hadronic matter 
consisting ones with $\alpha=3/2$ $(\gamma=0.625)$ undergoes 
a third order phase transition.
Therefore, the hadronic matter consisting color singlet bubbles, 
which has internal structure $\gamma=3/2$ for gluons 
and massless flavorless quarks undergoes 
a first order phase transition.  
For example, the bubbles with specific baryonic 
and mesonic internal structures have 
$\gamma=3/2$ and $\gamma=13/8$~\cite{Auberson86a}, respectively, 
and bubbles with 6 massless flavors and 3 colors with exact 
$SU_{6,t}\rightarrow SU(3)\times SU(3)$ have $\gamma=6$~\cite{Auberson86a}.
Therefore, it is expected that the strong color-flavor correlation 
will lead $\gamma\gg 1$ for three flavors with symmetry 
$SU(3)_c\times {SU(3)_f}_L\times {SU(3)_f}_R$ 
and for two flavors with symmetry 
$SU(3)_c\times {SU(2)_f}_L\times {SU(2)_f}_R$.

The basic question is we follow now at which extent do the color-flavor 
symmetry and the modification of the volume fluctuation affects the order 
and shape of the phase transition line. 
It is known from the nuclear shell 
theory~\cite{Strutinsky67a,Strutinsky67b,Balian70a,Balian70b,Balian70c}
that the deformation of the cavity boundary smears the density of states 
by smoothing the boundary surface.
Auberson {\em et. al.}~\cite{Auberson86b} 
have argued that within the MIT bag model 
the mass fluctuates around the initial bag volume 
and it is approximated by Eq.(\ref{emass_vol_aub}).
This procedure leads to the Gaussian-like volume fluctuation
(see Eq.(\ref{aub_v_fluct})).
The resulting density of states for Hagedorn bubbles is given 
in Eq.(\ref{aub_dens_m_v}).
The quark and gluon bubble's Gaussian volume fluctuation approach 
will be denoted as model (II).
The phase transition diagram in the $(T-\mu_B)$ plane 
is displayed in Fig.~\ref{Aub_phase1}
for various internal color-flavor structures 
$\gamma=0.5$, 1.0, 1.5, 2.0 and 3.0. 
The excluded volume for the hadronic mass spectrum particles 
is taken to be $r_{\mbox{H}}=0.30\mbox{fm}$.
The bubble's initial volume fluctuation starts 
from $r_{\mbox{FB}}=1.0 \mbox{fm}$.

Fig.~\ref{Aub_phase1} exhibits a first order phase transition for 
the colorless quark and gluon bubbles with $\gamma=\frac{3}{2}$ 
and also for bubbles with color nonsinglet components 
of $\frac{3}{2}>\gamma>\frac{5}{8}$,
a second order phase transition for bubbles $\frac{5}{8}\ge\gamma>\frac{1}{2}$. 
A third order phase transition is found for colored bubbles 
with $\gamma=\frac{1}{2}$. 
This means that the third order phase transition is the maximum order.
It takes place near the critical temperature $T\approx 155$ MeV
where the hadronic mass spectrum gas undergoes a third order phase
transition to real deconfined quark-gluon plasma.
%
Hagedorn bubbles with a strong color-flavor correlation 
$\gamma\ge\frac{3}{2}$ 
but with net color singlet states trigger always 
a first order phase transition. 
Furthermore,  the phase transition line for bubbles with color 
nonsinglet components and $\gamma<\frac{3}{2}$ are almost the same, 
although they undergo phase transitions of different order. 
The phase transition line drops at a chemical potential of
$\mu_B\approx 1350$~MeV for low temperatures, $T\le 100$~MeV. 
The interesting result to note in this scenario is, 
that the bubbles with specific internal color-flavor correlations 
and specific color components undergo a second order phase transition 
and that only colored bubbles undergo a third order phase transition.
We would like to remind the reader that a specific flavor correlation
has been introduced via the center of mass projection in order 
to reproduce the bootstrap density of states~\cite{Kapusta82b}. 
The Lorenz structure of the center of mass projection 
is not invariant in the hot bath and can be broken partially 
or completely. This mechanism might modify the density of states.
In the present model, we have assumed that this mechanism, in somehow,
is absorbed by the phenomenological color-flavor parameter $\gamma$.    
The story is rather different for color singlet quark and gluon 
bags (i.e. Hagedorn bubbles)  
but with complicated color-flavor internal symmetries.
Hagedorn bubbles with $\gamma=$~1.5, 2.0 and 3.0 modify 
the phase transition diagram significantly
and the system prefers to stay in the hadronic phase 
for large chemical potential and sufficiently low temperature. 
When the medium's temperature exceeds $T=100~\mbox{MeV}$, 
the bubble color-flavor symmetry becomes less important to modify 
the shape of the phase transition line although it plays the vital role
in determining the order of the phase transition.
Hence, the phase transition diagram appear to be almost the same 
for Hagedorn bubbles with various color-flavor internal symmetries.
This means that regardless of the bubbles internal color-flavor complications,
quark-gluon droplets or plasma is formed. 

We display in Fig.~\ref{Aub_pres1} the pressures for the gas of known
spectrum particles and the gas of Hagedorn bubbles versus the baryonic 
chemical potential $\mu_B$ at temperatures just below 
the phase transition line.
Hagedorn bubbles are strongly suppressed in diluted hadronic matter
and they don't appear at small baryonic chemical potential.
They likely appear just below the phase transition line
because of high thermal excitations of the medium.
Furthermore, when the baryonic chemical potential increases
and exceeds $\mu_B\approx 1000~\mbox{MeV}$, 
Hagedorn bubbles start to appear in the hadronic phase 
and their gas pressure increases significantly as 
the baryonic chemical potential increases. 
They becomes more dominant than 
the hadronic mass spectrum particle gas 
for highly compressed hadronic matter.

\begin{figure} 
\includegraphics{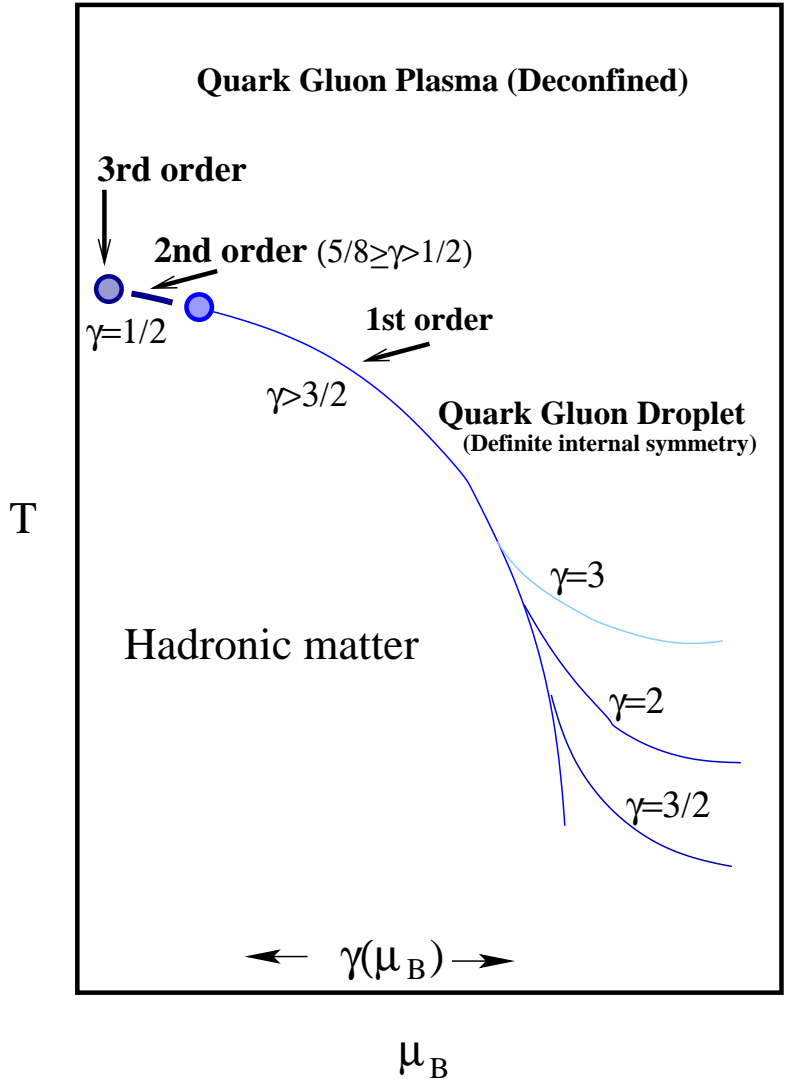} 
\caption{\label{phase_sketch_aub}
Sketch of the phase transition diagram as given in Fig.~\ref{phase_sketch_goren}
but for density of states for model (II) as given by Eq.(\ref{aub_dens_m_v}) 
(e.g. soft volume fluctuation). 
In this scenario the color-flavor correlation is determined by the power factor $\alpha=4\gamma-\frac{1}{2}$.
The points $\gamma=3/2$ and $1/2$ correspond to a second and 
a third order phase transition, respectively. 
A first order phase transition takes place only for $\gamma>3/2$.}
\end{figure}

The sketch of the phase transition diagram for model (II) 
is depicted in Fig.~\ref{phase_sketch_aub}.
The parameter $\alpha=\left(4\gamma-\frac{1}{2}\right)$ 
is assumed to be modified in the medium and increases as 
the hadronic matter is compressed.
In this scenario, the hadronic phase undergoes 
a third order phase transition to colored bubbles 
with $\alpha=1.5$ ($\gamma=0.5$) 
and forms real deconfined quark-gluon plasma in diluted and hot matter. 
The bubbles with color nonsinglet components with structure 
$2\ge\alpha>\frac{3}{2}$ (i.e. $\left(\frac{5}{8}\ge\gamma>\frac{1}{2}\right)$) 
have a second order phase transition 
while for $\frac{11}{2}>\alpha>2$ 
($\left(\frac{3}{2}>\gamma>\frac{5}{8}\right)$)
there appears a first order phase transition. 
The hadronic phase consisting Hagedorn states given by color singlet 
bubbles ($\gamma\ge 3/2$) undergoes a first order phase transition 
to quark-gluon droplets or plasma.
Increasing the value of $\gamma$, modifies
the phase transition diagram shape substantially in particular 
when the baryonic chemical potential becomes sufficiently large.  

The comparison between phase transition scenarios for models (I) and (II) 
shows that the modification of Hagedorn bubble's internal structure 
due to the interquark potential is important 
in order to draw and determine the order of 
the phase transition diagram.  
The interquarks potential modification could slightly 
deform the bag boundary and consequently the order 
of the phase transition 
even if the internal color-flavor symmetry remains intact. 

\begin{figure}
\includegraphics{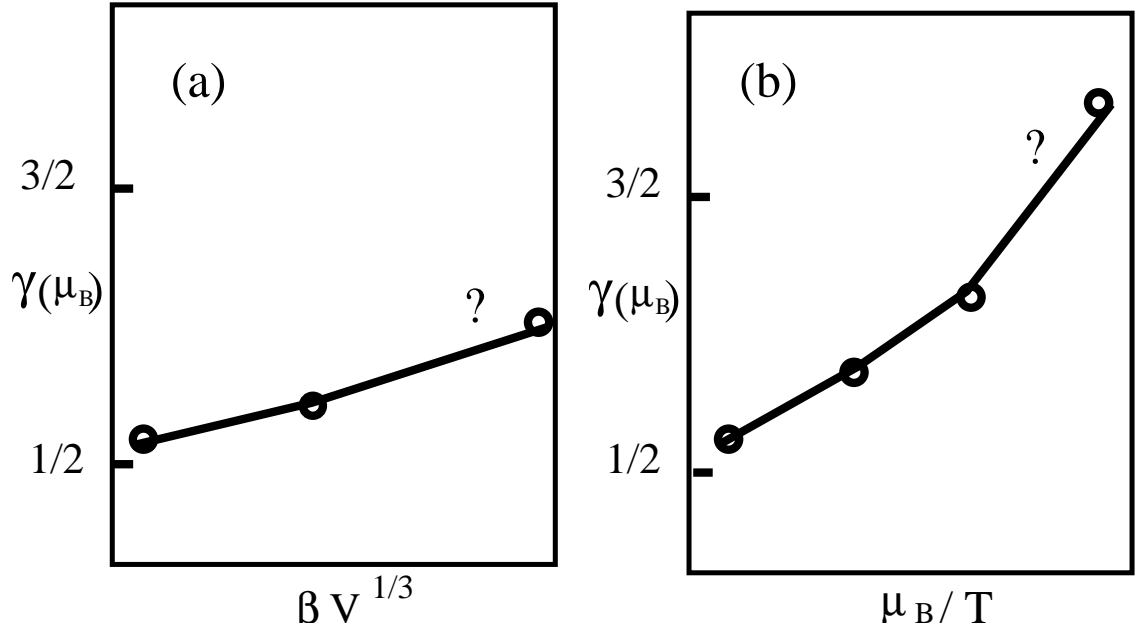}
\caption{\label{internal_sketch}
Rough Sketch of the dependence of Hagedorn bubble's 
phenomenological internal structure parameter 
$\gamma\equiv\gamma(\mu_B)$ on the volume, the temperature 
(or the energy in the microcanonical ensemble) 
and the chemical potential.}
\end{figure}

Naively, Hagedorn bubble's internal structure $\alpha$ (or $\gamma$) 
can also be modified in the medium.
The rough sketch for the dependence of the phenomenological Hagedorn 
bubble's internal structure $\gamma$ on the volume, 
temperature and chemical potential is depicted in Fig.~\ref{internal_sketch}.
Primarily insight on melting the frozen internal color degrees of freedom 
in the medium have been given by Elze and Greiner \cite{Elze84a}.

It is possible to go beyond the approximations of the previous models 
by considering higher order volume fluctuations or even finding 
a realistic hadronic bubble's wavefunction. 
In this context, it is reasonable to assume that 
Hagedorn bubbles are thermally excited 
to higher quantum states whenever they appear. 
These highly excited Hagedorn bubbles can then even evaporate or emit 
smaller hadronic bubbles. In this case, the volume fluctuation becomes 
stronger and the bubble's internal color-flavor structure is modified 
by smoothing the cavity boundary and consequently 
the density of states for Hagedorn states.  
When the phase transition takes place, the quark-gluon droplets continue 
to expand in the hot and diluted matter and then they eventually 
overlap with each other to form real deconfined quark-gluon plasma.
On the other hand, in dense and cold hadronic matter the situation 
is rather different, however. When the baryonic chemical potential 
increases, the hadronic bubbles agglomerate and merge to form bubbles 
which are likely to have higher color-flavor symmetry 
and larger baryonic number.  
Hagedorn bubbles with simple symmetries merge and 
form dense bubbles with higher 
color-flavor symmetries and soften the equation of state. 
They shift the phase transition to higher density and temperature.
At warm and large baryonic density, the system is dominated by Hagedorn 
bubbles.
When these dense Hagedorn bubbles expand and overlap with each other, 
the system will undergo a phase transition to quark-gluon droplets.
These droplets continue to expand and lose their internal color-flavor 
symmetry and then eventually they merge all together 
to form  quark-gluon plasma.

The baryonic number for each Hagedorn bubble 
increases when the system is compressed and cooled 
and more dense Hagedorn bubbles with larger baryonic 
numbers appear in the system.
At highly compressed matter, most of the baryonic 
density tends to concentrate in Hagedorn bubbles.
However, the bubble's size fluctuates smoothly 
and even shrinks in order to reduce 
the overlap effect with other hadronic bubbles.
Hagedorn Bubbles with low mass density merge 
with other bubbles to form denser ones.
When the maximum density is reached, 
the system undergoes a phase transition 
to Hagedorn bubbles foam.
For low temperature and very dense matter, 
the bubble's size can not shrink anymore due 
to the high constituent quarks and gluons pressure 
and the bubbles high surface tension 
in particular for cold matter. 
Hence the equation of state 
can be softened only by merging Hagedorn bubbles 
to form denser bubbles.
Furthermore, the thermal excitation dissolves 
the surface between the bubbles spontaneously 
and subsequently the Hagedorn bubbles foam collapses 
to form quark-gluon plasma.   

\section{Summary and conclusions}

We have studied the order, shape and critical point for 
the phase transition diagram for hadronic matter consisting of all 
the known nonstrange hadronic particles and the highly excited hadronic 
bubbles correspond Hagedorn states those existed in the extreme 
conditions. 
The basic assumption is that Hagedorn bubble's internal structure 
exponent $\gamma$ depends on the medium 
and may modify itself self-consistently with respect 
to $\mu_B$ in particular for temperature just below $T_c$.  
The exponent $\gamma$ is related to Hagedorn bubble's 
internal color-flavor potential interaction.
We have shown the order of the phase transition depends 
basically on the phenomenological exponent $\gamma$.   
We have demonstrated that both the volume fluctuation 
and the internal color-flavor structure for Hagedorn states 
play a crucial role in determining the order 
of the phase transition to quark-gluon droplets or plasma.
It is found that Hagedorn bubbles' quantum excitations
modify the volume fluctuation by smoothing 
the quarks and gluons cavity boundary and subsequently 
change the order of the phase transition. 
The excited bubbles with smoothed surfaces cause a higher order phase transition. 
The order of the phase transition can be changed 
from a lower one to higher ones for the dilute and hot matter.

The phase transition is understood in the following way.
Hagedorn bubbles are suppressed strongly 
in the dilute hadronic matter because the pressure 
of the hadronic mass spectrum particles is higher 
than the Hagedorn bubble's internal pressure. 
These bubbles appear, however, due to 
the high thermal excitations.
When their internal pressure exceeds the pressure of the gas
of hadronic mass spectrum particles and other Hagedorn bubbles 
they start to expand at temperature just below the critical one.
This expansion process depends on Hagedorn bubble's 
volume fluctuation and its internal color-flavor symmetry as well.
The bubbles with lower internal color-flavor symmetries 
are easily excited than those with more 
complicated internal color-flavor structures.
On the other hand, Hagedorn bubbles become more dominant 
for the large baryonic chemical potential. 
When the hadronic gas is heated, the system passes 
a phase transition to quark-gluon plasma.
 
Naively, the bubble's volume fluctuation depends on the reaction 
of the bubble's constituent quarks and gluons to the medium. 
It is expected the bubble's volume fluctuation becomes 
stronger in the dilute hadronic matter 
in particular near the critical temperature 
due to the high thermal excitations of the constituent particles. 
On the other hand, this volume fluctuation is supposed 
to be suppressed in the highly compressed matter 
in order to reduce the overlap effect among the hadrons. 
Two models for Hagedorn bubble's volume fluctuation 
have been considered for bubbles with various 
internal color-flavor symmetries. 
The first model is the Gaussian volume distribution function  
while the second one is the maximal volume distribution function, 
e.g. smearing the delta function $\delta(v-v_0)$ to volume independent 
function.   
It is shown that the strength of volume fluctuation 
is essential for a hadronic phase consisting bubbles 
with low internal color-flavor symmetries 
to undergo a higher order phase transition 
while the complexity of the bubble's color-flavor symmetry 
is essential to determine the shape 
of the first order phase transition line 
for the large baryonic chemical potential.   
The variation of the exponent $\gamma(\mu_B)$ in the medium 
for Hagedorn bubbles with specific internal color-flavor structures 
and the co-existence of the tri-critical point 
will be considered in the forthcoming work.

\begin{acknowledgments} 
I. Z. gratefully acknowledges support from the Alexander von Humboldt. 
He is also indebted to Walter Greiner for his encouragements and discussions.
The authors also thanks the Frankfurt Center for Scientific Computing.
The early stages of this work were supported by Bethlehem University. 
\end{acknowledgments}

\bibliography{harv_sept26}
\newpage 
\newpage 
\newpage
\newpage 
\newpage 
\newpage
\newpage
\newpage 

\newpage 

\newpage

\end{document}